1 DE DICIEMBRE DE 2018

# BRASIL

## INFORME PAÍS

JUAN B. GONZÁLEZ BLANCO
UNIVERSIDAD LOYOLA ANDALUCÍA
Grado en Economía y Relaciones Internacionales

# Índice

## 1. Introducción a la historia y demografía del país

Brasil es un país situado en Sudamérica, en la costa del Océano Atlántico. Es el quinto Estado con más territorio del mundo, con unos 8 millones y medio de kilómetros cuadrados. Al ser tan extenso, comparte frontera con casi todos los países de la Sudamérica continental: al norte con la Guyana Francesa, Surinam, Guyana y Venezuela; al oeste con Colombia, Perú y Bolivia; al sur con Paraguay, Argentina y Uruguay. Por otro lado, tiene unos 7.500 km de costa con el Océano Atlántico (CIA, 2018).

La historia de Brasil comienza mucho antes de la colonización y conquista por los europeos, pero es poco conocida porque se conservan pocas evidencias arqueológicas del período: sólo se sabe que estaba habitada por muchas poblaciones dispersas y étnicamente distintas, con gran variedad de lenguas. Según la mayoría de autores, el primer desembarco de europeos en Brasil se realizó en el 1500, aunque hay disputa historiográfica sobre si llegaron primeros los españoles o los portugueses (Izquierdo Labrado, 2003; Newitt, 2005). En 1494 se había firmado el Tratado de Tordesillas, que repartía los territorios de la América a conquistar en una parte de Castilla y Aragón y otra de Portugal: Brasil estaba dentro de la zona de influencia portuguesa (pero en una extensión mucho menor que la actual), y fue colonizada por este reino desde el siglo XVI. Tras dos siglos de intentos del resto de países europeos de conquistar territorios en el actual Brasil, en 1750 España y Portugal firman el conocido como Tratado de Permuta, que fijaría los límites de Brasil de forma parecida a las fronteras actuales.

En 1821, el príncipe heredero al trono de Portugal proclamó la independencia de Brasil y se coronó emperador del Brasil al año siguiente. El imperio duró hasta el año 1889, cuando un golpe militar impuso un modelo republicano federal, creando los Estados Unidos de Brasil.

Durante todo este período, Brasil se había convertido en una colonia azucarera y cafetera, basándose su economía en la agricultura de plantaciones esclavas, con mano de obra esclavizada en el interior del país por los llamados *bandeirantes* y con esclavos traídos a la fuerza de África. Se estima que fueron llevados unos 6 millones de esclavos africanos a Brasil hasta la abolición de la esclavitud en 1888, el doble de los que llevó Gran Bretaña a sus colonias en el mismo período (TASTD, n.d.).

En 1930, un nuevo golpe militar acaba con la república federal y coloca a Getulio Vargas en el poder. Al ver el progresivo avance de la izquierda en el país (liderado por el Partido Comunista Brasileño), en 1937 Vargas da un golpe de estado contra el diseño gubernamental que él mismo había creado, y comienza a gobernar como dictador. Otro golpe militar obliga en 1945 a Vargas a renunciar a sus plenos poderes y se declarará una nueva constitución, democrática, en lo que se llamó la República Nova. Esta República traerá mayores libertades individuales y progreso político, culminando en el proyecto de reforma agraria y social que pretendió llevar a cabo el presidente Joao Goulart. La oligarquía latifundista se opone a estas medidas, haciendo que los militares vuelvan a dar un golpe de estado en 1964 e instaurar un régimen de excepción que duró hasta 1985. Todo este tiempo gobernaron diferentes mariscales y generales del ejército brasileño, y la economía tomó gran impulso gracias a la inversión extranjera y a las grandes empresas nacionales (Skidmore & Fiker, 2003).

En 1984 cobra gran fuerza el movimiento Diretas Já, que luchó por las elecciones directas y libres y consiguió su objetivo en 1985. A partir de este año se articula un proceso constituyente que culminó en 1988 con la nueva Constitución Federal, que establece el sistema político brasileño tal y como lo conocemos hoy en día.

**Gráfica 1.** Desarrollo económico y demográfico

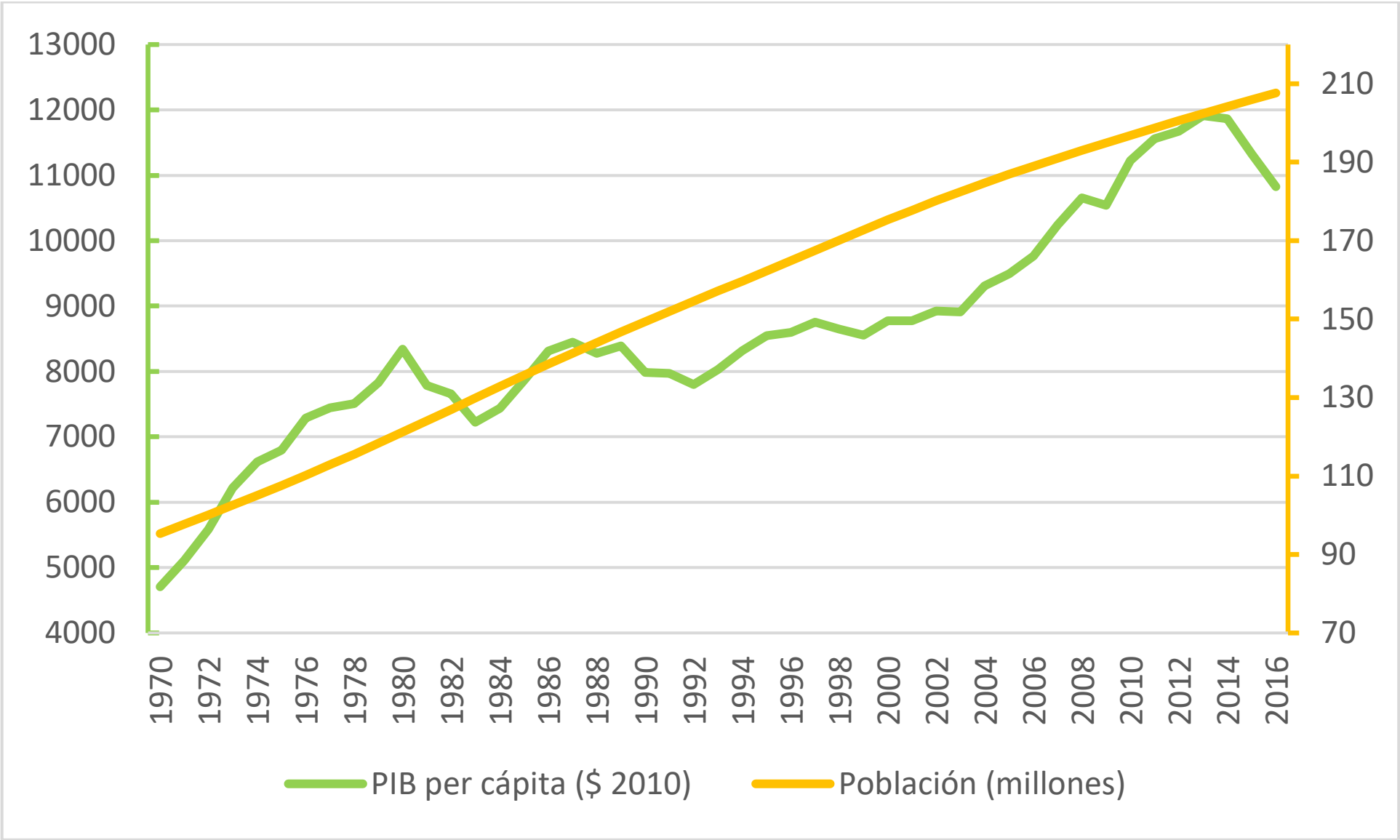

Fuente: elaboración propia a partir de (WB, 2018).

Desde 1970, Brasil se ha desarrollado con fuerza demográficamente: la mayor esperanza de vida y las altas tasas de natalidad durante el final del siglo XX han hecho que Brasil cuente hoy en día con más del doble de la población que tenía hace casi 50 años. Este desarrollo demográfico ha sido desigual en el territorio del país: la población urbana ha pasado de suponer un 55% del total a un 86%. Este hecho se ha notado principalmente en las grandes ciudades del país, como Sao Paulo, cuya área metropolitana alcanza los 20 millones de habitantes. Esto supone un gran desafío para el país, ya que la gran desigualdad que persiste en él hoy en día provoca que exista una gran densidad de población en las zonas más pobres de las ciudades: lo que se conoce como *favelas*.

Actualmente, más de 11 millones de brasileños viven hacinados en favelas[1], con unas condiciones muy precarias y altos índices de criminalidad y homicidios (IBGE, 2017a).

## 2. Marco institucional y socioeconómico.

### 2.1. Sistema político.

El sistema político de Brasil es un sistema presidencialista, organizándose el Estado en forma de República Federal. La República está formada por 26 estados y el Distrito Federal de Brasilia. La Presidencia de Brasil es elegida en las elecciones generales, que se realizan normalmente a dos vueltas, una vez cada 4 años.

El Estado divide sus poderes, como suele ser habitual, en 3 ramas. El poder ejecutivo es desempeñado por la Presidencia y su gabinete ministerial, con amplios poderes ya que se trata al mismo tiempo de la Jefatura del Gobierno y del Estado, de forma similar a los Estados Unidos. El poder legislativo está formado por dos cámaras de representantes (Congreso Nacional), a saber, el Senado Federal, con representación territorial (3 senadores por estado), y la Cámara de los Diputados, distribuidos de forma proporcional a la población de cada estado. Con respecto al poder judicial, caben destacar los dos principales tribunales: el Supremo Tribunal Federal (Tribunal Supremo-Constitucional) y el Tribunal Superior de Justicia.

En estos momentos, Brasil vive un momento bastante difícil políticamente. Tras 13 años de gobierno del Partido de los Trabajadores (PT), un partido de izquierdas liderado por un antiguo metalúrgico y líder sindicalista de las industrias de Sao Paulo, Luiz Inácio Lula da Silva, los partidos de la oposición llevaron a cabo en 2016 un *impeachment* contra la sucesora de Lula, Dilma Rousseff. El *impeachment* se produce por las acusaciones de la oposición de aumentar el gasto público en educación y políticas sociales por encima de lo acordado en los presupuestos (Bedinelli, 2016), y acaba con la subida al poder del Movimiento Democrático Brasileño de Michel Temer. Este movimiento de la oposición, sumada a su acusación de corrupción contra el expresidente Lula para frenar su vuelta a la presidencia (las encuestas, antes de que los juzgados le impidieran participar en las elecciones, le daban como ganador doblando en intención de votos a sus rivales (RTVE, 2018a)), provocaron un estallido social entre los defensores del expresidente y sus detractores. La situación es crítica en muchos puntos del país, especialmente en Río de

[1] Se puede ver la comparación entre la densidad de población en las favelas de Brasil y algunas de las 10 ciudades más densamente pobladas del mundo en la Figura 1 del Anexo Gráfico.

Janeiro, actualmente tomada por el ejército para mantener el orden frente a la violencia reinante (Londoño, 2018).

En medio de esta situación, el presidente Temer ha aprovechado para iniciar una serie de reformas estructurales en el país: privatizaciones a gran escala (incluso de la eléctrica pública brasileña Eletrobras), imposición de un techo de gasto muy ajustado, una reforma laboral que debilita los derechos laborales y a los sindicatos (y que ha provocado 3 huelgas generales), y una frustrada reforma de las pensiones para llevarlas a un régimen de capitalización (EFE, 2018a; Hermida, 2018).

Esta situación se espera que se estabilice tras unas elecciones en las que el Partido de los Trabajadores partía con desventaja al haber sido anulada la candidatura de Lula, que como hemos visto contaba con un apoyo mayoritario. Frente al PT, se encontraba Jair Bolsonaro, el homófobo y ultraderechista candidato por el Partido Social Liberal, llamado por los medios el "Trump brasileño". Este era el candidato favorito por los mercados (Avendaño, 2018), y finalmente el vencedor de los últimos comicios. Jair Bolsonaro será el presidente de Brasil hasta 2022, con un programa político no muy definido más allá de la intención de dar marcha atrás a las medidas sociales del PT y de incluir en su gobierno a "cuatro o cinco generales" y poner la gestión económica en manos del neoliberal Paulo Guedes (EO, 2018; San Román, 2018).

## 2.2. Principales reformas estructurales y políticas económicas recientes

Las principales políticas económicas llevadas a cabo por los últimos gobiernos (Lula da Silva y Rousseff, en los últimos 15 años) suponen un viraje con respecto a las políticas liberalizadoras de los años 90. Con la inflación controlada mediante unos altos tipos de interés y una política monetaria restrictiva a través del llamado Plan Real desarrollado por los anteriores gobiernos (ver Gráfica 4), el gobierno *lulista* propuso un plan de reindustrialización basado en una mayor inversión en infraestructuras (*Política de Aceleraçao do Crescimento*), en la recuperación del papel del Estado Federal como organizador e impulsor de la inversión industrial (recuperando agencias gubernamentales que se habían eliminado durante la década anterior, como el *Conselho Nacional de Desenvolvimento Industrial* y el *Conselho de Desenvolvimento Econômico e Social*) y la búsqueda de una mayor competitividad internacional y un aumento de las exportaciones (Bachiller Cabria, 2012).

Los resultados de estas políticas de industrialización y de mejora de la balanza comercial las discutiremos más adelante. No obstante, cabe destacar otro de los objetivos de la política económica del PT: la lucha contra la pobreza en Brasil. Para su consecución, el gobierno aumentó en gran medida el gasto social, especialmente en educación y en sanidad, además de aumentar el salario mínimo y el presupuesto destinado a programas de transferencia directa de renta. El más destacado de estos es la Bolsa Familia, que transfiere cierta renta a las familias más pobres si cumplen con las condiciones de escolarizar a sus hijos (con el fin de reducir el trabajo infantil, que en los años 90 llegaba a la cifra de 1 de cada 5 niños de entre 10 y 14 años (SEDLAC, 2018)) y vacunarles (Binder, 2014).

**Gráfica 2.** Algunos indicadores sociales durante el gobierno del PT

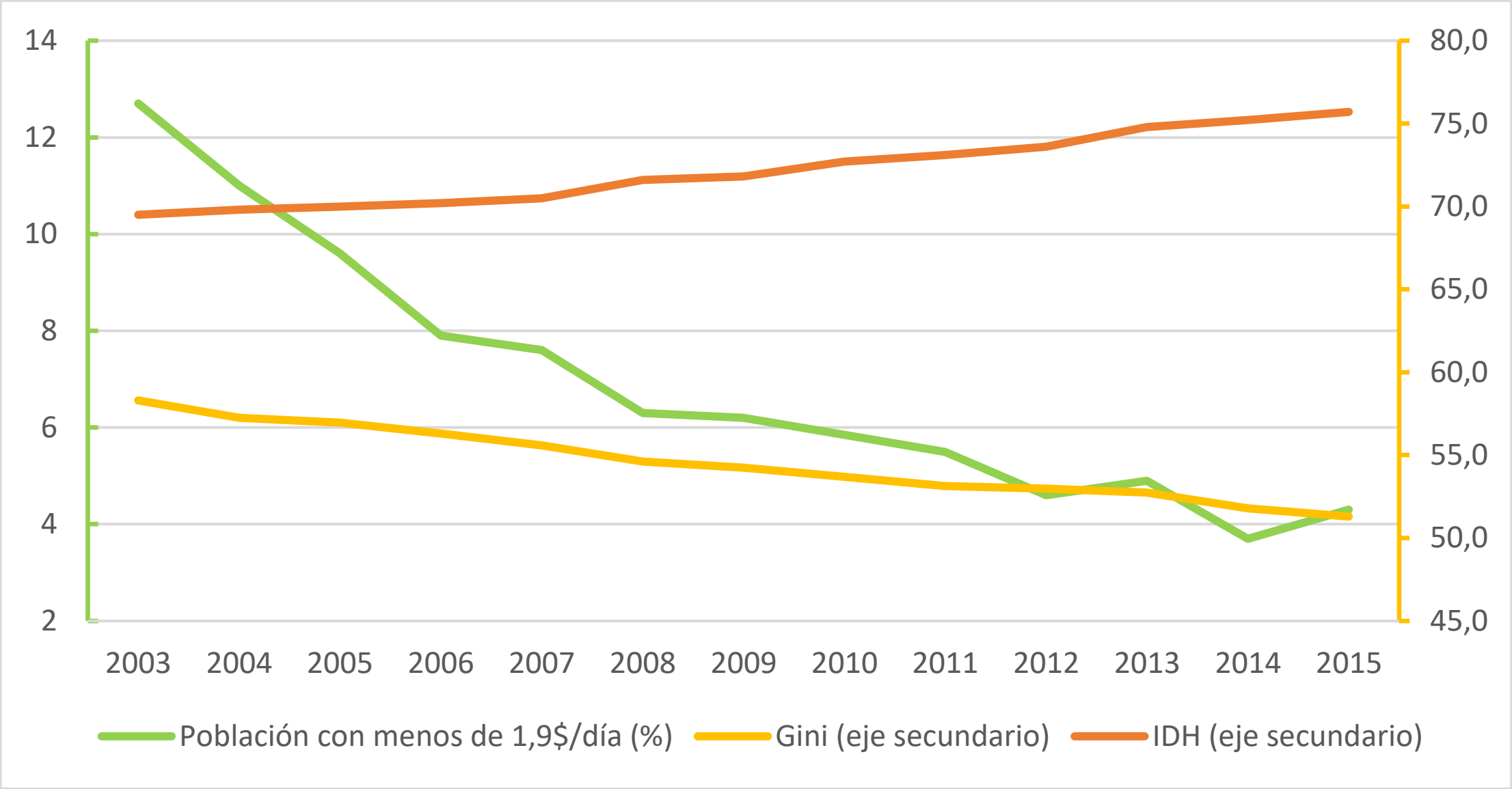

Fuente: elaboración propia a partir de (IPEA, 2018; UNDP, 2018; WB, 2018)

El gran desarrollo económico de Brasil durante los últimos años, además de las medidas redistributivas que impulsó el gobierno del PT, ha resultado en una gran reducción de la pobreza: unos 30 millones de personas salieron de la pobreza durante este período (BBC, 2018b; IPEA, 2018). La desigualdad medida con el Gini también se ha reducido ligeramente, pero no ha pasado de una bajada 'cosmética' ya que el llamado "pacto social lulista" implicaba intentar mejorar la situación de los más pobres del país pero sin tocar los intereses de la clase dominante: el 10% más rico sigue teniendo más del 40% de la renta nacional, aunque se haya reducido desde un 46% (WB, 2018). Cabe recordar que, pese a esta leve mejoría, en 2015 Brasil seguía siendo el tercer país más desigual del mundo según el índice de Gini.  A su vez, el Índice de Desarrollo Humano ha crecido unos 6 puntos, pero esto no se debe a medidas del gobierno ya que llevaba creciendo desde los años 90 (ver Figura 2 Anexo Gráfico). De todas formas, Brasil sigue estando en este indicador multidimensional por detrás de países como Chile o Argentina, e incluso de economías mucho más reducidas, como Cuba o Uruguay (UNDP, 2018).

**Gráfica 3.** Salario mínimo mensual y desempleo durante el gobierno del PT

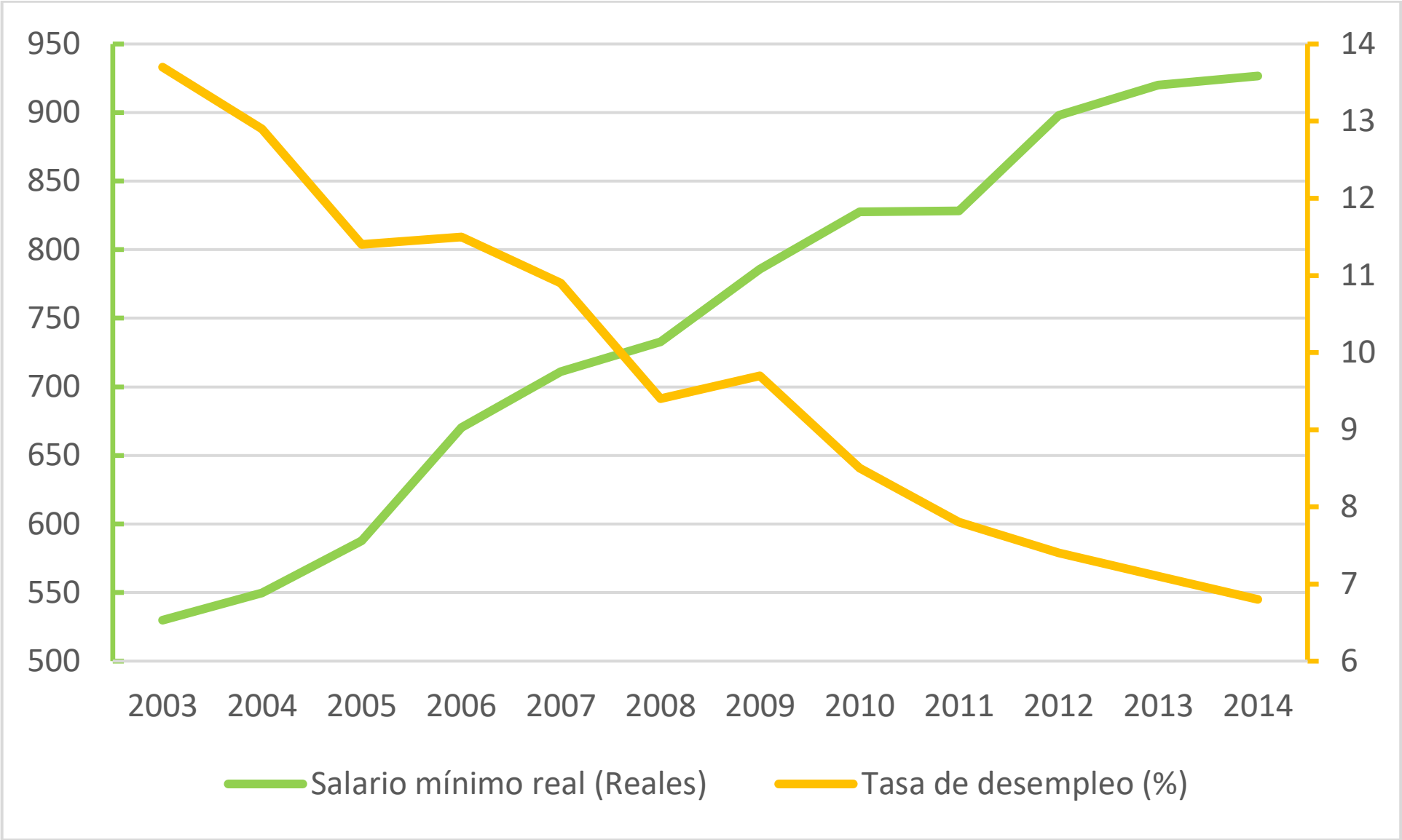

Fuente: elaboración propia a partir de (IPEA, 2018; WB, 2018)

Como podemos observar en la Gráfica 3, durante los últimos años, el empleo ha mejorado considerablemente en los últimos años, prácticamente reduciéndose a la mitad la tasa de desempleo en una década. A su vez, el salario mínimo real ha crecido un 69% en ese mismo período. La gran coyuntura económica de Brasil durante este tiempo, cuando las exportaciones crecen de forma continua y la inserción exterior mejora, como veremos más adelante, permitió un crecimiento de los salarios y del empleo sostenido, que llevó a las reducciones de desigualdad que antes comentamos.

Esto puede resultar sorprendente si tenemos en cuenta que la teoría económica ortodoxa considera que el aumento del salario mínimo suele traer aparejada una reducción del empleo. En el caso de Brasil, algunos economistas han explicado que se debe a la particular coyuntura económica, con un optimismo en las expectativas bastante alto (Joao Saboia, 2016; João Saboia & Hallak Neto, 2018). A ello se le suma que una gran parte del empleo creado durante esta época estaba localizado en la esfera legal, con lo que la economía sumergida pasó de un 21% a un 16% sobre el PIB en esta época ("América Económica | La economía sumergida de Brasil supone más del 16% del PIB del país," 2014). No obstante, esta reducción de la desigualdad y el crecimiento del empleo está limitado por una serie de factores, como el reducido crecimiento económico y la alta tasa de interés del país (Piketty, 2014).

Con la llegada al poder de Michel Temer, como hemos visto, se ha dado marcha atrás a buena parte de las medidas sociales del PT, con el objetivo de reducir el gasto público, flexibilizar el mercado de trabajo, y abrir el país a la inversión extranjera mediante la privatización de gigantes empresariales como Eletrobras o Petrobras (no llevada a cabo finalmente). Sin embargo, no se redujo el gasto en el programa Bolsa Família debido a su alta popularidad.

## 2.3. Descripción de la organización económica y regulación de los mercados de bienes y servicios y de trabajo.

La organización económica brasileña es incluso más compleja que la de sus vecinos sudamericanos. El intrincado sistema político-económico de la República Federativa, con varios niveles de gobierno muchas veces solapados entre sí y con grandes diferencias regulatorias entre estados, dificulta en gran medida el funcionamiento del mercado de manera uniforme. Además, la gran importancia de la empresa pública en sectores clave como son la banca y la energía, sin entrar a enumerar las más de 140 empresas participadas mayoritariamente por el Estado, provocan grandes distorsiones en la competencia de las empresas, con monopolios en algunos sectores como el petrolero (OECD, 2018).

Los precios, como hemos visto anteriormente, son uno de los grandes problemas estructurales que afronta Brasil. La inflación lleva golpeando al país desde hace medio siglo y, aunque se haya estabilizado tras el Plan Real como veremos más adelante, desde 2010 a 2016 no bajó del 5%, situándose en más de un 9% en 2015.

**Gráfica 4.** Inflación (% anual) (escala logarítmica)

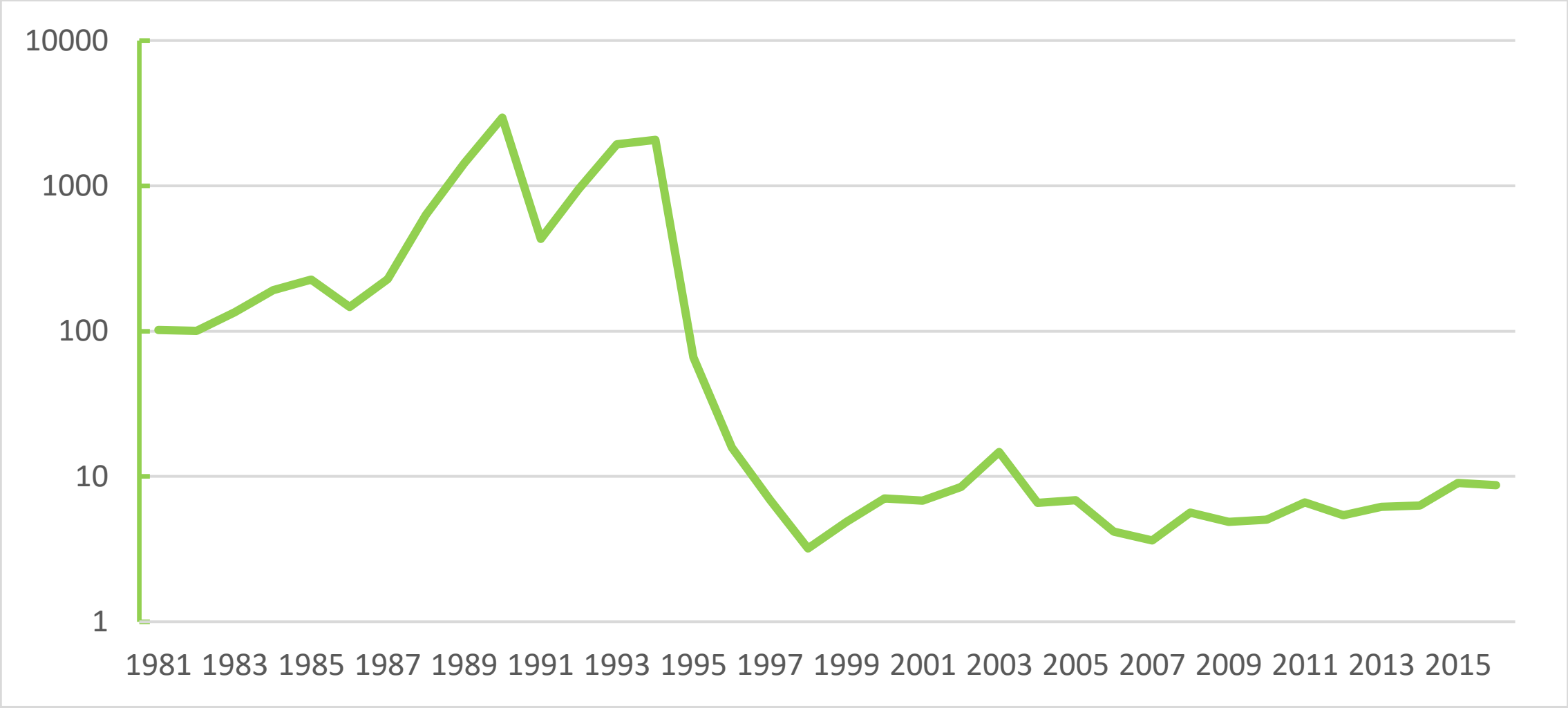

Fuente: elaboración propia a partir de (WB, 2018)

Con respecto a la estructura de la propiedad, habiendo tratado ya la gran desigualdad que se impone en Brasil, destaca en especial la estructura agraria de la propiedad de la tierra. Esta desigualdad es aún mayor que la de la renta, porque el 0,006% de los propietarios controla el 20% de la superficie cultivable del país. La mitad de ella está en manos del 0,76% de los propietarios, mientras que más de 4 millones de familias de campesinos no tienen tierra, en el país que concentra una quinta parte de todas las tierras de cultivo del mundo. De esta forma, el coeficiente de Gini agrario (distribución de la tierra) se situó en 86 puntos en 2014 (Dataluta, 2017; Fernandes, 2015; Lucía, 2016).

La pésima distribución agraria ha llevado a la lucha campesina desde finales del siglo XX, con grandes movimientos de campesinos sin tierra, entre los que destaca el Movimiento de los Trabajadores Rurales Sin Tierra, vinculado a la Comisión Pastoral de la Tierra. Estos movimientos, con gran influencia entre los campesinos, llevan a cabo ocupaciones de tierras y luchan por una reforma agraria que no llega (ver Figura 3 del Anexo Gráfico). Los conflictos por la propiedad de la tierra se han saldado con la muerte de cientos de militantes de estos movimientos y el asesinato de algunos de sus líderes (Aranda, 2013).

Siguiendo con la lucha de los más desfavorecidos del país, la organización de la negociación colectiva y los sindicatos es un modelo híbrido entre la libertad sindical y el corporativismo. La legislación sindical en Brasil no ha variado demasiado a lo largo del último siglo, siendo la base legal de la regulación laboral la Carta del Trabajo promulgada en 1943 (que a su vez toma como ejemplo la Carta del Lavoro de Mussolini de 1927). No se reconoce más que un sindicato por sector y por región (municipal), lo que evita que haya competencia entre sindicatos pero también imposibilita que estos se organicen a nivel nacional (Bragança de Vasconcellos, 2013). Así, los trabajadores parten en una situación aún más desfavorable en las negociaciones laborales, y muchos de ellos ni siquiera se afilian por estos motivos, haciendo que Brasil sea uno de los países con menor densidad de trabajadores sindicados, como vemos en la Gráfica 4. De hecho, la regulación sindical es tan negativa que Brasil no ha podido ratificar el Convenio sobre Libertad Sindical de la Organización Internacional del Trabajo (ILO, 2018a).

**Gráfica 5.** Porcentaje de trabajadores sindicados (2016).

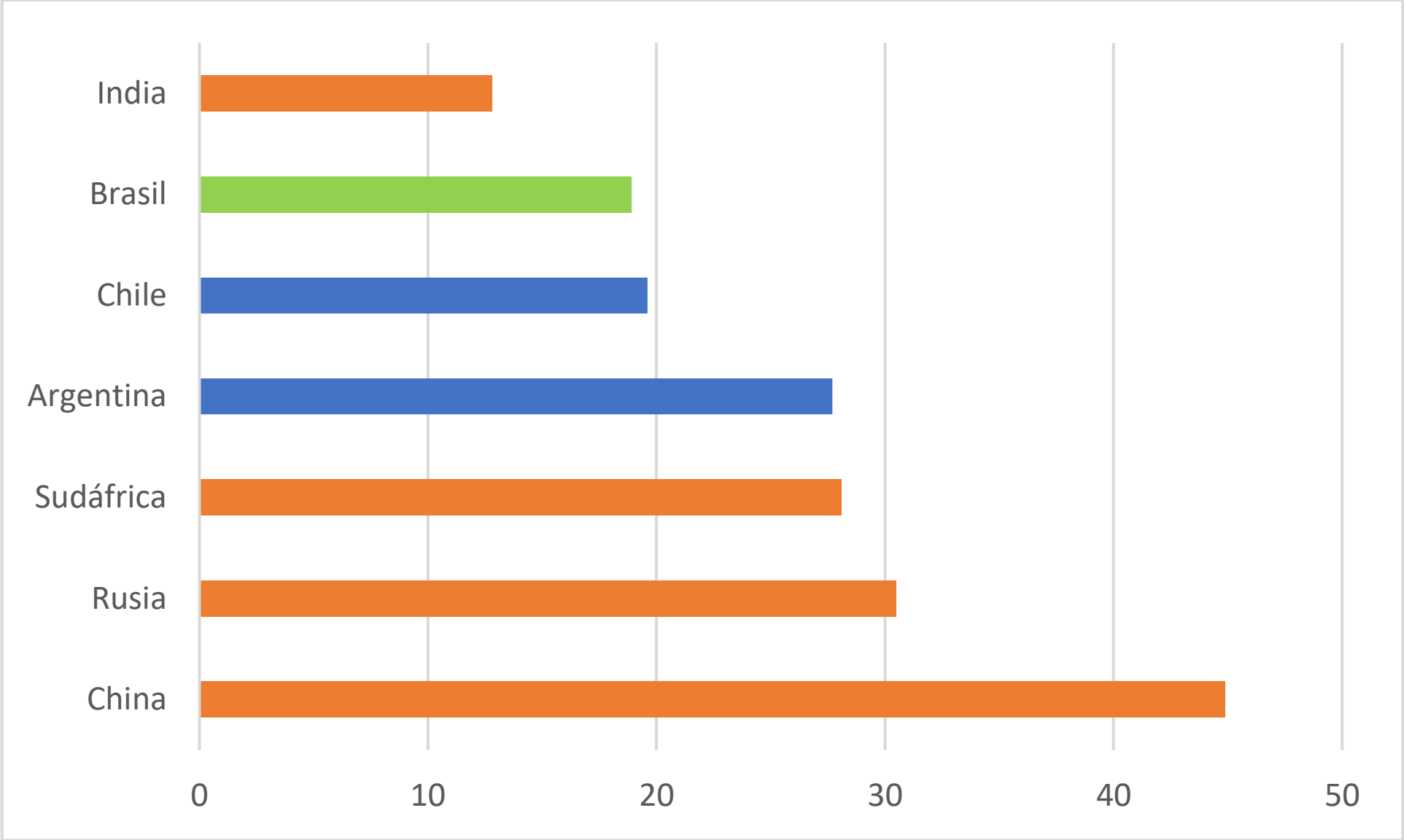

Fuente: elaboración propia a partir de (ILO, 2018b)

*Nota: de aquí en adelante, los BRICS aparecen en naranja, los países sudamericanos en azul, y Brasil en verde.*

# 3. Análisis desde el lado de la oferta.

## 3.1. Dotación de recursos naturales, trabajo y capital.

Brasil tiene una gran dotación de recursos naturales. Veamos en primer lugar la estructura de la producción energética del país.

Actualmente, Brasil es el noveno máximo productor de petróleo del mundo, por encima de países como Kuwait, Qatar o Venezuela (CIA, 2018). Durante 2017 produjo cerca de mil millones de barriles de petróleo, y cuenta con unas reservas probadas de 12 mil millones de barriles más (ANP, 2018). De toda su producción del último año, Brasil dedicó a la exportación más de 380 millones de barriles, con lo que consiguió divisas por valor de más de 16 mil millones y medio de dólares. Estos números están lejos del récord alcanzado en 2011, cuando las exportaciones de crudo superaron los 20 mil millones de dólares, debido a los altos precios (*op. cit.*). Actualmente, el petróleo (crudo y refinado) es el tercer producto en volumen de exportación en Brasil (OEC, 2018).

Esto no siempre fue así. De 1970 a 1989, la principal fuente de la producción energética de Brasil era la leña, que se obtenía tanto del Amazonas como de las grandes selvas atlánticas (*matas* *atlánticas*). La transición (no precisamente hacia energías renovables) en la oferta energética brasileña ha sido gradual pero imparable: en 2017, la producción de petróleo ya suponía el 45% del total.

**Gráfica 6.** Fuentes de producción de energía primaria en 1970 y en 2017.

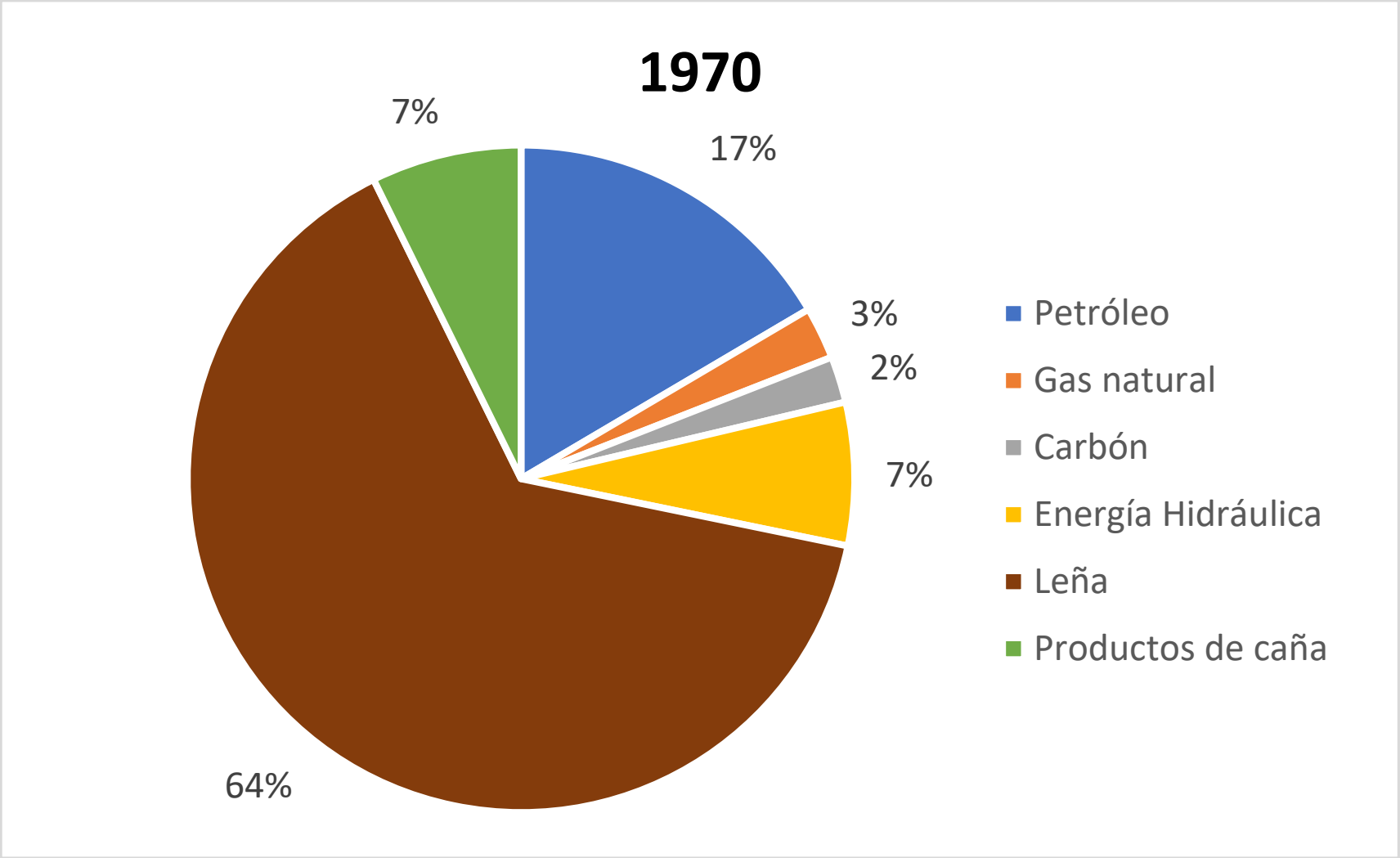

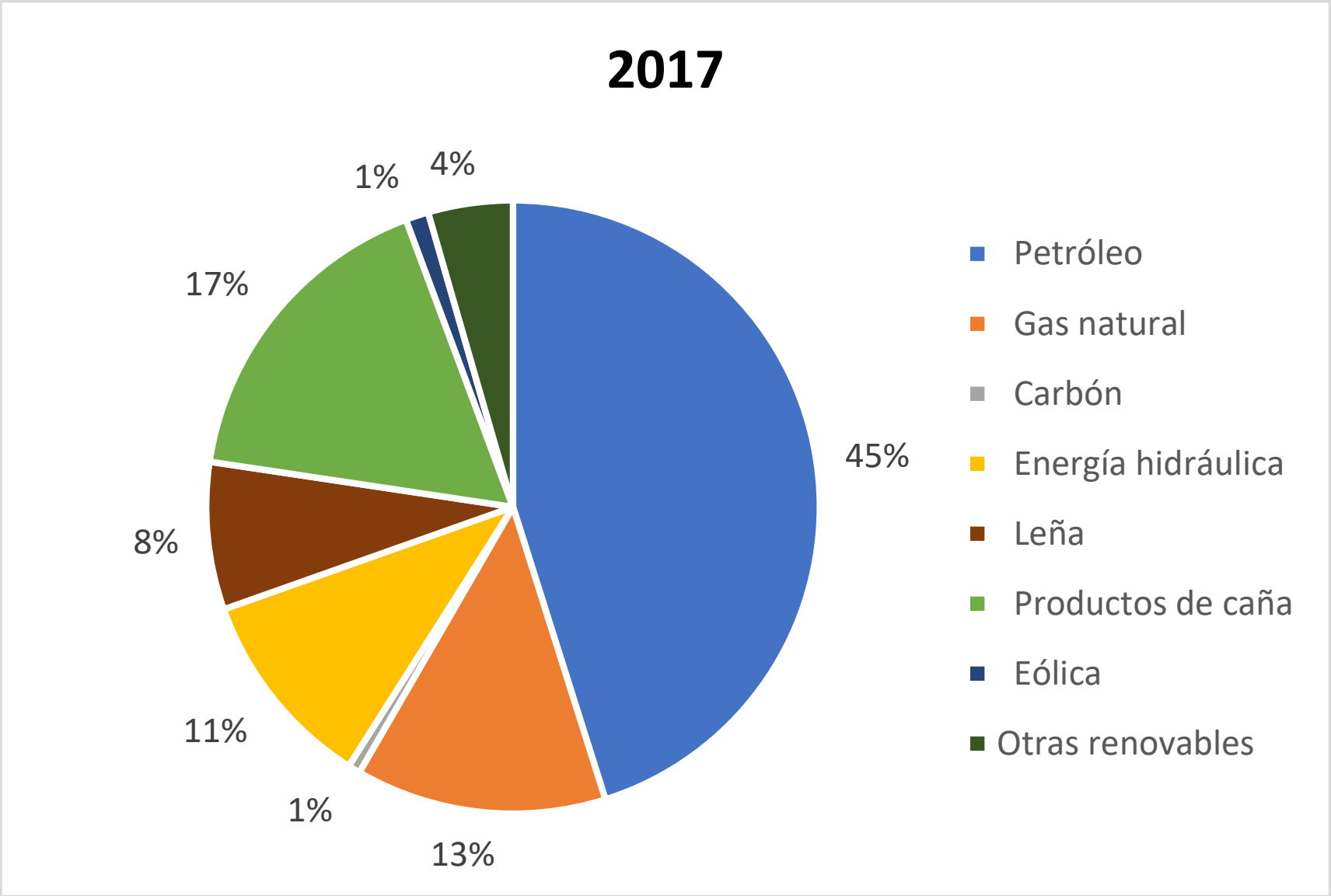

Fuente: elaboración propia a partir de (MME, 2018)

Esta transición hacia una economía productora y exportadora de energía ha acabado con uno de los problemas que lastraban la balanza comercial del país, que le hacía muy vulnerable ante fluctuaciones de los precios energéticos y contribuía a la hiperinflación que sufría. Como podemos ver en la Gráfica 7, el incremento que se observa en la producción petrolífera ha conseguido acabar con la gran dependencia energética que arrastraba el país desde los años 70.

**Gráfica 7.** Reducción de la dependencia energética gracias al petróleo.

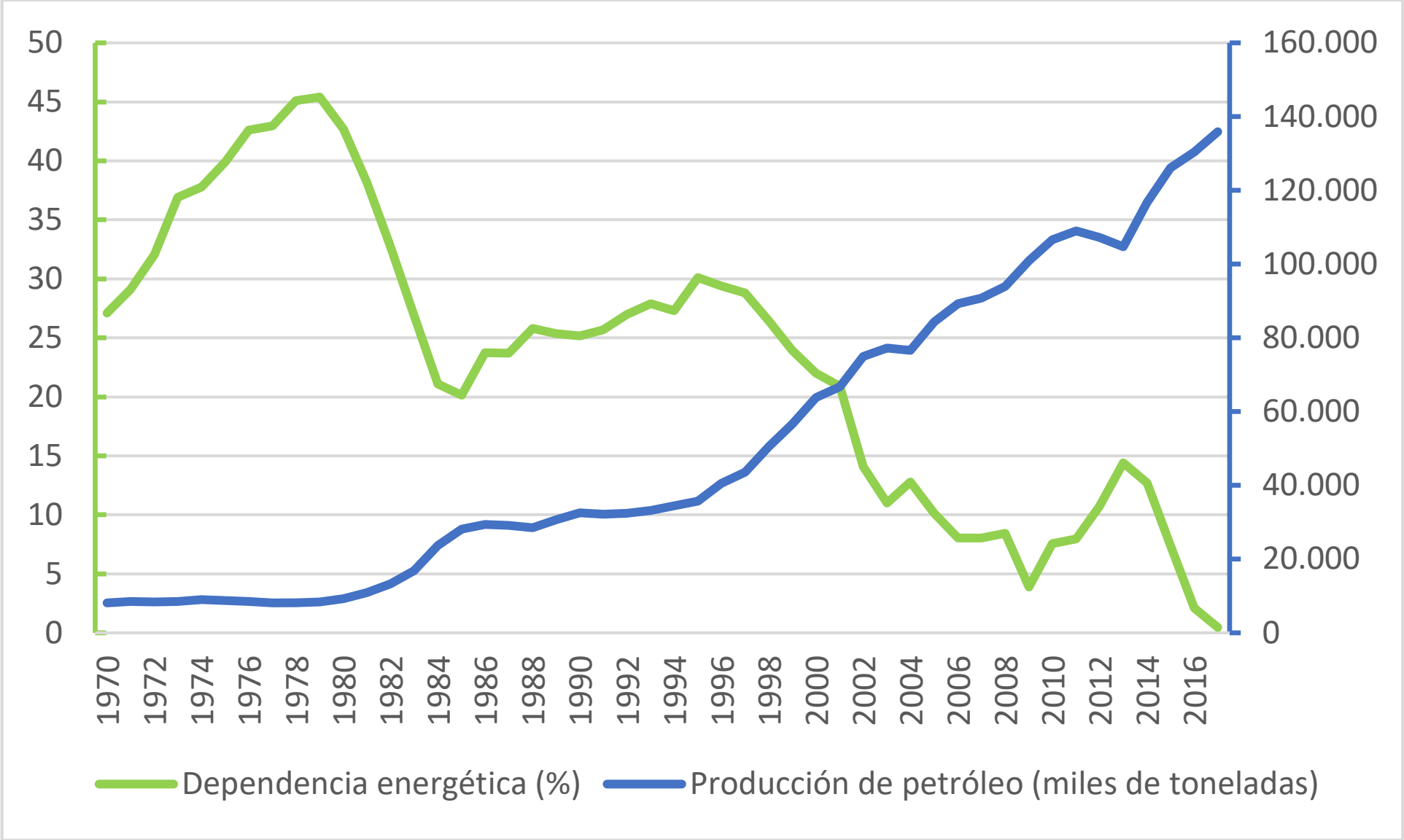

Fuente: elaboración propia a partir de (MME, 2018)

Además, es necesario remarcar que esta no es la única transición energética que ha llevado a cabo Brasil. En 1970, el 35,5% del consumo energético lo realizaba el sector residencial (principalmente de leña y electricidad), mientras que la industria y el transporte no llegaban al 50% (MME, 2018). En 2017, el panorama ha cambiado por completo: sólo el 9,6% del consumo lo realizan los hogares, mientras que dos tercios del total es consumido por la industria y el sector del transporte (ver Figura 4 del Anexo Gráfico).

Como hemos visto con la producción energética, Brasil siempre ha explotado sus recursos naturales, sin tener en cuenta en la mayoría de los casos las repercusiones ecológicas de sus acciones. Por ejemplo, la deforestación, que durante los años 70 y 80 se realizó con el objetivo principal de obtener madera (para leña y para producción de carbón vegetal y papel), pero continuó tras el auge de la extracción de petróleo con un objetivo diferente: el aumento de la superficie cultivable. Como podemos observar en la Gráfica 8, desde 1990 (desde que existen datos en el Observatorio de la Tierra del gobierno brasileño), se han deforestado más de 550.000 kilómetros cuadrados de selva (la inmensa mayoría en la Amazonia), más que la superficie de países como España o Alemania (OBT, 2018) [2].

[2] En la figura 5 del Anexo Gráfico se puede ver la deforestación amazónica en imágenes por satélite.

**Gráfica 8**. Deforestación y producción de soja.

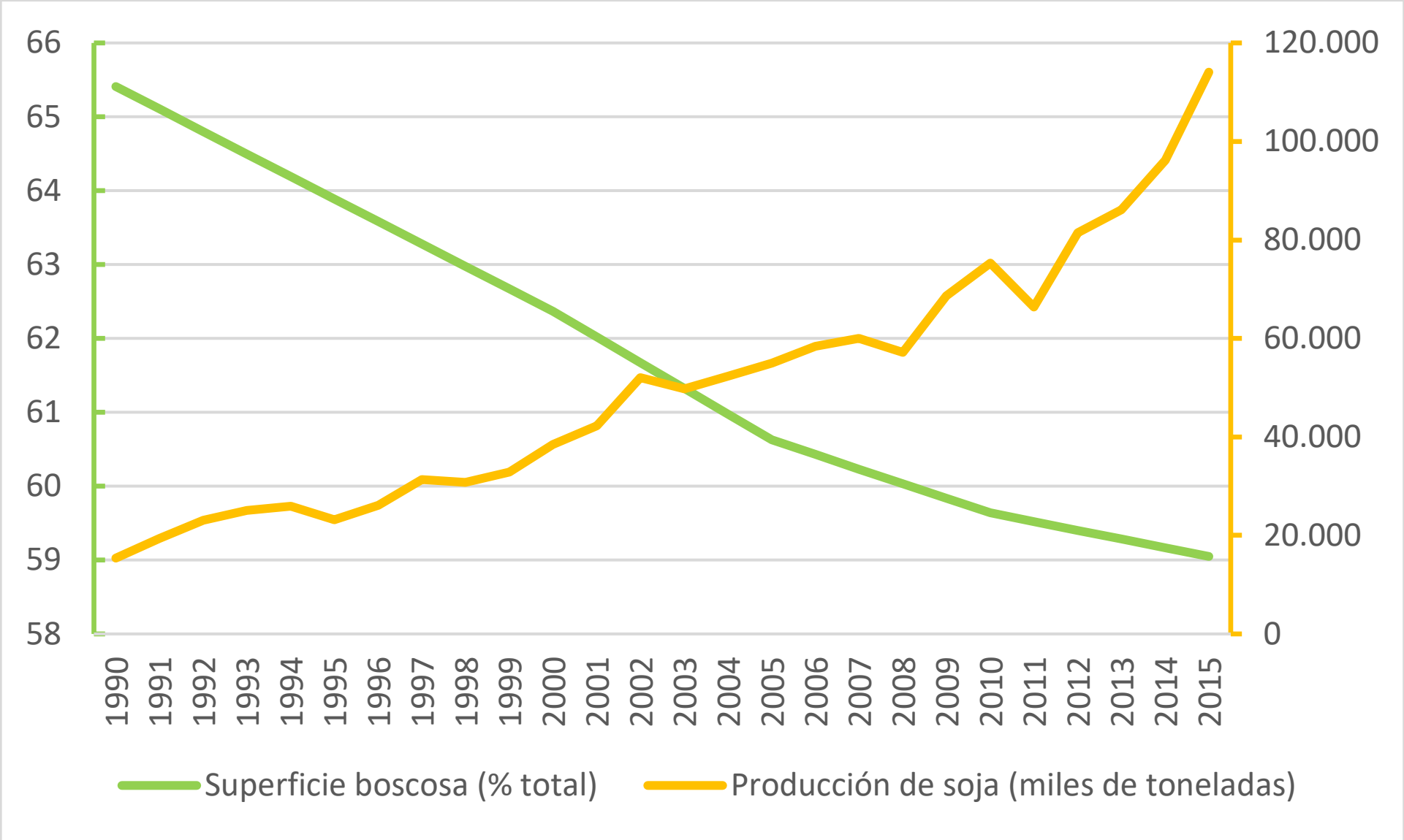

Fuente: elaboración propia a partir de (Conab, 2018; WB, 2018)

Este aumento de la superficie cultivable se ha dedicado principalmente a desarrollar la producción de 3 cultivos: el maíz, la caña de azúcar y, por encima de todas ellas, la soja. La soja es el producto agrario estrella de Brasil en la actualidad: tras más que triplicar el área dedicada a este cultivo desde 1991 (aumento del 266%), en 2017 su producción alcanzó los 34 mil millones de dólares (MAPA, 2017). Le sigue el maíz, cuya producción se ha duplicado en los últimos años impulsada por el crecimiento de su precio debido a la demanda para la producción de "biocombustibles": mientras que en el año 2000 el precio se situaba en 75$ por tonelada métrica, en 2012 se alcanzó el máximo histórico con unos precios de 333$ (IM, n.d.). Finalmente, la caña de azúcar, el producto estrella junto con el café durante los siglos XIX y XX, sigue teniendo gran importancia para el país, con unos números similares al maíz.

**Gráfica 9.** Producción de los principales cultivos, 2017 (millones de US$)

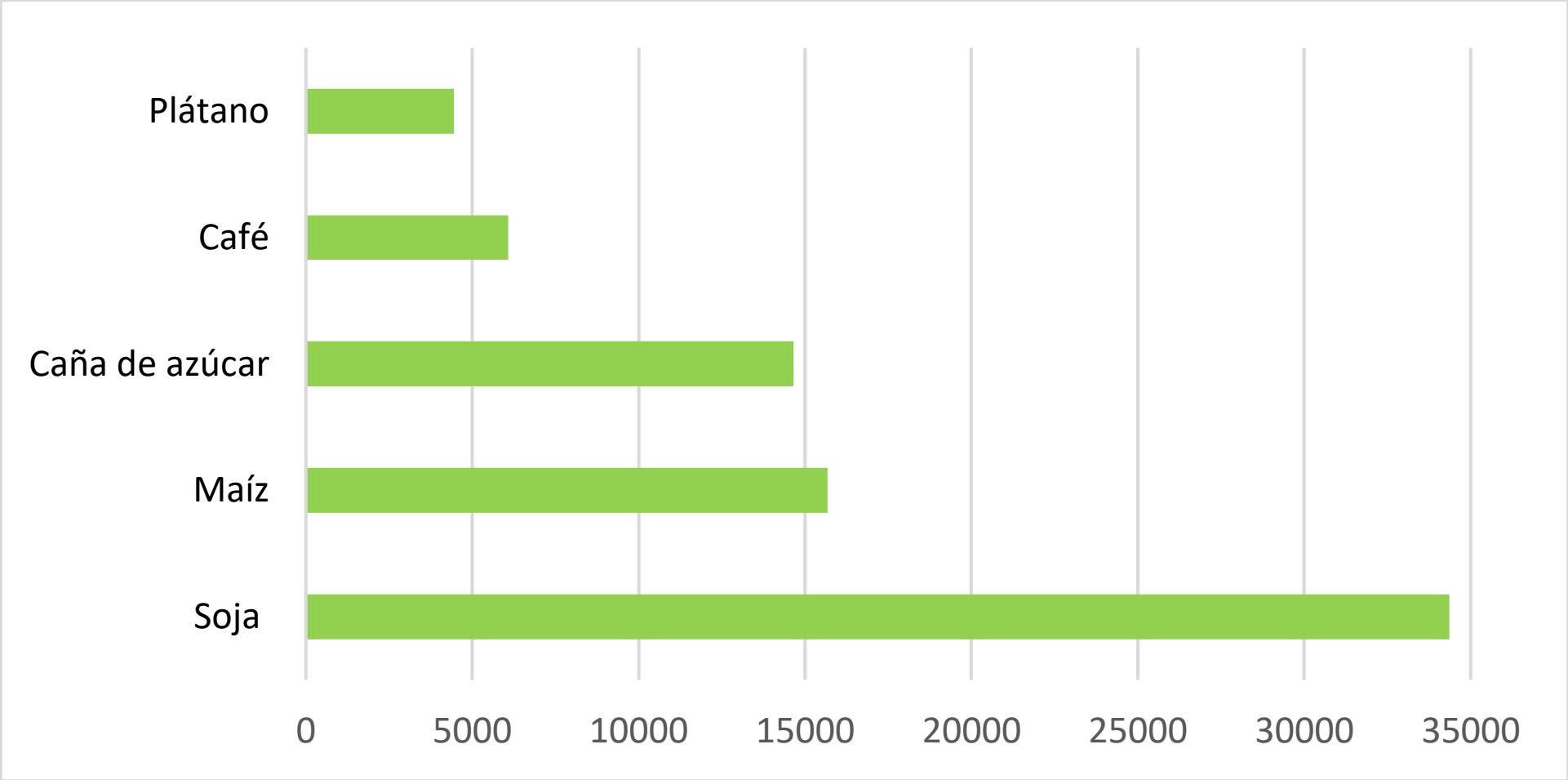

Fuente: elaboración propia a partir de (MAPA, 2017)

La producción agrícola brasileña es más que suficiente para cubrir la demanda interna, por lo que una parte considerable se destina a las exportaciones: los productos agrícolas y alimenticios suponen un 29% de las exportaciones, con un valor de más de 56 mil millones de dólares (OEC, 2018). Si a estos le sumamos las exportaciones de productos animales, se alcanza más del 36% del total. Sólo en el mercado de la soja, Brasil es el segundo máximo exportador, por detrás de los Estados Unidos, llevando a los mercados internacionales el 38% del total [3]. De la misma forma, la soja compone el 39% de las exportaciones que hace Brasil con destino a China, su principal socio comercial (*op. cit.*).

Siguiendo con las dotaciones naturales de Brasil, cabe destacar sus recursos mineros. Ya hemos hablado del petróleo con anterioridad, pero también resalta su producción de otros minerales. La explotación minera es muy antigua en Brasil, de hecho, el estado de Minas Gerais debe su nombre a la riqueza minera de la región. Aun así, en los últimos tiempos, la explotación minera se ha intensificado: en 1971 las rentas de la minería suponían el 0,4% del PIB del país, comparado con el récord del 3,3% del PIB que se alcanzó en 2008. En 2015, la proporción de la minería sobre la producción total había bajado a un más discreto 1,4% (WB, 2018).

Actualmente, Brasil es el segundo exportador mundial de mineral de hierro, con el 20% de la exportación mundial (que de nuevo va principalmente a parar a China) (OEC, 2018). En 2016 los trabajadores mineros consiguieron llevar al mercado más de 11 mil millones de dólares en hierro, como vemos en la Gráfica 10. Tras el hierro, su principal producto es el oro, con grandes minas que llevan operando muchos años (ver Figura 6 del Anexo Gráfico) y que obtuvieron una producción de más de 2 mil millones de dólares.

**Gráfica 10**. Producción comercializada de minerales, 2016 (millones de US$)

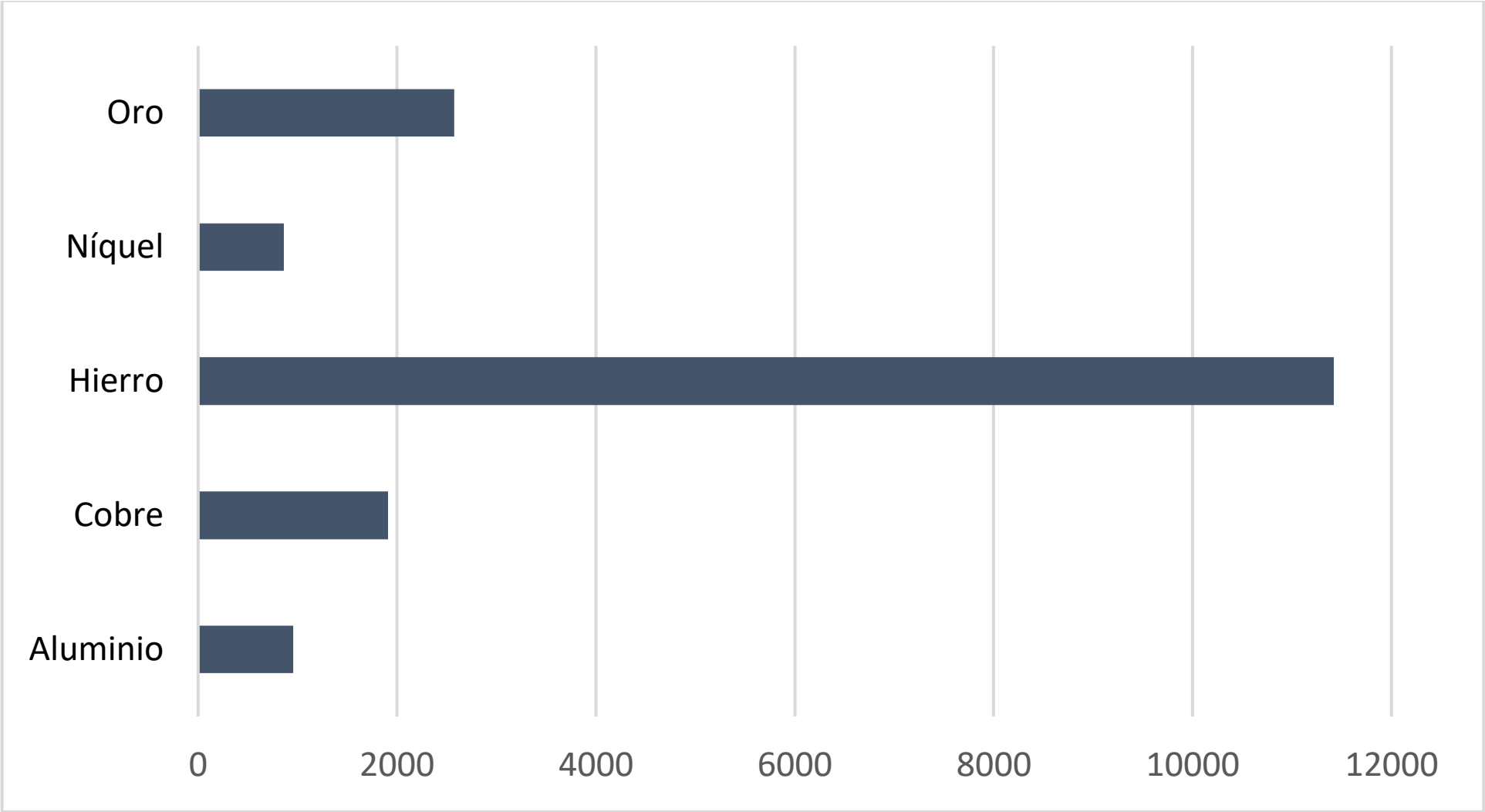

Fuente: elaboración propia a partir de (DNPM, 2017)

[3] España, después de China, es el principal destino de la soja brasileña.

Para concluir con los recursos naturales, es necesario destacar que la tercera mayor reserva de agua dulce del mundo se encuentra principalmente en territorio brasileño. El Acuífero Guaraní está compartido con Argentina, Paraguay y Uruguay, pero algo más del 70% del mismo se encuentra bajo territorio brasilero. Es conocido que este acuífero será un recurso estratégico de primer orden en el futuro, cuando debido al cambio climático empiece a escasear el agua dulce en otras zonas del planeta, por lo que su gestión ha captado la atención incluso del Banco Mundial (Foster et al., 2006).

## 3.2. Evolución de la estructura sectorial de la economía.

**Gráfica 11.** Evolución de la estructura sectorial del valor añadido

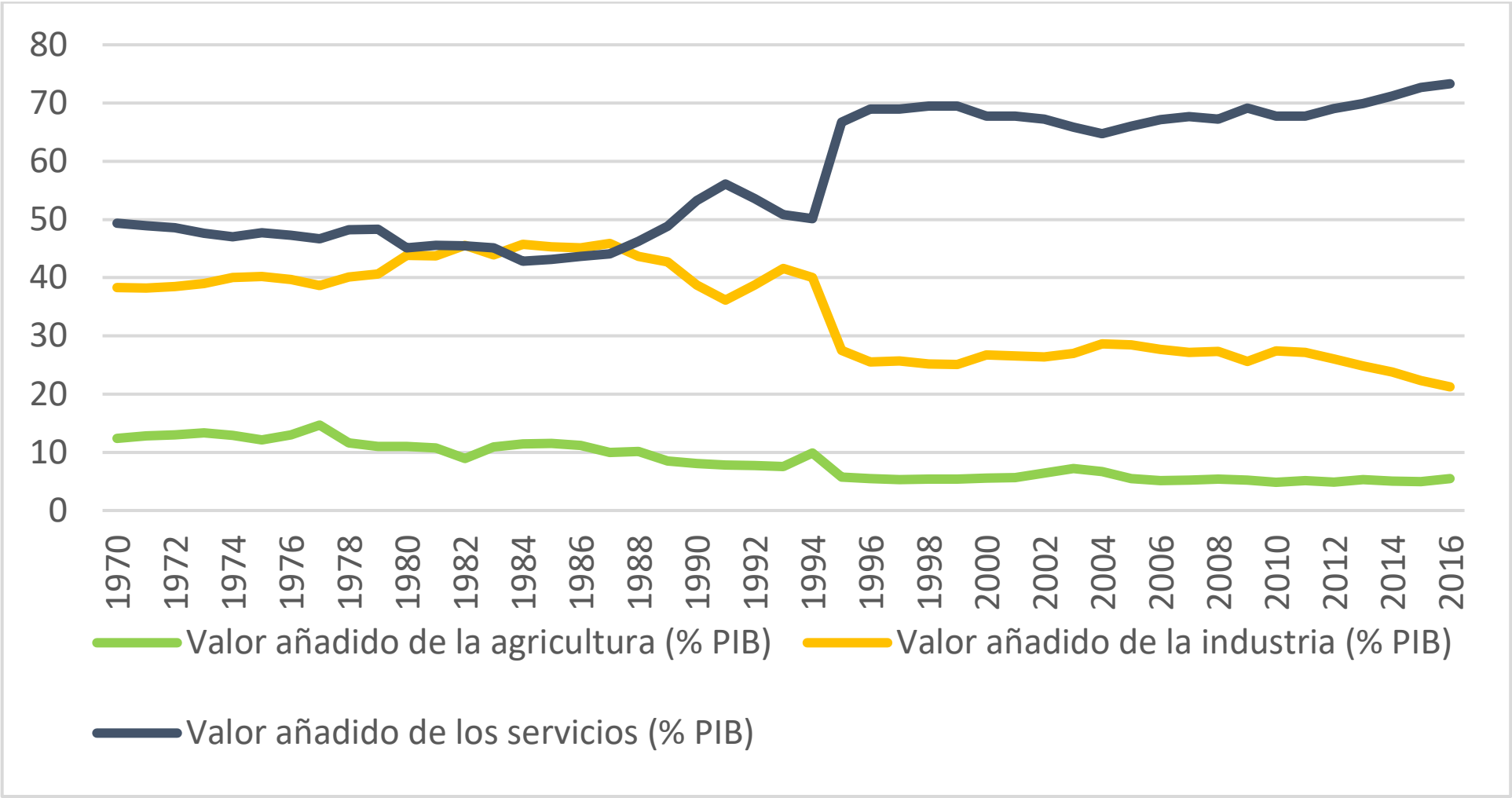

Fuente: elaboración propia a partir de (WB, 2018)

Tal y como se observa en la Gráfica 11, la estructura sectorial de Brasil ha cambiado considerablemente a lo largo del período estudiado. La industria y los servicios han sido desde 1970 los dos principales motores de la economía brasilera, siendo predominante el valor añadido del sector terciario exceptuando algunas ocasiones a lo largo de los años 80. A partir de la caída de la sangrienta dictadura militar, el turismo volvió a afluir hacia Brasil, y las mejores expectativas incrementaron la confianza de los inversores y del sector financiero, haciendo que el sector servicios creciera sobre la industria hasta llegar a una diferencia de 20 puntos entre ambas sobre el PIB. A esto se le sumaron las privatizaciones y las políticas de desindustrialización llevadas a cabo por los gobiernos del PMDB y PSDB a lo largo de los años 90, que consagran la terciarización de la economía brasileña: la industria pasa de suponer el 42,6% del PIB en 1988 a un 26,7% en el año 2000, mientras que los servicios pasan del 46,2% al 67,7% en el mismo período. Mientras tanto, la agricultura ha estado en declive desde los años 70, pese a ser Brasil como hemos visto una gran potencia agropecuaria. Actualmente, la producción del país se debe en un 5,4% a la agricultura, un 21,2% a la industria y un 73,3% a los servicios.

Una evolución similar pero diferenciada se puede observar en la distribución sectorial del empleo de la Gráfica 12. Desde el año 1990, el líder claro es el sector servicios, que concentra siempre más del 50% del empleo total hasta situarse en un 63% en 2016. El ascenso del sector servicios ha sido a costa del sector primario, ya que el sector industrial se mantiene en un 20% durante todo el período. Resulta esto llamativo, ya que mientras se ha mantenido el peso de la industria en el empleo, el peso en el valor añadido ha caído hasta la mitad de lo que era en 1989. Esto refleja un claro problema de productividad por trabajador en el país, problema que ha ido acentuándose a lo largo de estos años.

**Gráfica 12.** Evolución de la estructura sectorial del empleo.

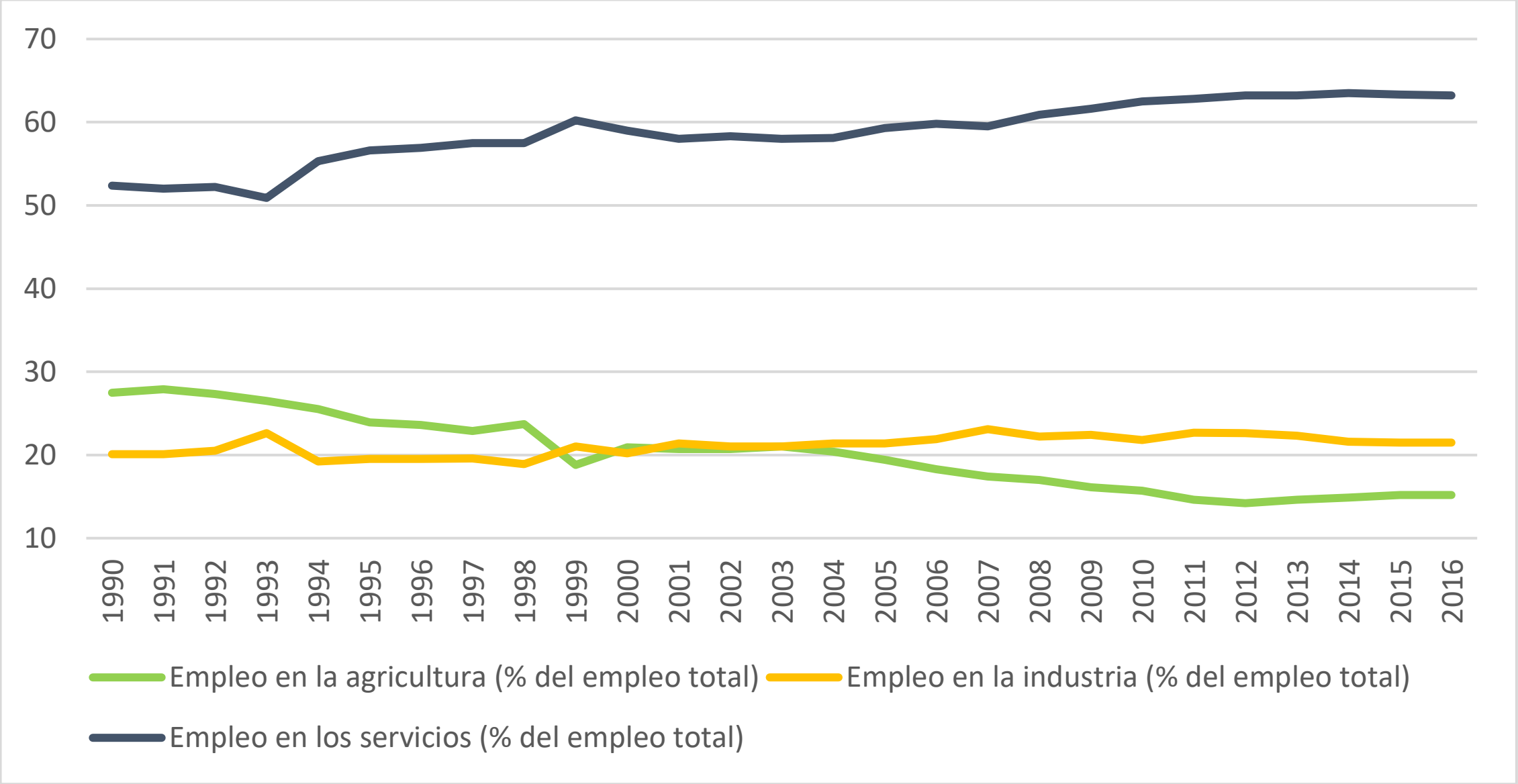

Fuente: elaboración propia a partir de (WB, 2018)

La importancia de los servicios en Brasil es diferencial con respecto a las otras potencias económicas sudamericanas y con el resto del grupo de los BRICS, como se aprecia en la Gráfica 13. El principal problema de esta terciarización es la baja productividad y calidad del empleo que se genera en el sector brasilero de los servicios. Esto supone una grave limitación para el crecimiento del país, que necesita desarrollar más los servicios de alta cualificación y alto valor añadido para poder mantener un desarrollo económico estable (WB, 2016). De hecho, la productividad es el principal reto al que se enfrenta Brasil en los años venideros, según la mayoría de informes económicos de las instituciones especializadas (IMF, 2017a; OECD, 2018).

**Gráfica 13**. Valor añadido de los servicios, 2016 (% PIB)

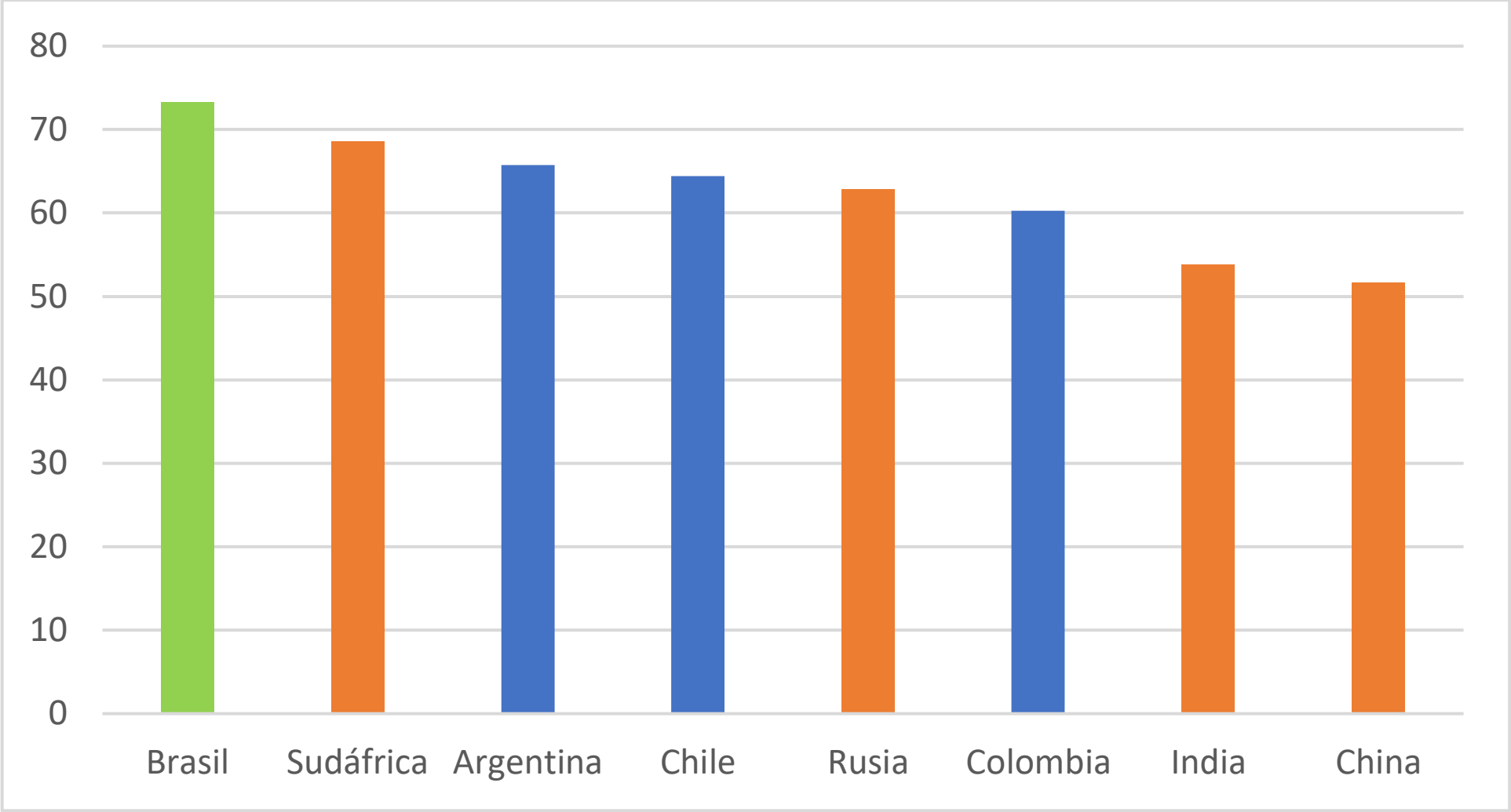

Fuente: elaboración propia a partir de (WB, 2018)

### 3.3. Desarrollo tecnológico y eficiencia de los factores productivos.

Uno de los grandes impulsores de la productividad laboral es sin duda el desarrollo tecnológico. Brasil en este aspecto ha mejorado en los últimos años, ocupando el lugar que se merece como líder en Sudamérica en investigación y desarrollo. El acceso más generalizado a la educación y el intento de algunas empresas (especialmente en el sector de transformación de productos (IBGE, 2014)) de mejorar su productividad mediante el I+D ha contribuido a que el número de personas empleadas en la investigación y el desarrollo haya crecido en la última década, como se aprecia claramente en la Gráfica 14.

**Gráfica 14.** Evolución reciente del empleo en investigación y desarrollo.

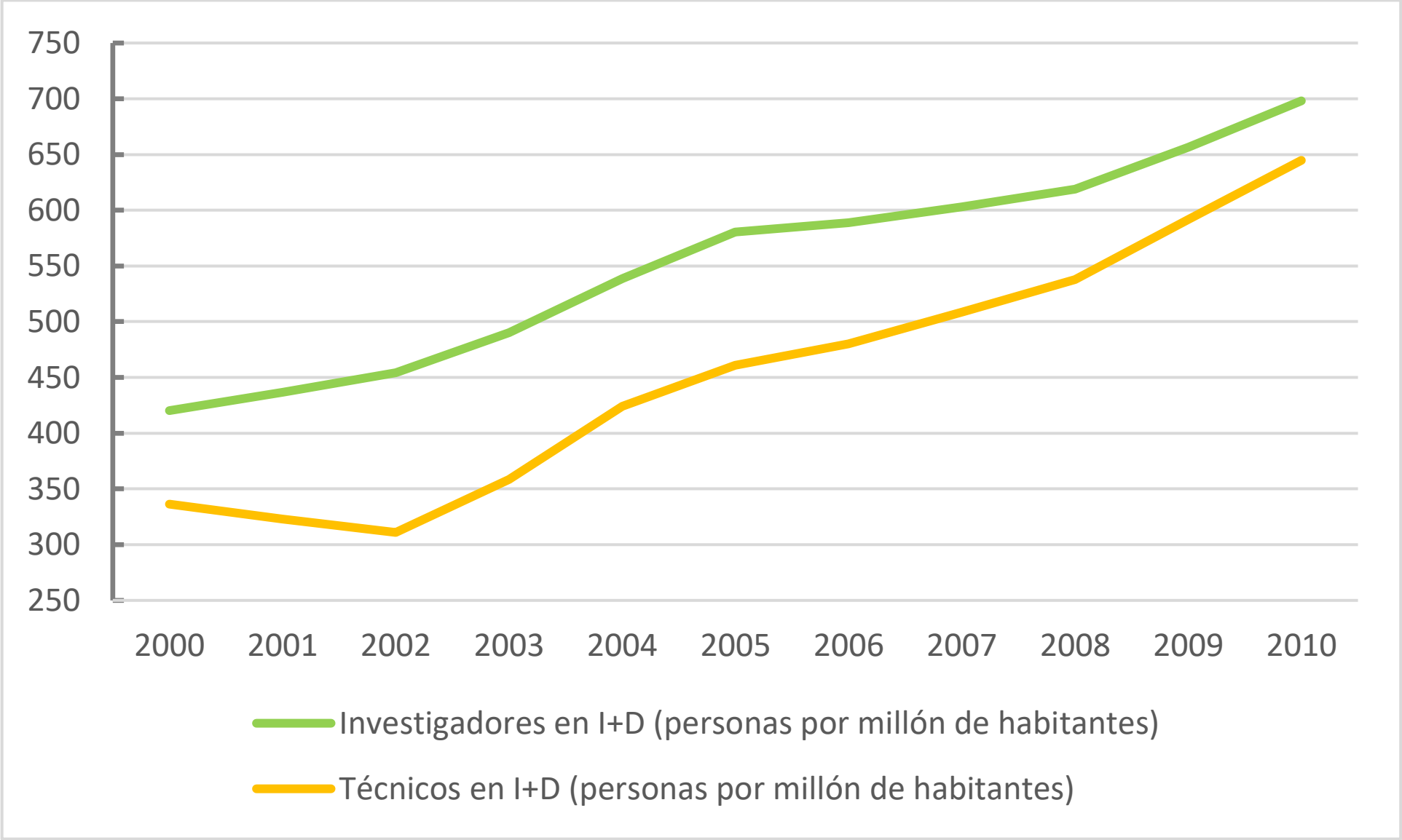

Fuente: elaboración propia a partir de (WB, 2018)

Un mayor número de personas trabajando en la I+D también multiplica el descubrimiento de procesos innovadores a través de la producción académica. Brasil es ahora mismo la mayor potencia sudamericana en producción de artículos científicos y académicos (WB, 2018), superando la producción conjunta del resto de países. También es líder en el continente en el número de patentes solicitadas por residentes, rozando el 80% de las patentes de toda Sudamérica, como se observa en la Gráfica 15. Además, Brasil es el segundo país de los BRICS y de las potencias sudamericanas en gasto en I+D con respecto al PIB, sólo por detrás de China (ver Figura 7 del Anexo Gráfico)

Uno de los principales motores de este crecimiento es su industria farmacéutica, que mediante la intervención estatal (implementando *compulsory licences* [4]) y la producción de medicamentos genéricos ha mejorado de forma exponencial el acceso a los medicamentos en el país (WHO, 2011). La ambición de ser el centro de la producción regional de medicamentos ha llevado a Brasil a invertir con fuerza en el futuro de su industria farmacéutica, con planes de colaboración y transferencia tecnológica con Cuba, líder en la investigación biomédica en Latinoamérica y el Caribe (WHO, 2015).

**Gráfica 15.** Patentes solicitadas por residentes en Sudamérica (2013)

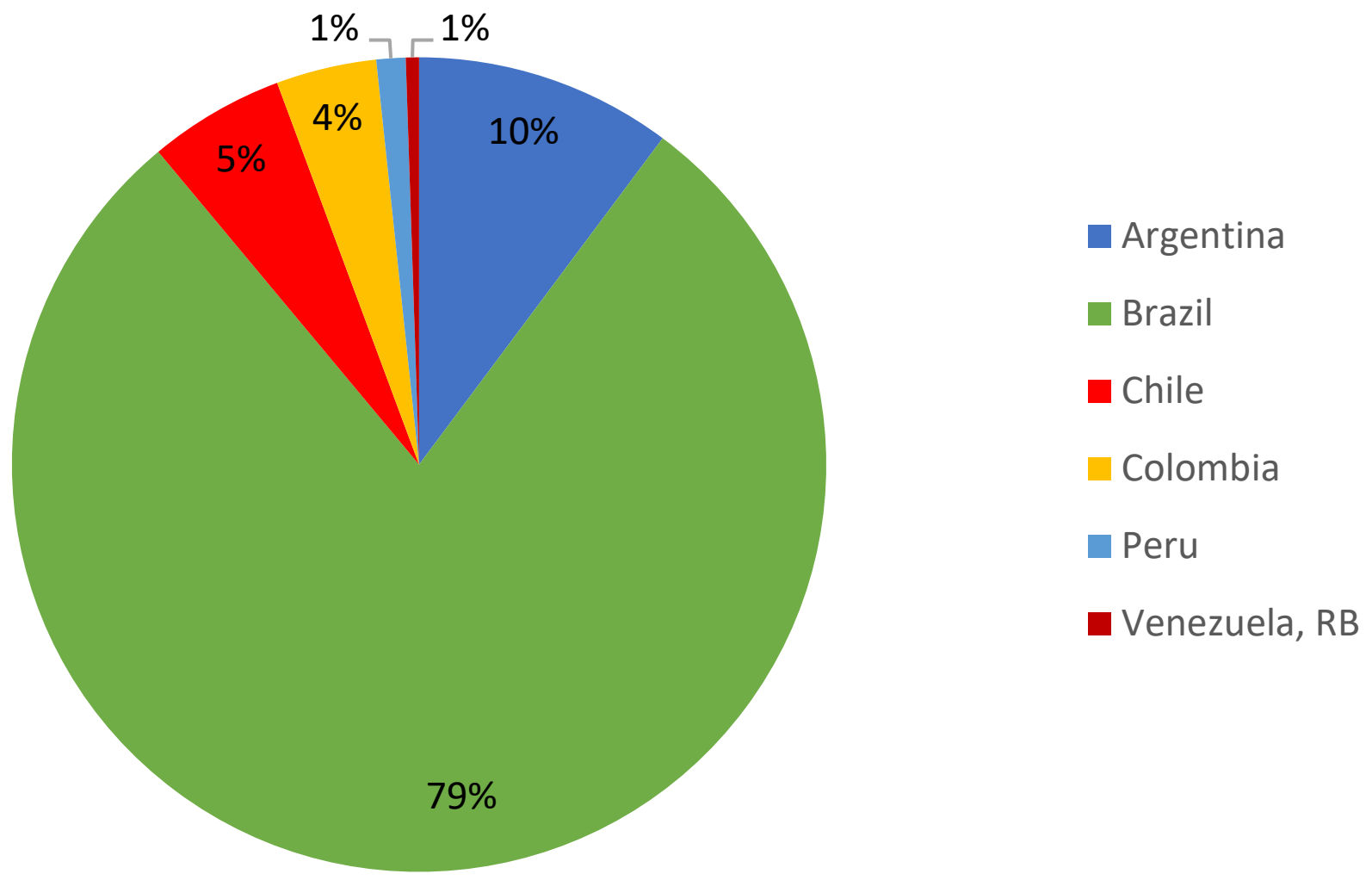

Fuente: elaboración propia a partir de (WB, 2018)

El otro motor del desarrollo tecnológico en Brasil son las TICs. Durante los últimos años (2000-2016), la población con acceso a internet ha pasado del 2,9% al 59,7% (WB, 2018). El desarrollo de las TIC ha hecho de Brasil uno de los grandes exportadores tecnológicos de los BRICS. Como se observa en la Gráfica 16, Brasil es el segundo país con mayor peso de las exportaciones de bienes de alta tecnología sobre el total de bienes manufacturados, además de ser el segundo en la proporción que representan los servicios TIC sobre sus exportaciones de servicios.

[4] Las *compulsory licences* son licencias que puede decretar cualquier gobierno en vista de alguna emergencia de salud nacional, para permitir la producción genérica de medicamentos protegidos por patentes a nivel internacional. Esta es una de las flexibilidades que se incluyeron en los acuerdos TRIPS para mejorar el acceso a los medicamentos, con escasos resultados (WTO, 1994).

**Gráfica 16.** Tecnología exportada por los BRICS (2016)

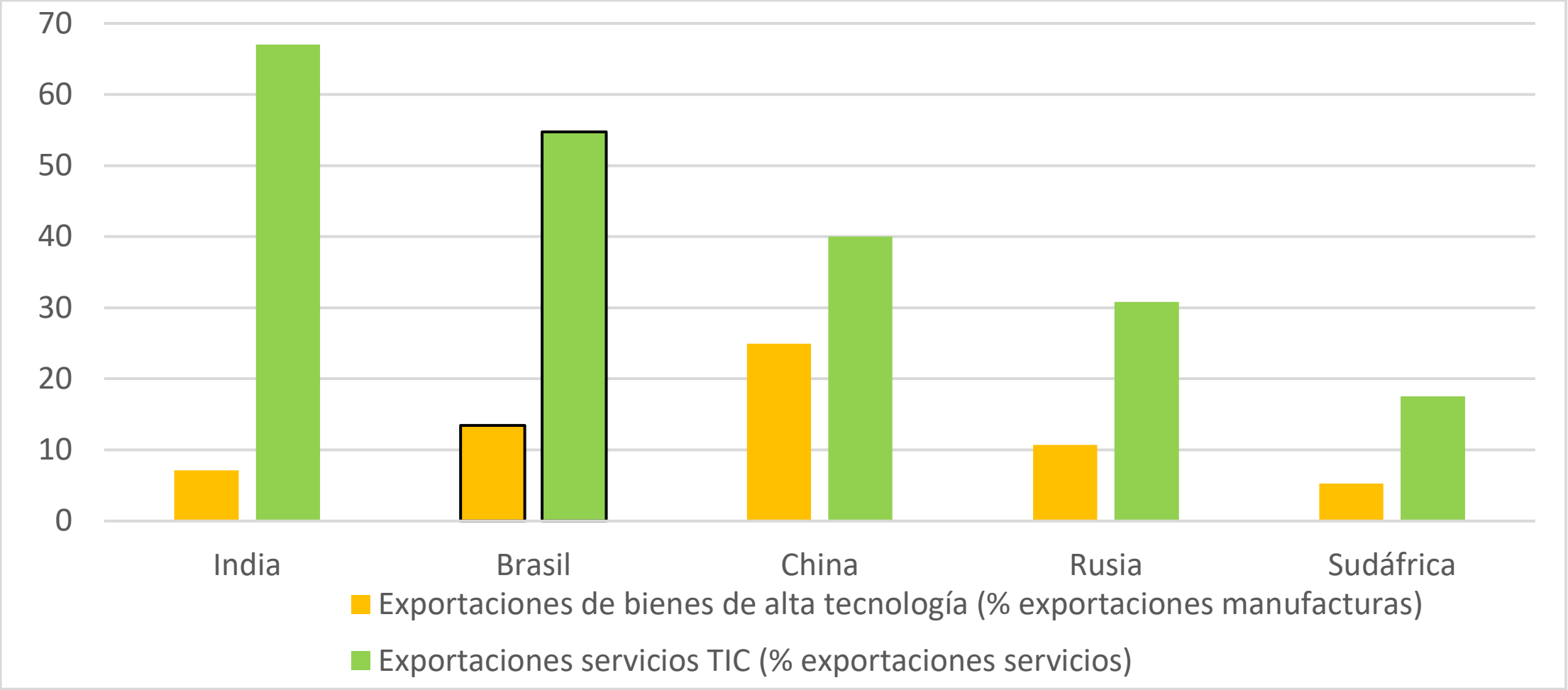

Fuente: elaboración propia a partir de (WB, 2018)

Sin embargo, esta capacidad en las tecnologías de la información y de la comunicación, claves para el desarrollo productivo de un país, no han contribuido especialmente a la mejora de la productividad que el país necesita.

Y es que, desde 1991, la productividad de los trabajadores brasileños no ha crecido como se esperaba, comparando al país con los otros grandes sudamericanos y los BRICS, como podemos ver en la Gráfica 17. De hecho, si observamos la gráfica, se puede ver que Chile y Brasil parten de la misma productividad del trabajo, unos 26.000$ por trabajador, y en 2018 Chile supera los 50.000$ mientras que Brasil se sitúa en unos 32.000$. Si comparamos las tasas de crecimiento anuales de la productividad en la última década, los resultados son aún peores: entre 2001 y 2013, la productividad china creció a un ritmo del 9,6% anual, la de Rusia a un 3,5%, la de Colombia a un 2,1%, mientras que Brasil se estancaba en un 1,6% anual (WB, 2016).

**Gráfica 17.** Productividad por trabajador (dólares internacionales de 2011, PPA)

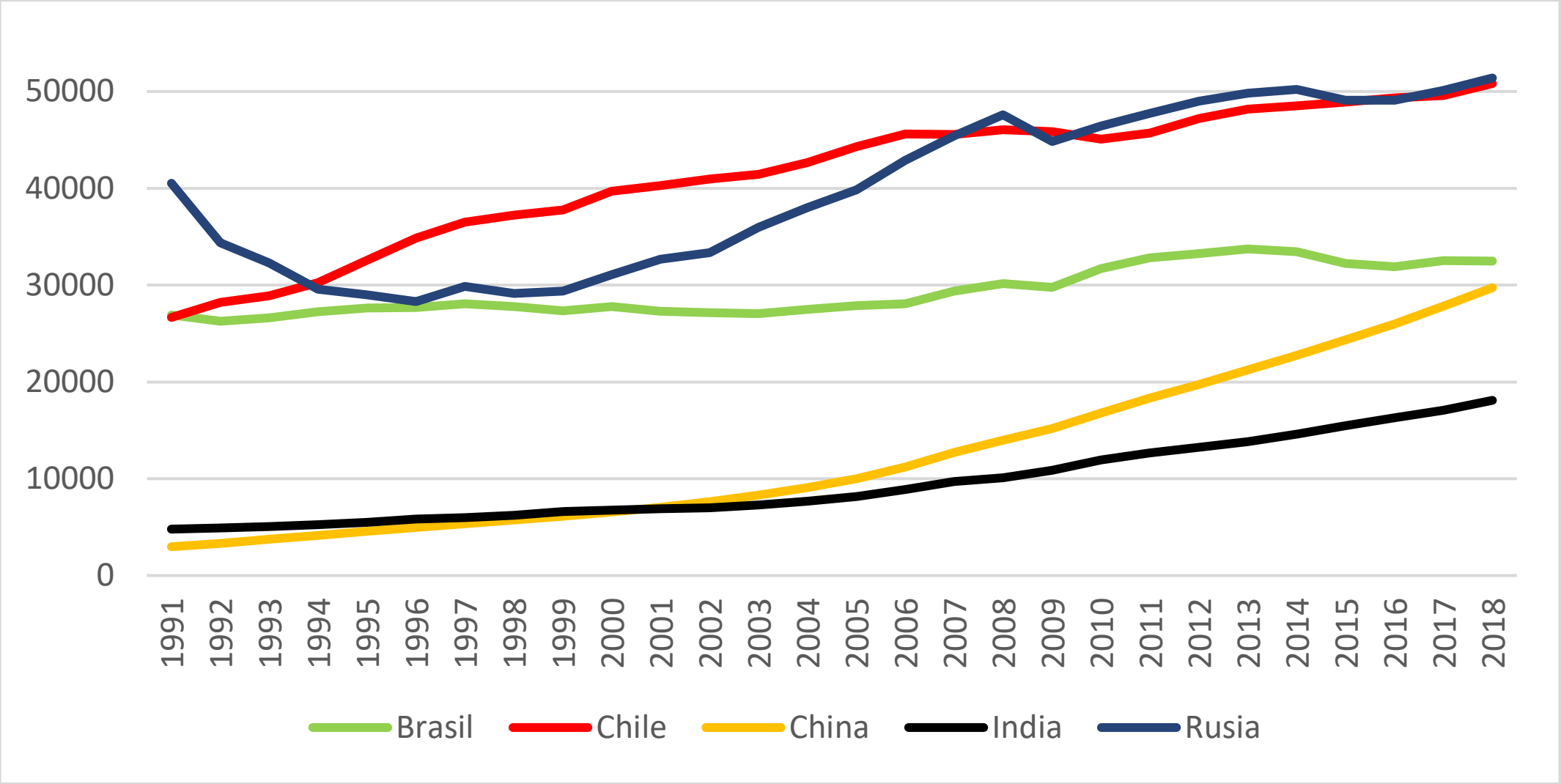

Fuente: elaboración propia a partir de (ILO, 2018b)

Como vimos anteriormente, los sueldos mínimos (además de los medios) no dejaron de subir desde 2003 hasta la crisis que golpeó Brasil en 2015. De hecho, los salarios medios han tenido un crecimiento superior al de la productividad del trabajo desde 2003 hasta la crisis (ver Figura 8 Anexo Gráfico), haciendo que la competitividad de los productos brasileños se resienta. Este ha sido uno de los grandes problemas de fondo del desarrollo brasilero a partir del año 2000: el crecimiento ha sido impulsado principalmente por el crecimiento del consumo de las familias debido a la mejora de los salarios y la salida de millones de personas del umbral de la pobreza. Sin embargo, como más adelante veremos, la inversión no creció en la misma medida, y la productividad de los factores tampoco.

Pese a que la economía brasileña ha ido incrementando su stock de capital desde los años 70, más que duplicando la intensidad de capital de la producción (stock de capital por trabajador) durante este período, el crecimiento se ha frenado en los últimos años: de 1998 a 2014, la intensidad de capital sólo ha aumentado un 6,2% anual. Además, si vemos el indicador TFP, que mide la productividad del total de los factores, vemos que no sólo no ha crecido desde los años 70, sino que en efecto se ha ido reduciendo.

**Gráfica 18.** Productividad total de los factores (TFP) e intensidad del capital

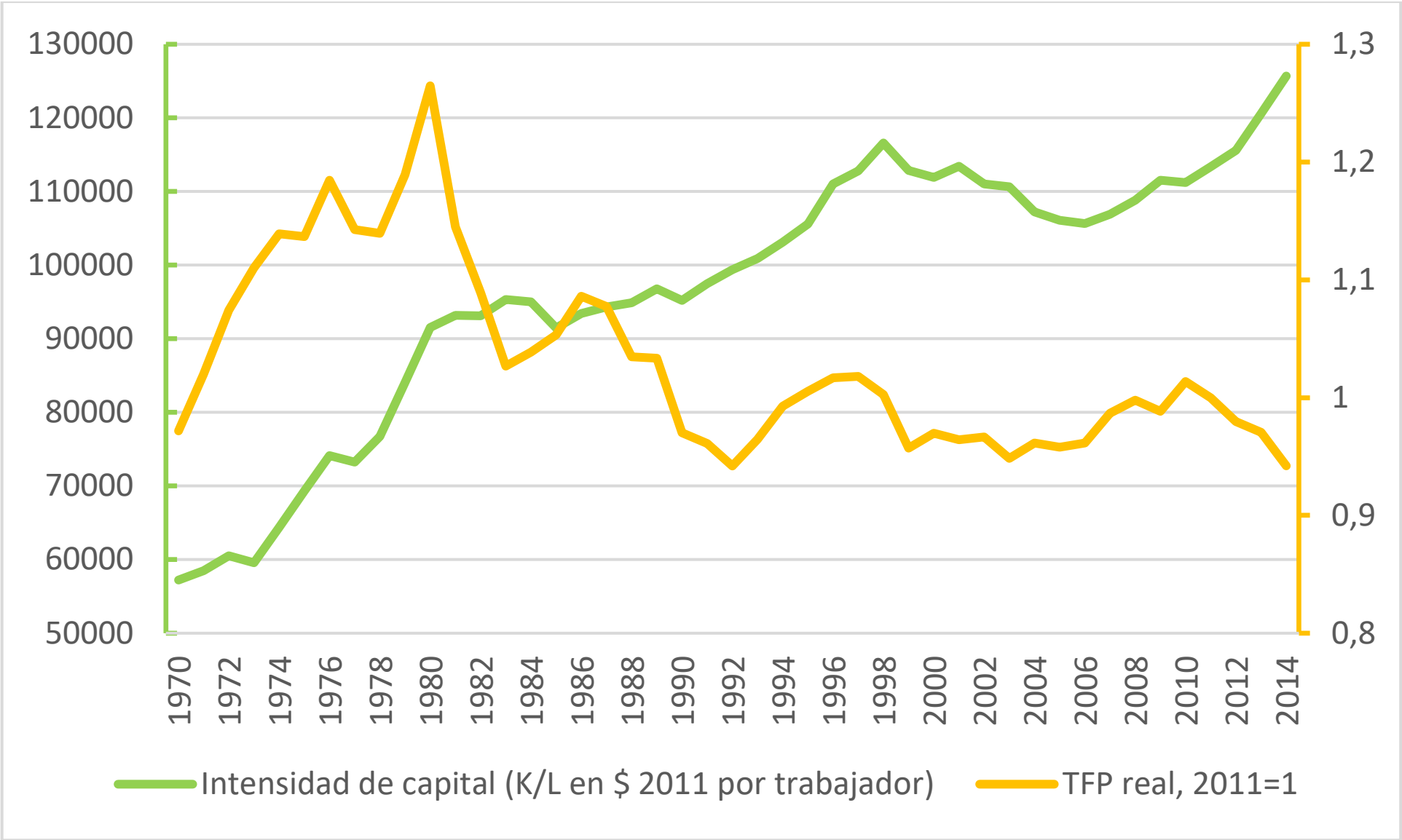

Fuente: elaboración propia a partir de (Feenstra, Inklaar, & Timmer, 2015)

Esta baja productividad fue uno de los factores que propiciaron la crisis brasileña en 2015. Con la caída de los precios de las materias primas, que suponían alrededor del 60% de las exportaciones (OEC, 2018), la demanda externa cayó, aumentó el desempleo y bajó a su vez la demanda interna. Un crecimiento que se debía principalmente al aumento del empleo formal y mayores sueldos, se frenó en seco cuando estos decayeron, y la demanda externa era menor al ser tan baja la competitividad brasilera.

¿Por qué es tan reducida la productividad brasileña? Según el Banco Mundial, existen múltiples razones complementarias. En primer lugar, aunque las inversiones en infraestructuras han aumentado a partir de 2003, no han conseguido superar la tasa de depreciación natural de aquellas, por lo que se han ido deteriorando con el tiempo. En segundo lugar, los altos tipos de interés impuestos por el Banco Central para controlar la inflación encarecen la financiación de las empresas brasileñas, haciendo que sea difícil financiar las inversiones. A esta dificultad de acceso al capital se le suma que el complejo sistema administrativo de la República Federal implica en ocasiones duplicidad de impuestos y regulaciones, que limita la entrada al mercado de pequeñas empresas (WB, 2016).

Cuando decimos que el complejo sistema institucional de Brasil dificulta el funcionamiento empresarial no es ninguna exageración. La OCDE estima que una empresa media en Brasil tiene que dedicar 6 veces más tiempo a preparar el pago de los impuestos que en Argentina o Chile, 9 veces más que en China, Sudáfrica o India, y más de 10 veces el necesario en Rusia.

**Gráfica 19.** Tiempo medio preparando papeles para el pago de impuestos (horas)

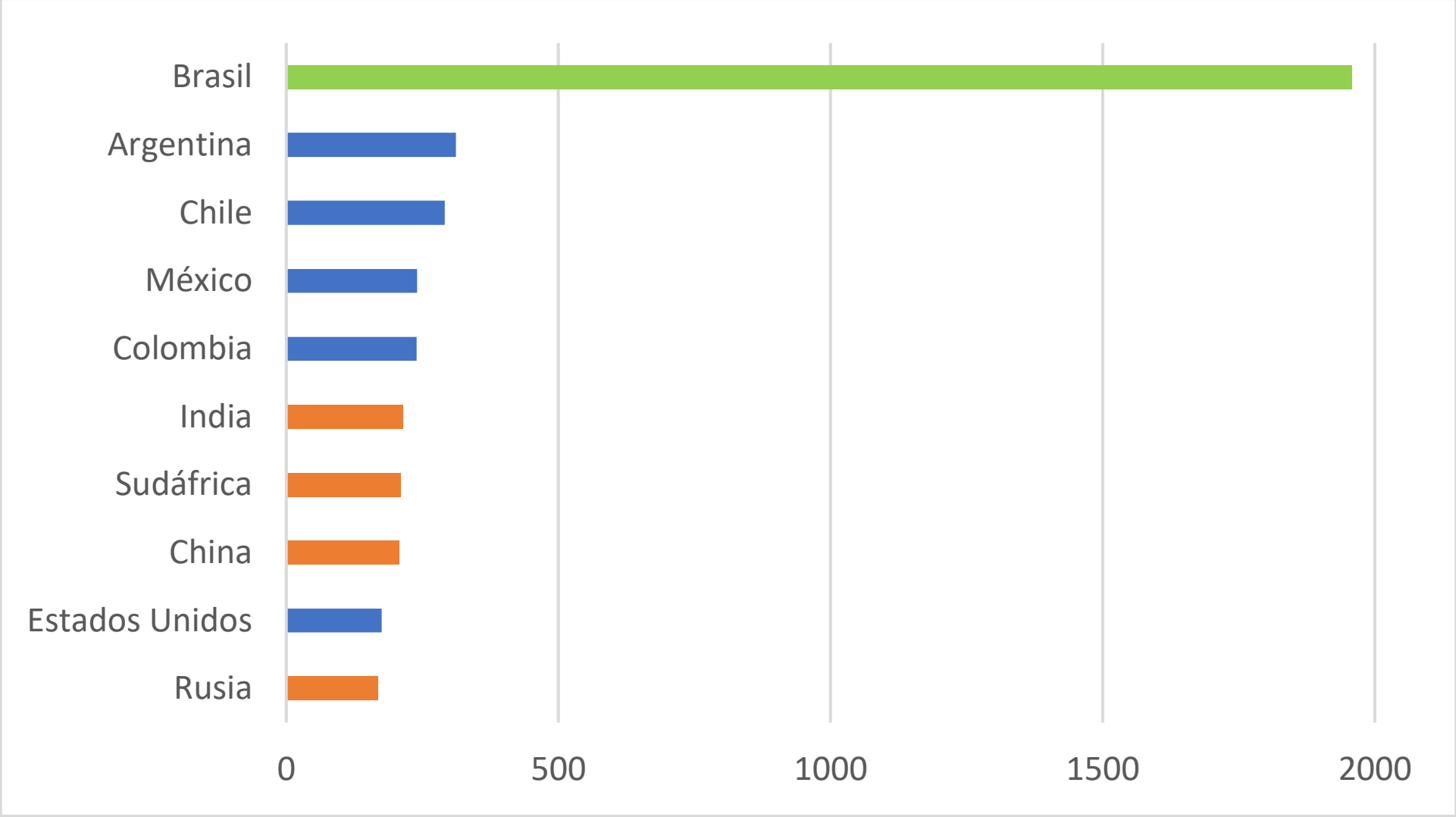

Fuente: elaboración propia a partir de (OECD, 2018)

Igualmente, si analizamos otros indicadores del ambiente empresarial en Brasil, vemos que es una de las economías donde es más difícil empezar un negocio del grupo de los BRICS y Sudamérica. Es el país donde más tiempo es necesario invertir para poder montar una empresa, unas 8 veces más que en Colombia o Rusia, y tiene un sistema regulatorio que dificulta en exceso la posibilidad de hacer negocios, que se sitúa en 125, siendo un índice de 1 la regulación que lo facilite más.

**Gráfica 20.** Dificultad para el emprendimiento y los negocios, 2017

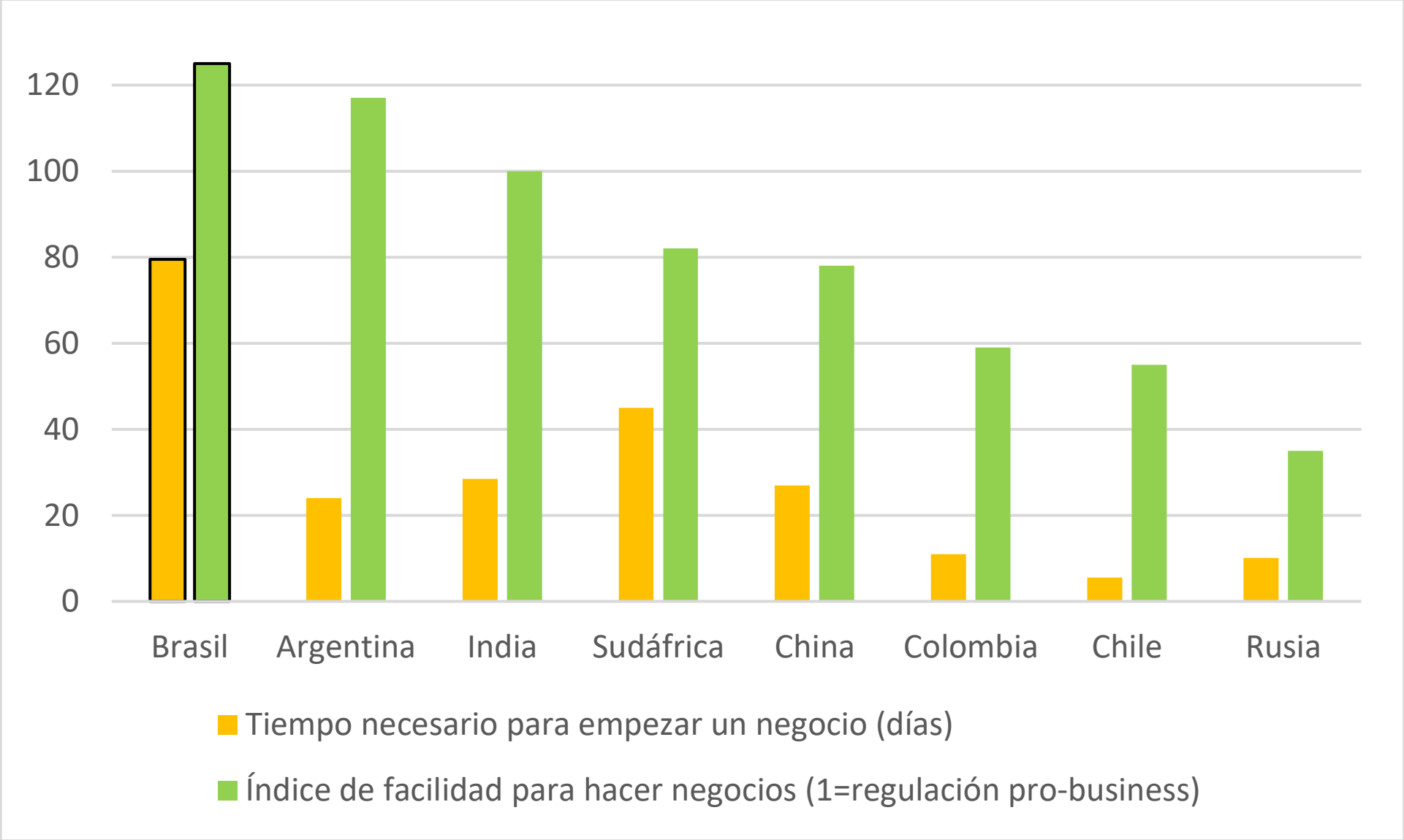

Fuente: elaboración propia a partir de (WB, 2018)

Además de la dificultad de la actividad empresarial, se le suman los aranceles de los productos extranjeros y, exceptuando las industrias extractivas de materias primas, que las empresas están centradas en la demanda interna, con lo que la economía brasileña tiene la peor inserción externa de todas las economías con las que se la han comparado, como luego veremos.

## 4. Análisis desde el lado de la distribución de la renta.

Ya vimos anteriormente que en las últimas décadas la desigualdad se ha reducido en Brasil. Sin embargo, esta reducción no ha conseguido cambiar la estructura de la economía, que sigue siendo profundamente desigual. El 'pacto lulista' ha mantenido el modelo productivo y las bases materiales de la sociedad intactas, por lo que los beneficios empresariales siguen siendo superiores a los salarios en un país con más de 100 millones de asalariados.

**Gráfica 21.** Distribución de la renta nacional entre agentes, 2017

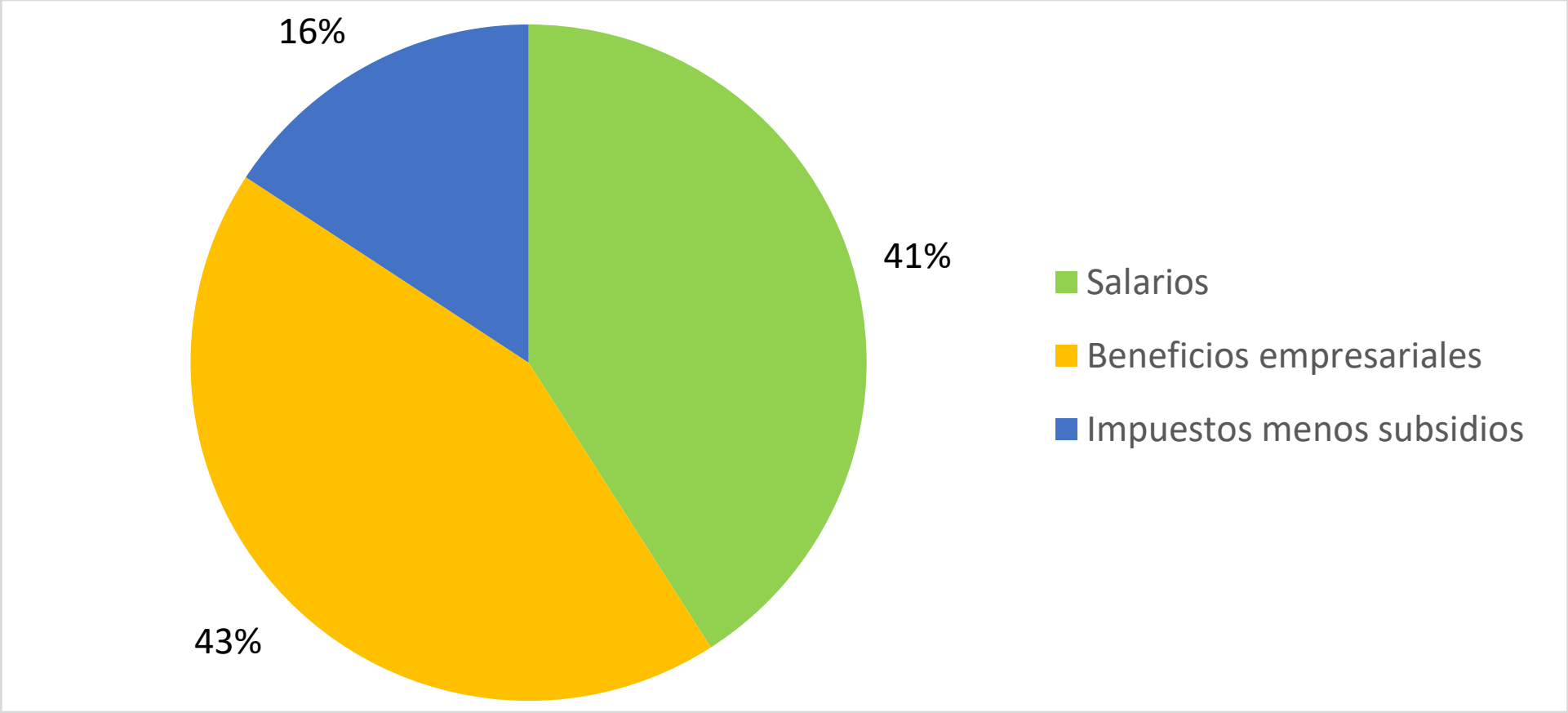

Fuente: elaboración propia a partir de (IBGE, 2017b)

Pese a las políticas redistributivas de los últimos años (y desmanteladas con la caída de Dilma Rousseff), la renta se sigue distribuyendo de una forma abismalmente desigual. Como podemos ver en la Gráfica 22, incluso tras 12 años de gobierno del Partido de los Trabajadores, el quintil más rico seguía acaparando el 56% de la renta del país y el quintil más pobre sólo recibe el 3% de lo producido en el país.

**Gráfica 22**. Distribución de la renta por quintiles de ingreso, 2017

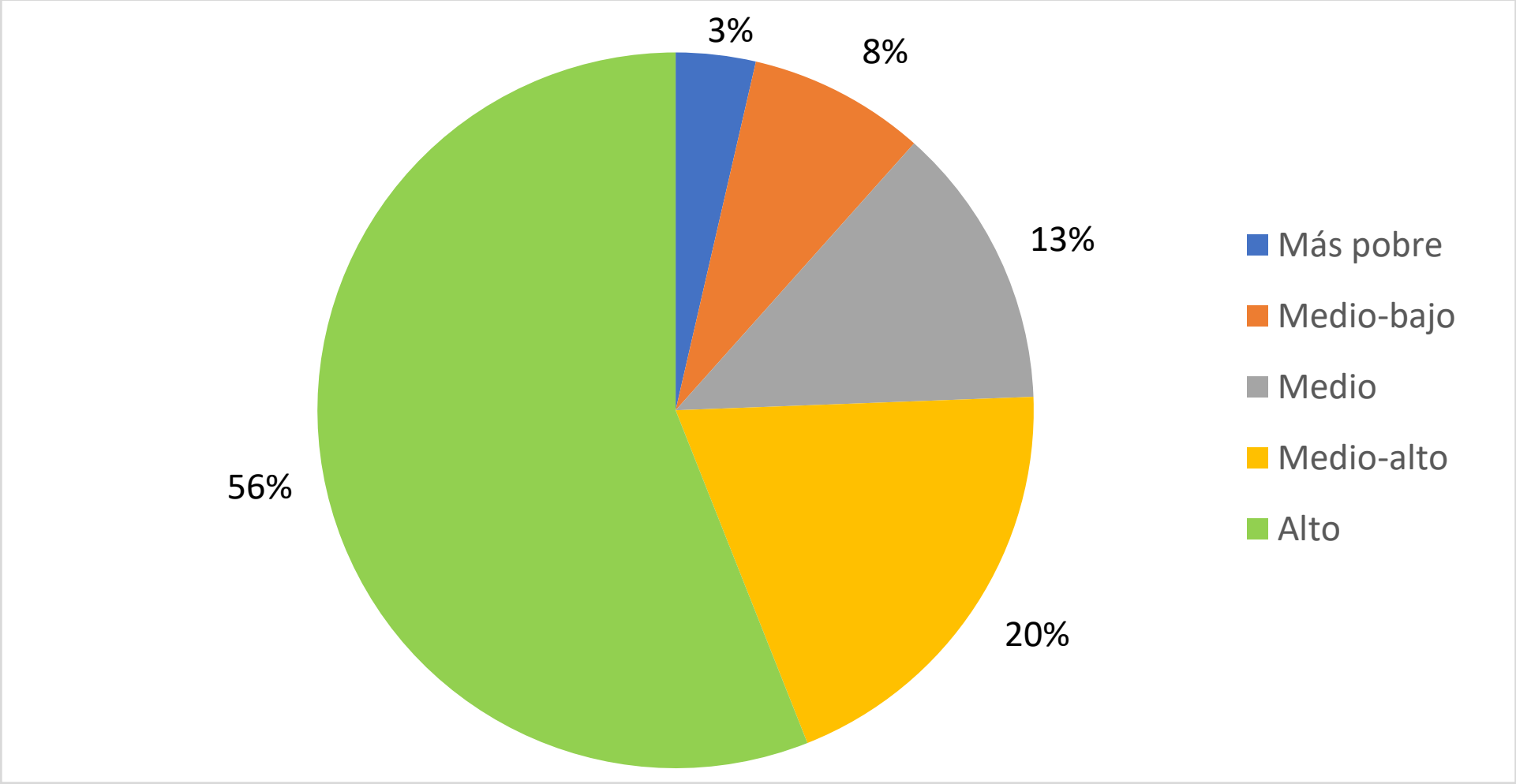

Fuente: elaboración propia a partir de (WB, 2018)

Durante los últimos años, las políticas fiscales han tenido un ligero éxito en la reducción coyuntural de la desigualdad: el Gini se reduce de media algo más de 3 puntos tras impuestos y transferencias (Lustig, Pessino, & Scott, 2013). No obstante, la fiscalidad en Brasil es menos progresiva que en otros países, especialmente si lo comparamos con los BRICS: el tipo máximo del IRPF es casi 20 puntos menor que en China, o 14 puntos menos que en India.

**Gráfica 23.** Tipo impositivo máximo del equivalente al IRPF, 2017

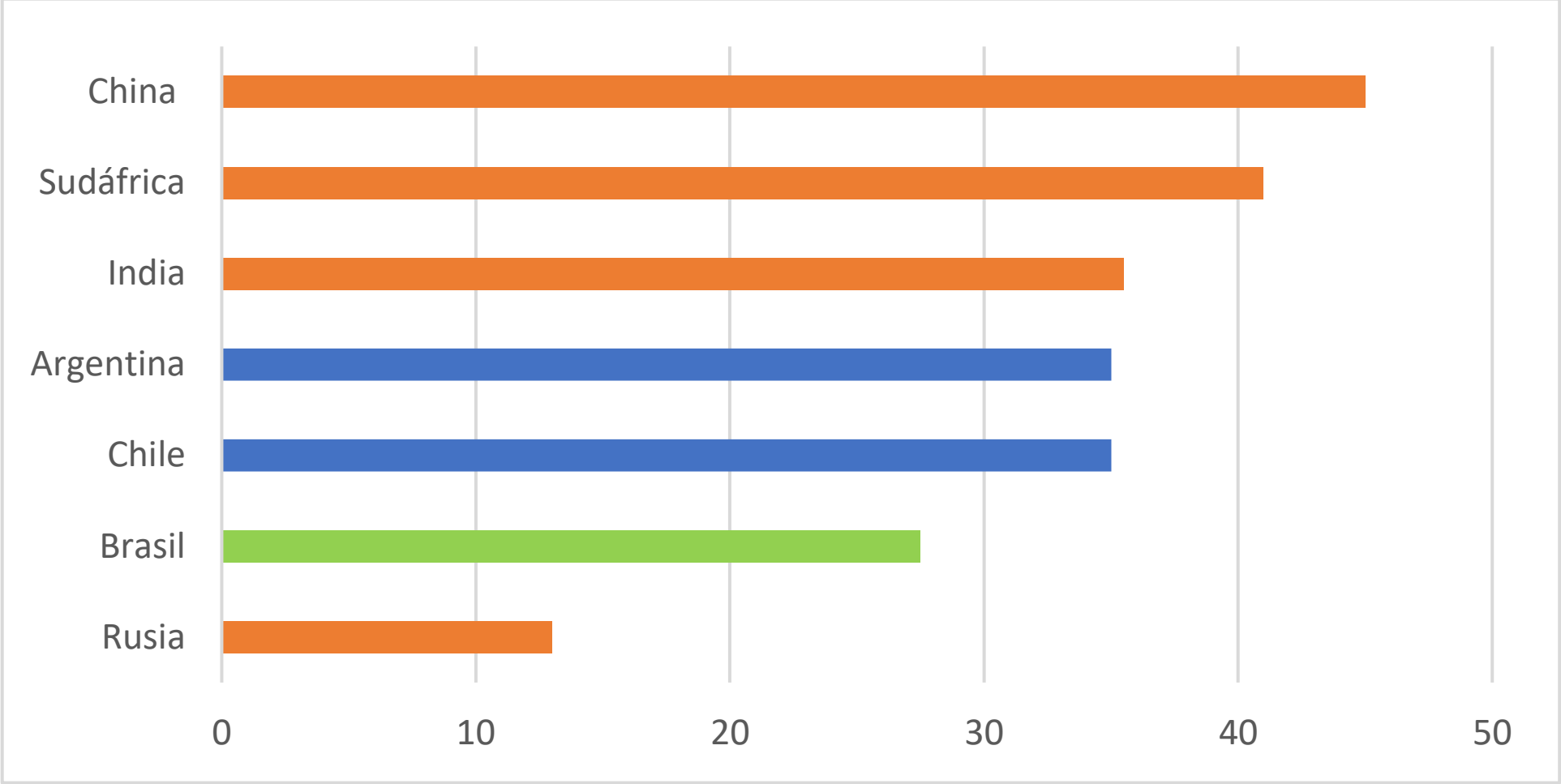

Fuente: elaboración propia a partir de (Expansión, 2017)

Es probable que la progresividad fiscal empeore con la entrada del gobierno de Bolsonaro, previsiblemente neoliberal, especialmente teniendo en cuenta que el presidente electo llamaba hace unos años a la evasión de impuestos, dando ejemplo evadiendo "todo lo posible" (BBC, 2018a).

Dado el rumbo que se espera que tome el ejecutivo brasileño a partir de 2019, con la reducción del gasto público como uno de los objetivos principales, podemos suponer que muchos de los programas de transferencias que se impulsaron durante las últimas décadas sean ahora, cuanto menos, reformados.

## 5. Análisis desde el lado de la demanda.

Con respecto a los componentes de la demanda, cabe analizar su evolución a lo largo del tiempo, que a diferencia de otros países no ha variado demasiado. Mientras que en 1970 el consumo privado suponía cerca del 70% del PIB, en 2016 suponía el 64%. A su vez, la inversión se ha reducido de un 21% a un 15% del PIB, mientras que el Estado ha ido cobrando cada vez más importancia en la demanda agregada. Con la caída de la dictadura militar, la participación del gobierno en la demanda pasó de un 10% a un 20% en apenas 5 años, con la creación del aparato estatal regional y el comienzo de las políticas a nivel regional. Pese a todo ello, la distribución de la demanda no ha variado en exceso en los últimos 50 años, exceptuando un período de continuo descenso de la participación de la demanda privada, que se vio acompañado de un crecimiento muy rápido de la inversión por las buenas expectativas tras el fin del régimen y del mencionado crecimiento del papel del Estado.

**Gráfica 24.** Distribución de la demanda entre los agentes (% PIB)

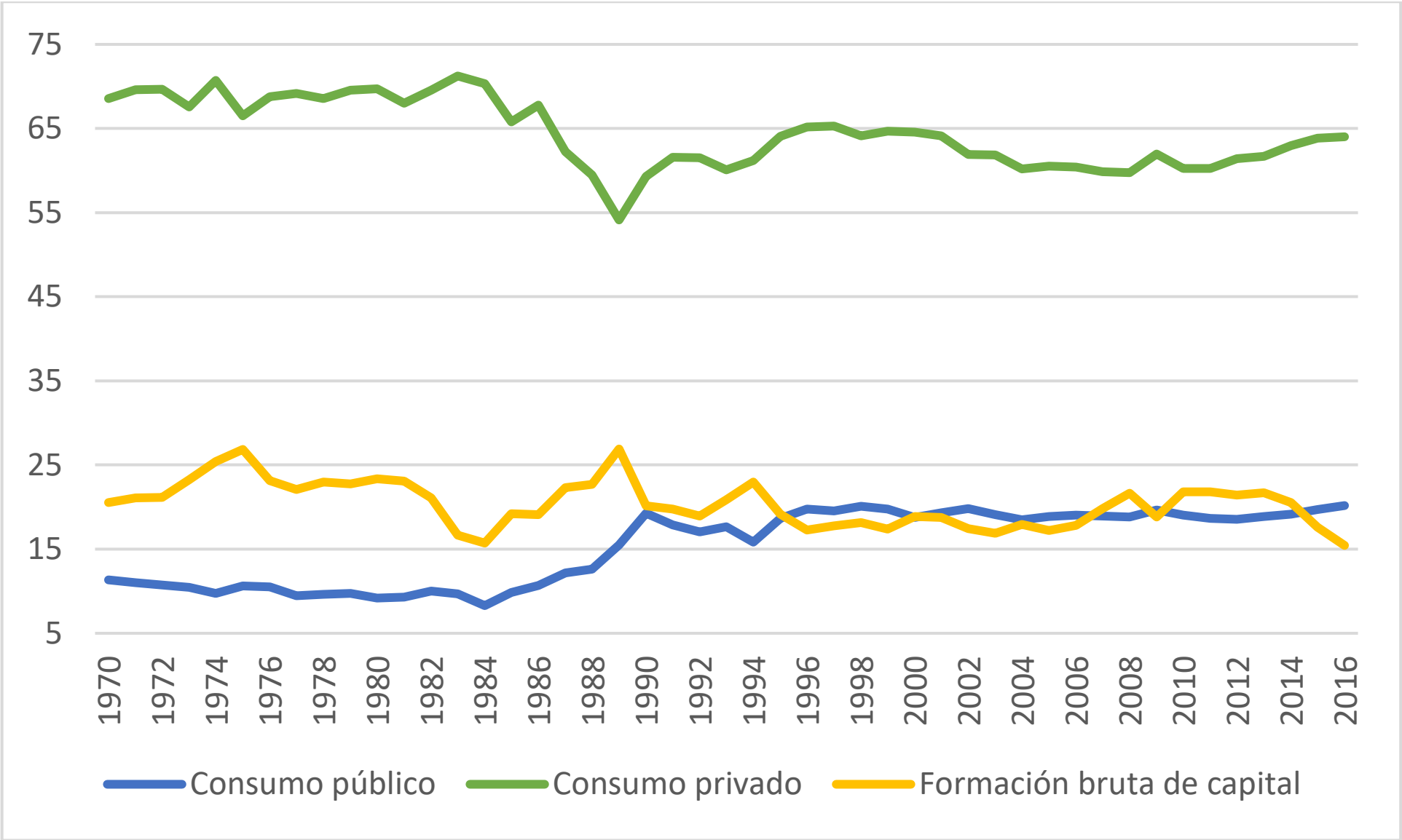

Fuente: elaboración propia a partir de (WB, 2018)

No obstante, el crecimiento de la inversión no se ha mantenido, y en 2016 se situó en un 15% tras una bajada de 5 puntos a partir de la crisis de 2014. La situación de Brasil es atípica si la comparamos con el resto de países de los BRICS y las otras grandes economías sudamericanas. El país carioca es el segundo en importancia del sector público en la demanda agregada, y al mismo tiempo el país donde la inversión representa la menor cuantía (ver Gráfica 25). Debemos recordar que Petrobras, la mayor empresa de Brasil por sus beneficios en 2017, es de propiedad estatal, al igual que la banca pública del país (Banco do Brasil), que se sitúa en cuarto lugar (Fortune, 2018). La sexta mayor empresa brasilera y la mayor eléctrica de Sudamérica, Eletrobras, también es de propiedad pública (aunque haya sido anunciada su privatización).

Así, nos encontramos con que el Estado controla un gran número de empresas, además de mantener bajo su control las principales empresas estratégicas (la mayor petrolera, la mayor eléctrica y la tercera mayor financiera del país). Mientras tanto, la inversión en Brasil cada vez representa una menor parte de su producto interior bruto, lo que suma a la pérdida de competitividad y productividad de la que hablábamos anteriormente. Se hace necesario un plan de fomento de la inversión en el país, independientemente de si se reduce el papel estatal en la economía nacional.

**Gráfica 25.** Consumo público y formación bruta de capital, 2016 (% PIB)

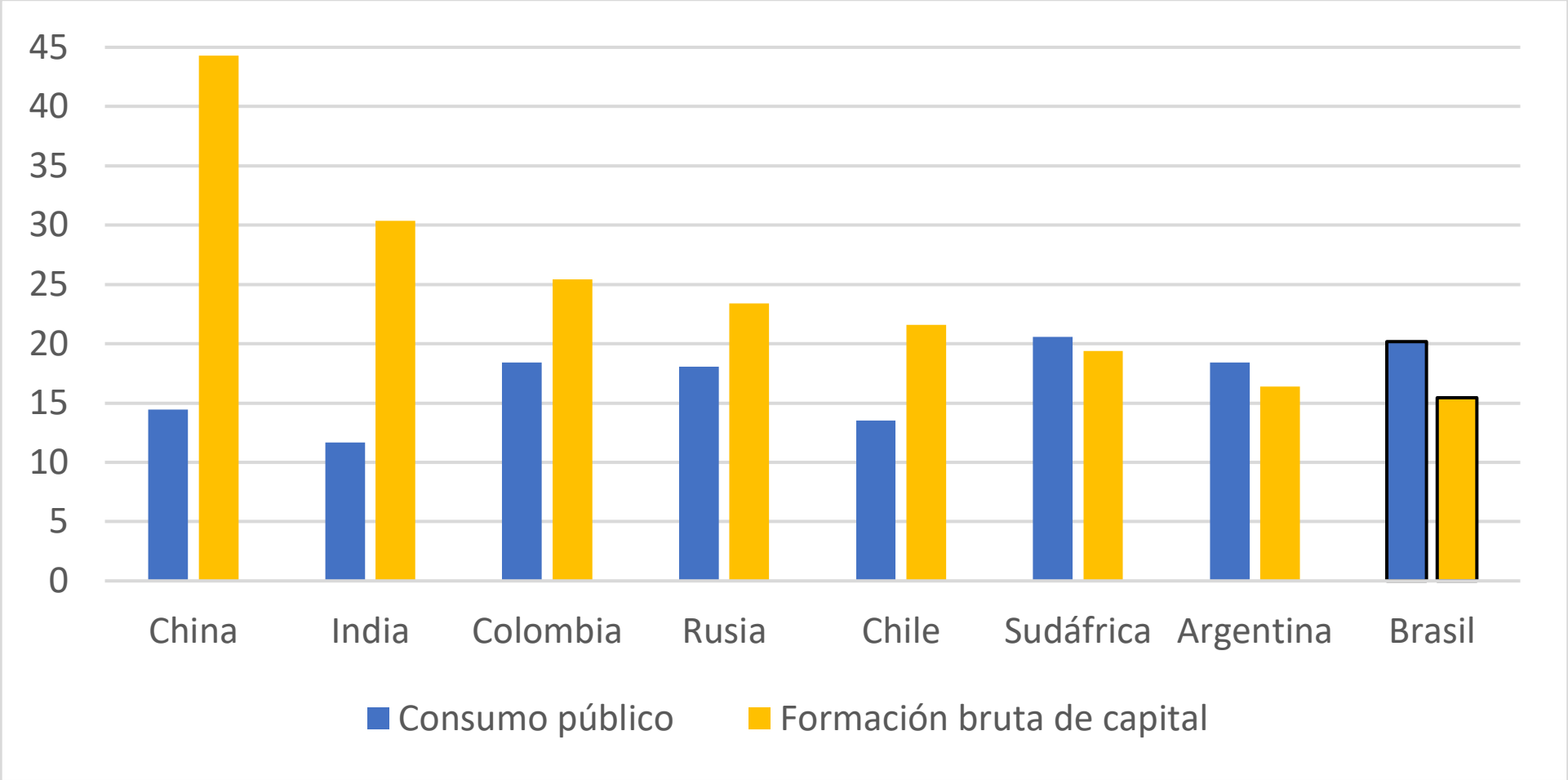

Fuente: elaboración propia a partir de (WB, 2018)

# 6. Análisis de los sectores público y financiero.

## 6.1. Saldo fiscal público, deuda pública y prima de riesgo.

El comportamiento del sector público brasileño nos revela algunas de las deficiencias de la economía del país. Con un déficit público estructural de un 3,5% del PIB de media desde el año 2000, el Estado brasileño revela una constante necesidad de financiación y una inadecuada gestión de la deuda pública y de la política fiscal.

No obstante, la entrada de miles de personas al mercado de trabajo formal y el aumento del consumo debido a la mejora en los salarios y las transferencias públicas durante los últimos 15 años, mejoró la situación de los ingresos públicos y permitió reducir la deuda pese al aumento del gasto público. Además, a la reducción de la deuda se le suma el crecimiento del PIB del país durante estos años. De esta forma, la deuda pública se redujo de un 74% del PIB en 2003 a un 61% en 2011.

**Gráfica 26.** Deuda pública y déficit fiscal

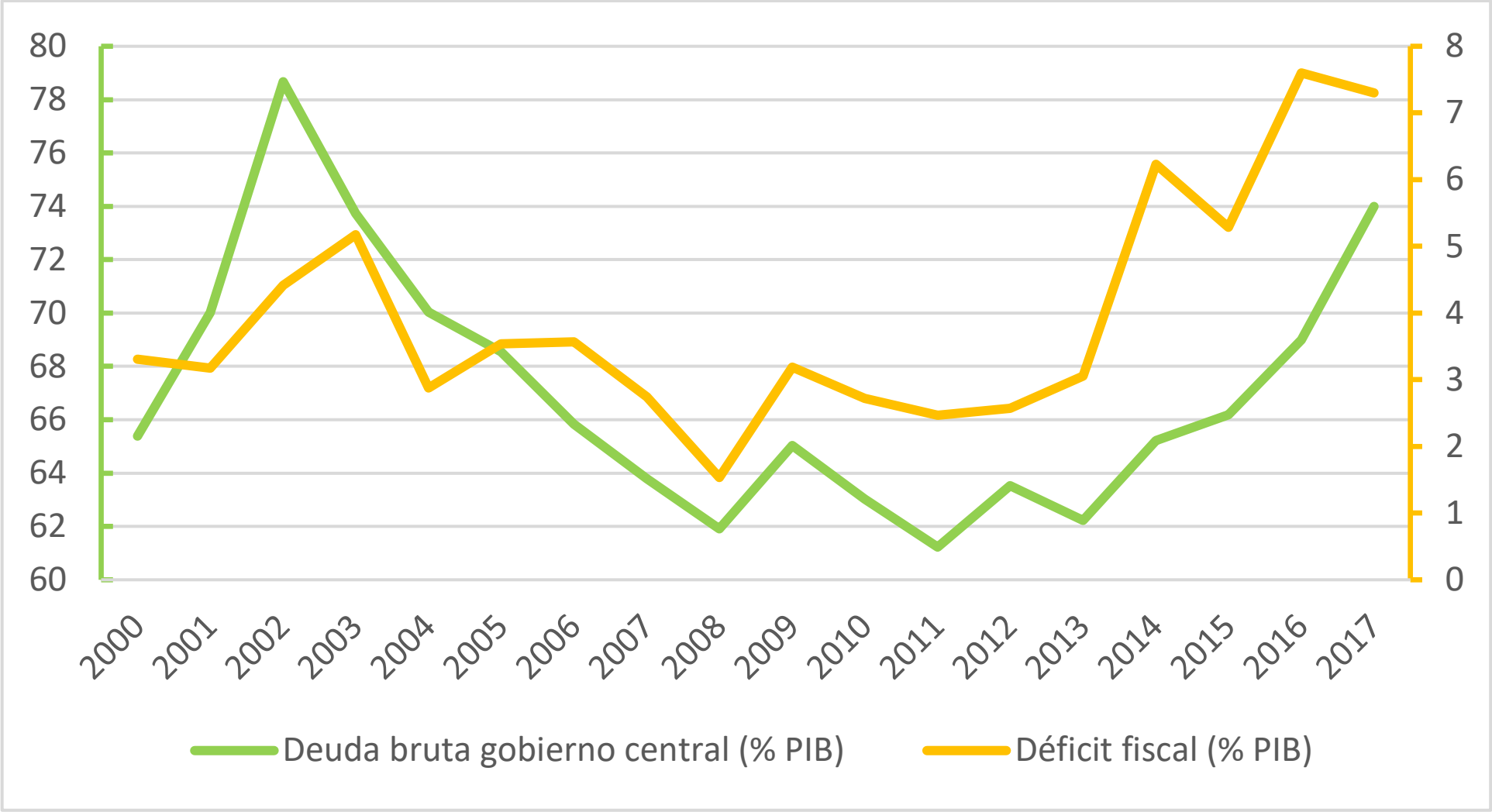

(CEPAL, 2018; IMF, 2017a)

Por otra parte, es necesario resaltar que la estructura estatal de Brasil contribuye al déficit público. La duplicidad de gastos gubernamentales entre Estado Federal y los Estados regionales suma a los gastos públicos más que los ingresos de los impuestos establecidos por las regiones, haciendo que estas tengan un déficit de más del 2% del PIB, y supongan la mitad del déficit primario que ha registrado el país en los años postcrisis.

La mejoría en el ingreso público no ha conseguido acabar con el déficit fiscal secundario, pero sí ha permitido una época de superávit fiscal primario continuado desde el año 2000 hasta 2014. Durante esta época, Brasil mantuvo un superávit primario medio del 2,6%, lo que muestra que sus ingresos superaban a sus gastos sin contar el servicio a la deuda. A pesar de este buen resultado, en 2014, con la llegada de la crisis, el ingreso público se hundió, situándose en números negativos (exceptuando un superávit del 0,2% en 2015) desde entonces. Por tanto, hay que destacar que esta situación podría mejorar con una reforma fiscal que situara la base de la recaudación en los capitales y no en el trabajo y los impuestos directos al consumo, evitando así que la recaudación estatal dependiera tanto del ciclo económico y la demanda privada (CEPAL, 2018).

La gran diferencia que se observa entre déficit fiscal primario y secundario es fruto del pago de la deuda. Los gobiernos brasileños en la actualidad están todavía pagando las deudas del pasado y los intereses que la acompañan, y este servicio resulta cada vez más costoso: de un 5% del PIB en 2008 a un 8% en 2015, duplicando la media de América Latina y el Caribe (*op. cit.*).

Si comparamos la situación de Brasil con el resto de los BRICS y las principales economías sudamericanas, obtenemos una conclusión similar. Desde el año 2000, Brasil ha sido el país que más recursos ha tenido que destinar al servicio de la deuda, situándose el servicio total en más del 40% de las exportaciones e ingresos primarios durante este período, como podemos ver en la Gráfica 27. Estas cifras son 8 veces mayores que las de China o 5 veces mayores que las de Sudáfrica, lo que refleja el alto coste de la deuda brasileña.

**Gráfica 27.** Servicio total a la deuda (% de las exportaciones e ingresos primarios)

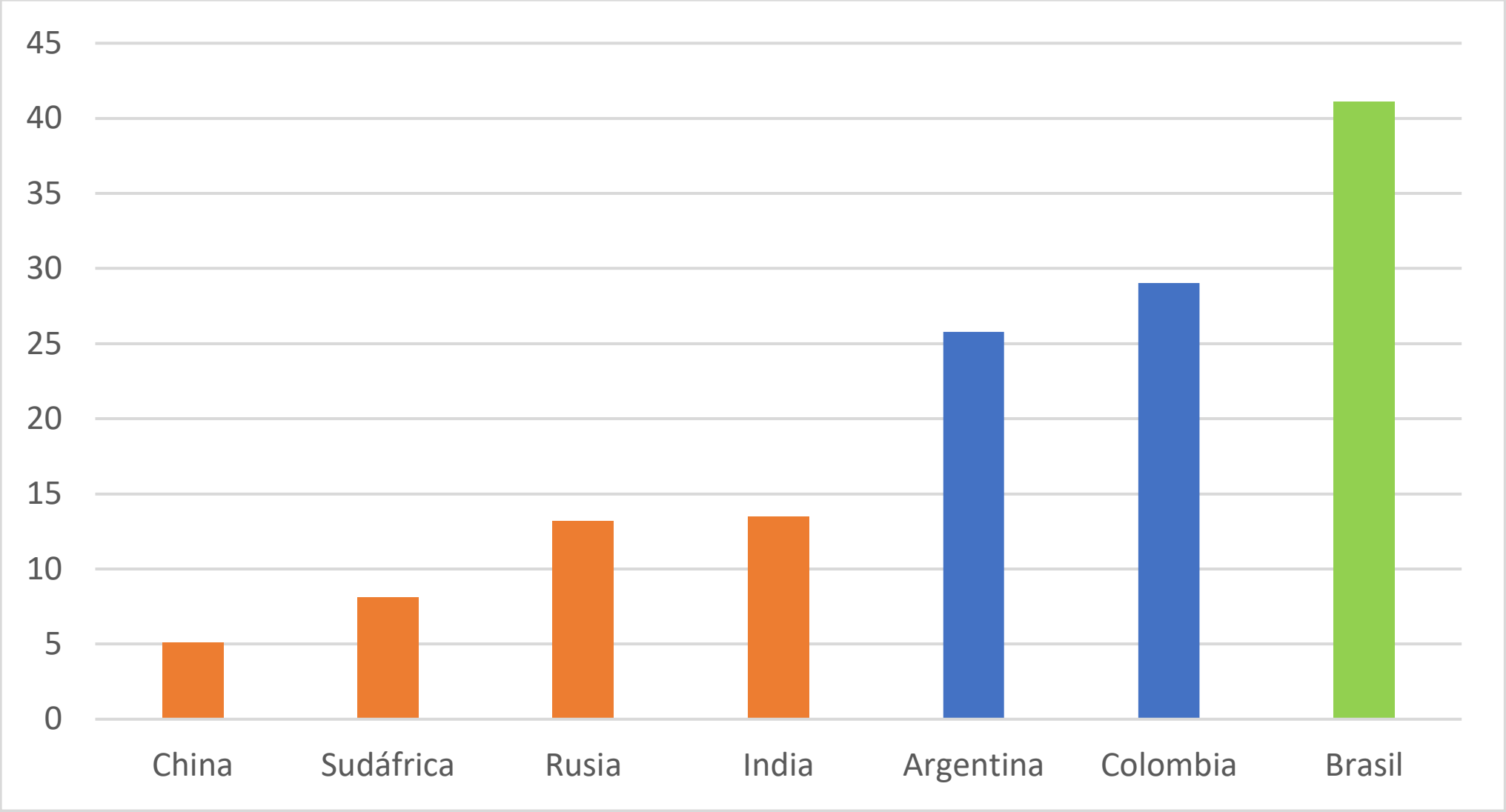

Fuente: elaboración propia a partir de (WB, 2018)

Esto hace que, dentro de este mismo grupo de países, Brasil sea el que acumula un mayor déficit fiscal y una deuda pública más alta con respecto a su PIB, como se manifiesta en la Gráfica 28. Esta deuda acumulada y las expectativas de que el Estado no sea capaz de cambiar su posición estructural hace que la deuda siga aumentando y los costes se recrudezcan cada vez más, poniendo en peligro el crecimiento futuro de la economía. La deuda pública es hoy uno de los principales riesgos de la economía mundial, especialmente para los países emergentes (IMF, 2018). La única noticia positiva sobre la deuda brasileña y la estabilidad del país es que la mayoría de ella (el 94%) está denominada en reales y pertenece a personas del país: en 2015, sólo el 6,5% de su deuda total era externa, siendo este la mejor cifra de toda América Latina (CEPAL, 2016).

**Gráfica 28.** Deuda pública y déficit fiscal en comparativa, último año disponible (2014-2017)

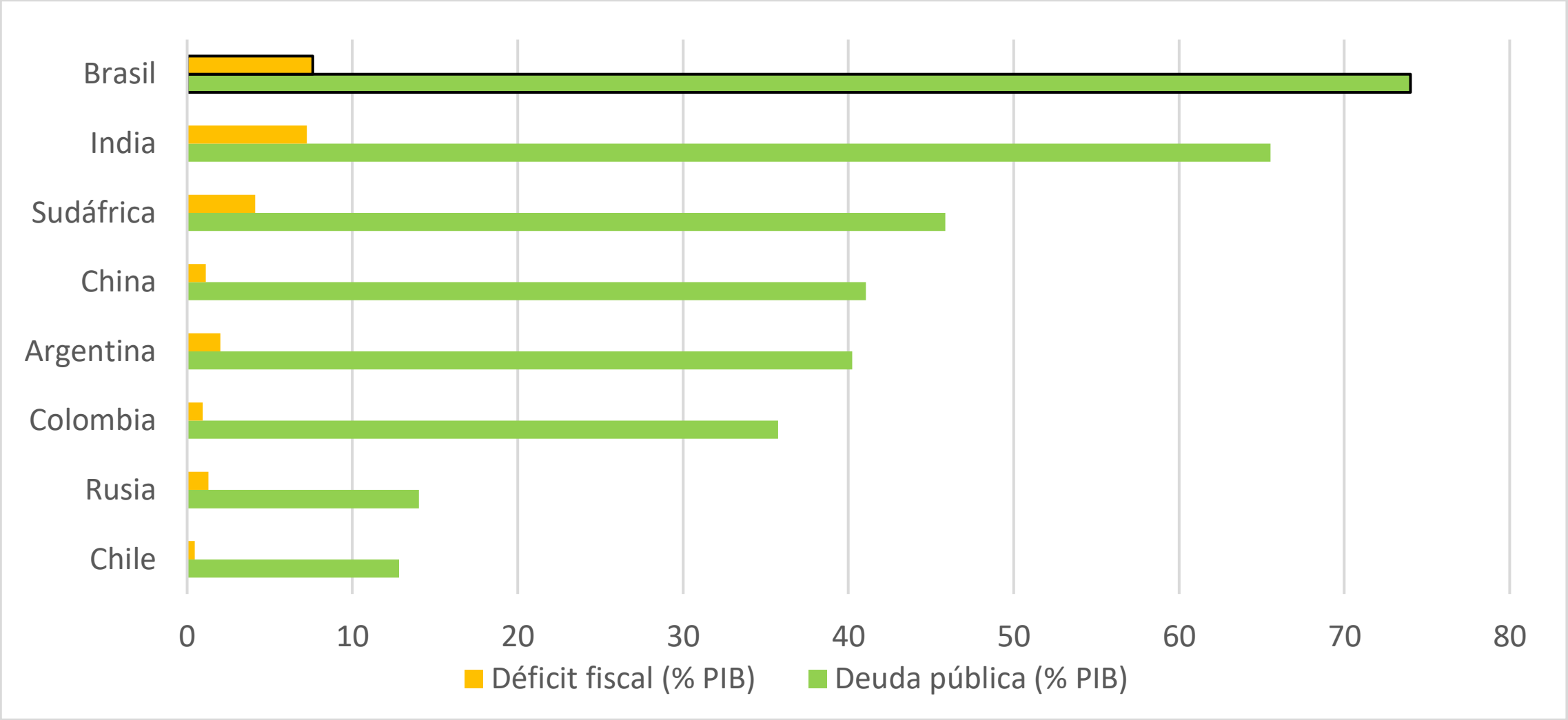

Fuente: elaboración propia a partir de (CEPAL, 2018; IMF, 2015)

## 6.2. Banco central.

La mayor intervención del Banco Central de Brasil y el Ministerio de Hacienda en las últimas décadas ha sido el Plan Real, que frenó la hiperinflación que experimentaba Brasil en la primera mitad de la década de los 90. Para ello se tomaron diferentes medidas en los años 1993 y 1994, principalmente un aumento de los tipos de interés y la sustitución de la moneda nacional. Se sustituyó el *cruzeiro real* por el *real brasileiro*, y el tipo de cambio de este se fijó alrededor de la paridad con el dólar estadounidense, siguiendo un camino similar al emprendido unos años antes por Argentina (Calcagno & Sáinz, 1999). Se redujo así la oferta monetaria, y se redujo la inflación de un 2.075% en 1994 a un 66% en 1995 y un 16% en 1996.

**Gráfica 29.** Inflación y oferta monetaria.

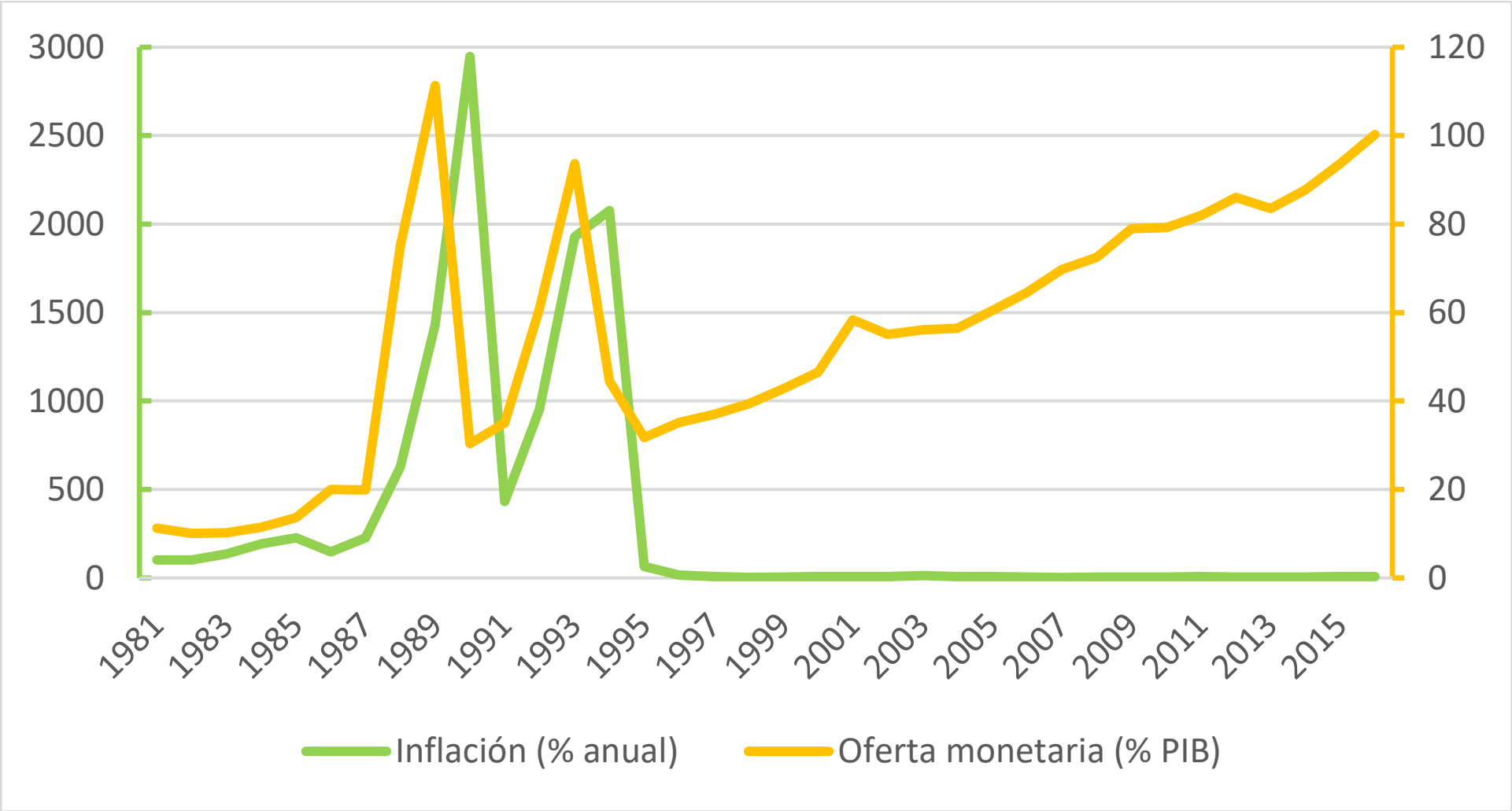

Fuente: elaboración propia a partir de (WB, 2018)

Tras el Plan Real la inflación se ha mantenido relativamente estable entre el 4 y el 10% anual, mientras que las políticas expansivas han aumentado considerablemente la oferta monetaria a través de transferencias directas e inversiones públicas, situándose en el volumen del PIB en 2016.

Todas estas medidas atajaban la situación coyuntural de crecimiento de los precios, y permitieron que los salarios reales crecieran y mejorara la demanda interna y el crédito a las personas físicas. Sin embargo, el Plan no se quedó ahí, sino que intentó atajar uno de los orígenes de la inflación: el sistema financiero público y privado. Durante años, el Estado había financiado el gasto público utilizando el crédito de los bancos públicos, especialmente los gobiernos de los estados habían utilizado los bancos estaduales para su financiación (de forma parecida a la situación de las cajas de ahorro españolas con respecto a las comunidades autónomas).

Así, el Ministerio de Hacienda emprendió una ambiciosa reestructuración del sistema financiero. Muchos de los bancos estatales se privatizaron, pero antes de eso el Estado Federal se hizo cargo de sanearlos, desembolsando a lo largo del proceso unos 100 mil millones de reales. A esto se le suma el aumento del tipo de interés fijado por el Banco Central de Brasil, con el objetivo de frenar la inflación, pero que resultó en un encarecimiento del servicio a la deuda pública, en un momento en que esta no dejaba de subir (Calcagno & Sáinz, 1999). Como resultado, la deuda pública se triplicó en los 4 años que duró el Plan Real y la reestructuración bancaria, hasta situarse en más de 300 mil millones de reales en 1998.

Por otra parte, como vemos en la Gráfica 30, la privatización (bancaria especialmente) atrajo un gran número de capitales extranjeros [5], disparándose la inversión extranjera directa. Dados los altos intereses en el sistema bancario, los sectores productivos domésticos prefirieron invertir en activos financieros antes que pedir créditos para financiar sus inversiones productivas, dejándose la inversión cada vez más en manos de capital extranjero, como luego veremos. De hecho, los créditos domésticos a la industria pasaron de suponer el 32% del total a un 20% en apenas 2 años. Estos cambios estructurales, provocados por las medidas asociadas al Plan Real, siguen estando presentes en la economía brasileña.

[5] Los grandes bancos españoles, como Santander y BBVA, se beneficiaron del desembolso público de las arcas brasileñas para hacerse con numerosos activos en el país (Calcagno & Sáinz, 1999).

**Gráfica 30.** Efectos del Plan Real.

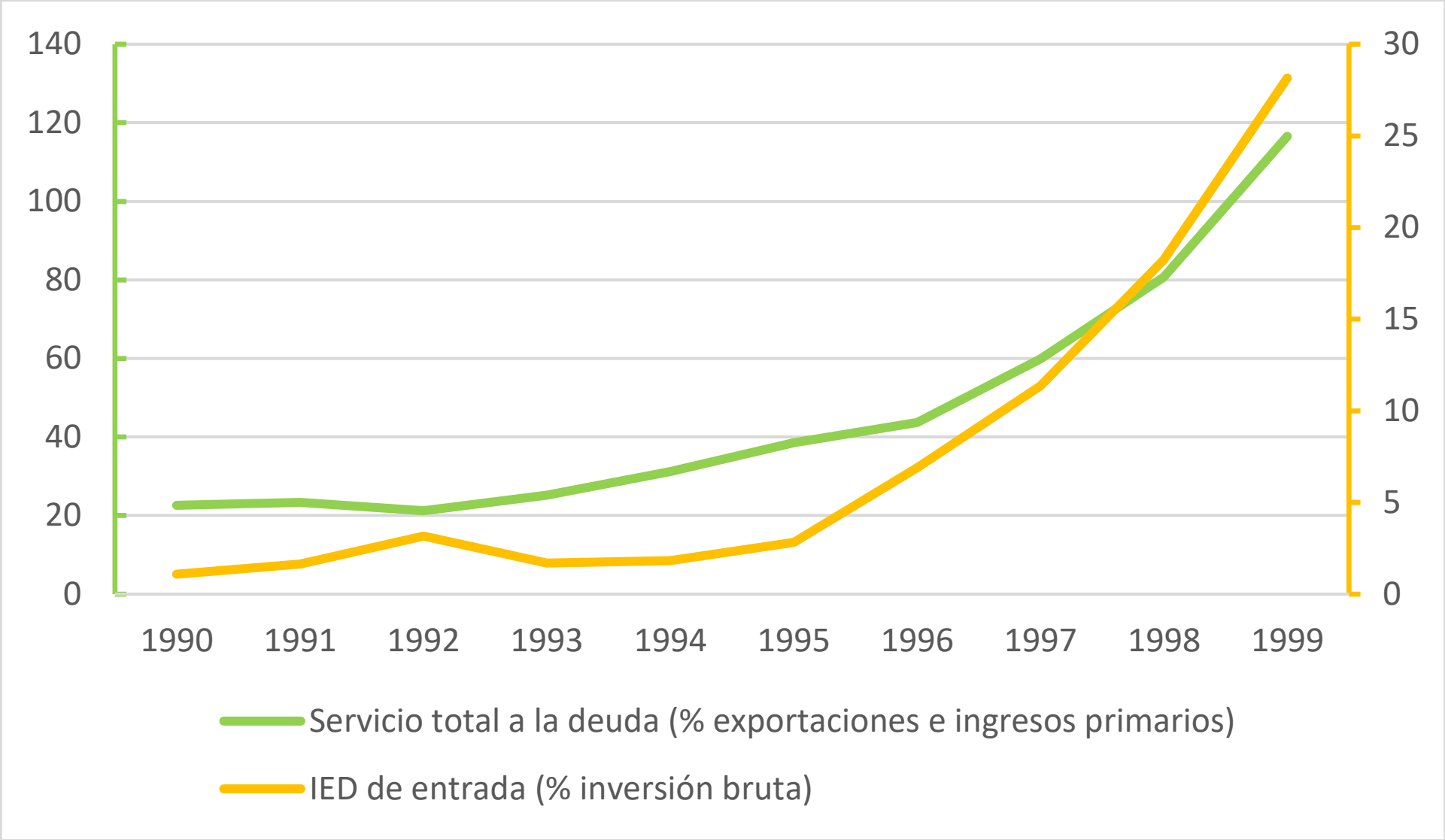

Fuente: elaboración propia a partir de (UNCTAD, 2018; WB, 2018)

Tras la entrada en circulación del real en 1994, se mantuvo en torno a la paridad con el dólar americano unos años, hasta ser sufrir una fuerte depreciación entre 1999 y 2002, por ataques especulativos contra la moneda brasileña, que llegó a cotizar a 4 reales por dólar. La actuación del FMI, asegurando con un préstamo de 30 mil millones el real consiguió frenar la espiral especulativa (Nepomuceno, 2002). Esta depreciación, que encendió todas las alarmas, mejoró las exportaciones, lo que, sumado a los altos precios de las *commodities* y la buena coyuntura internacional, hizo que el valor de las exportaciones se situara en 2008 en un 360% del valor del año 2000.

Esta es una relación bidireccional, es decir, el aumento de las exportaciones también supuso un aumento del valor de la moneda brasileña, que acabó llegando a los niveles previos a los movimientos especulativos del año 1999. Sin embargo, la crisis económica de 2014 ha provocado una nueva devaluación, con lo que volvió el tipo de cambio a rozar los 4 reales por dólar (Mendoça, 2015).

**Gráfica 31.** Evolución del tipo de cambio y exportaciones

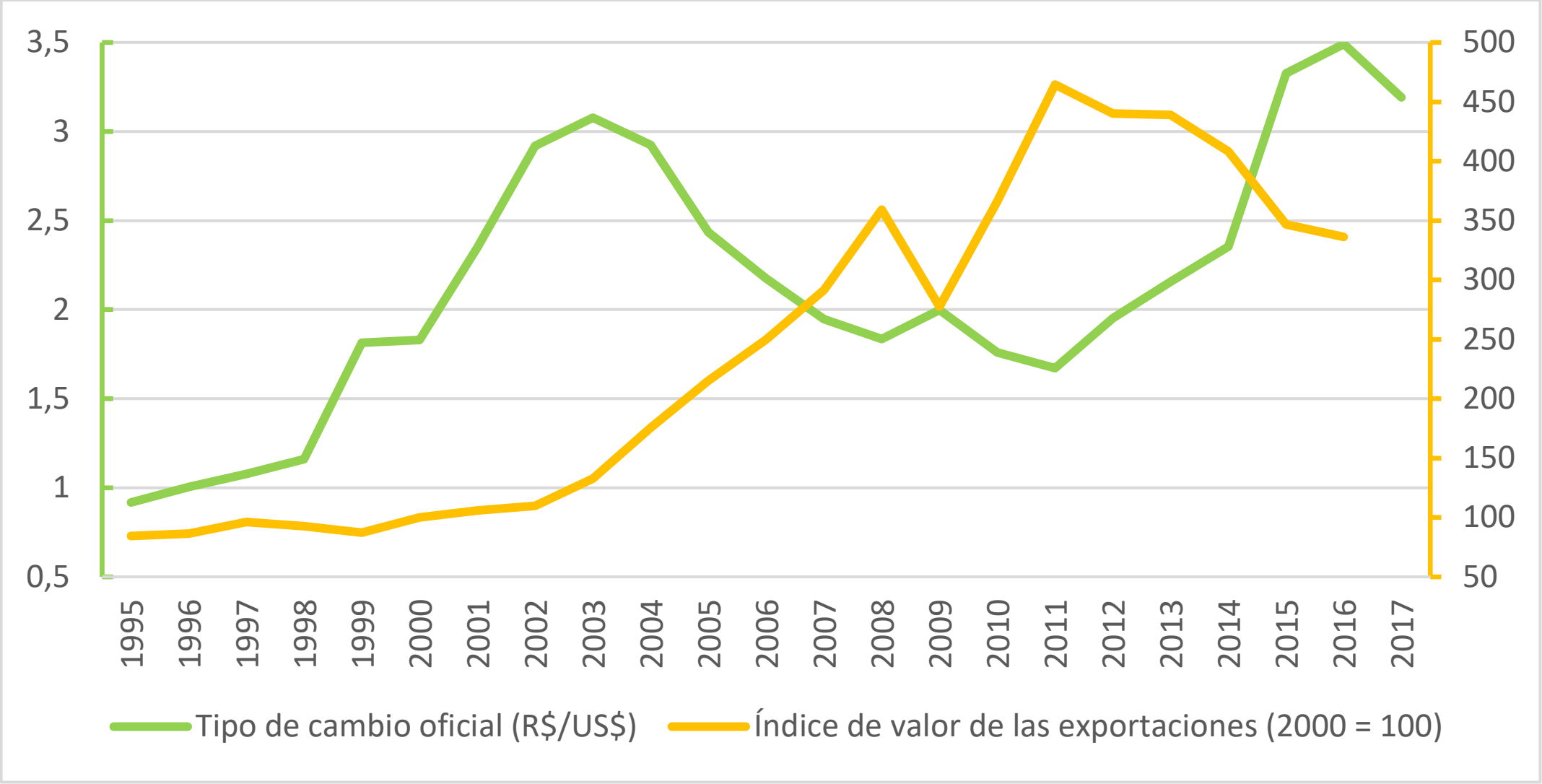

Fuente: elaboración propia a partir de (WB, 2018)

Esta devaluación ha ido en aumento en los últimos meses, hasta cotizar en los mercados a un máximo histórico de 4,2 reales por dólar ante la incertidumbre previa a las elecciones presidenciales de octubre de 2018 (RTVE, 2018b). En este momento, tras la victoria de Bolsonaro en las elecciones, la depreciación se ha frenado y la moneda ha cogido algo de fuerza con respecto a septiembre, volviendo al entorno del 3,6-3,7.

Se espera que con las próximas privatizaciones y la entrada de más capitales externos con la mejora de la estabilidad del país tras las elecciones de octubre la moneda se vaya apreciando ligeramente en los próximos meses.

## 6.3. Sistema bancario y mercados financieros.

En el sistema bancario de Brasil destacan principalmente 5 empresas que controlan el 80% de los créditos del país "El Cronista | Los cinco principales bancos de Brasil concentran el 80% de los créditos" y que son las que más clientes y beneficios acumulan. Como podemos ver en la Gráfica 32, es la banca pública la que más volumen de crédito presenta, dado que tanto la Caixa Econômica Federal como el Banco do Brasil son de propiedad estatal. Bradesco es el banco con más clientes (depósitos y créditos), Itaú Unibanco es el banco más importante tanto por activos como por beneficios de toda América Latina, y Santander ha conseguido estar entre las 5 primeras tanto en créditos, como en beneficios y clientes, siendo el único banco extranjero entre los 8 más importantes de Brasil. Esta importancia del Banco Santander ha quedado clara al anunciar el presidente electo Bolsonaro que Roberto Campos Neto, actual director del Santander Brasil, presidirá bajo su mandato el Banco Central do Brasil (Benites, 2018).

**Gráfica 32.** Principales entidades bancarias de Brasil, 2017.

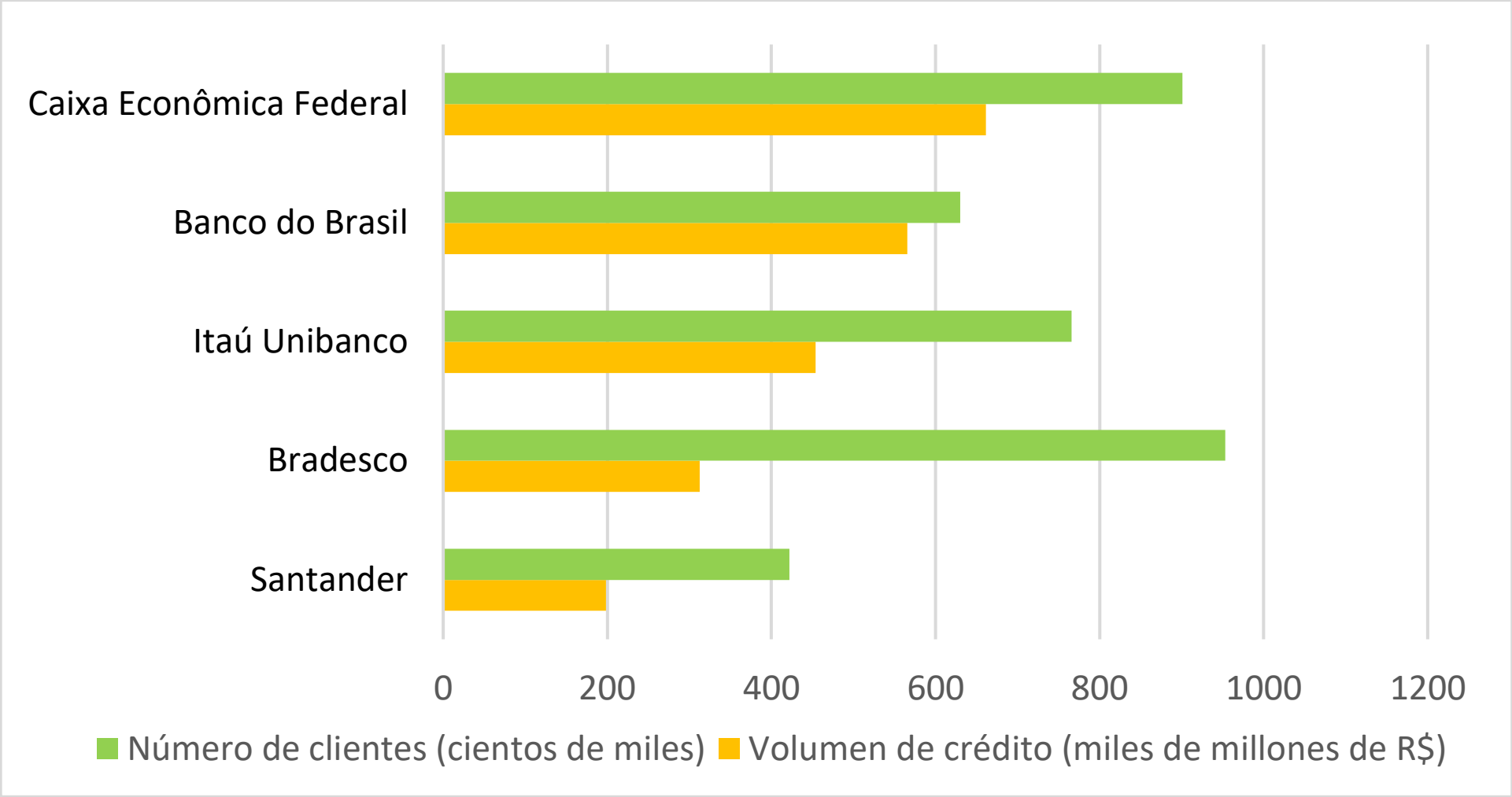

Fuente: elaboración propia a partir de (BCB, 2018a; Valor, 2017)

Estos bancos son los que controlan el sistema financiero de Brasil, con la particularidad de que dos de ellos son públicos. Además, todos dependen de los tipos de interés que fije el Banco Central, que han ido bajando desde un 14,3% en 2016 a un 6,5% en 2018, donde previsiblemente se estabilizarán debido a la rigidez de los planes para mantener controlada la inflación (Expansión, 2018). Estos tipos de interés condicionan el comportamiento del sector financiero, haciendo que los tipos de interés de crédito sean muy elevados. Si lo comparamos con los BRICS y las potencias sudamericanas, el tipo de interés de Brasil es cuatro veces mayor que la media del grupo. De hecho, en 2016 y 2017, este tipo de interés fue el segundo más alto del mundo, sólo por detrás de Madagascar.

**Gráfica 33.** Tipo de interés de crédito (%), 2016

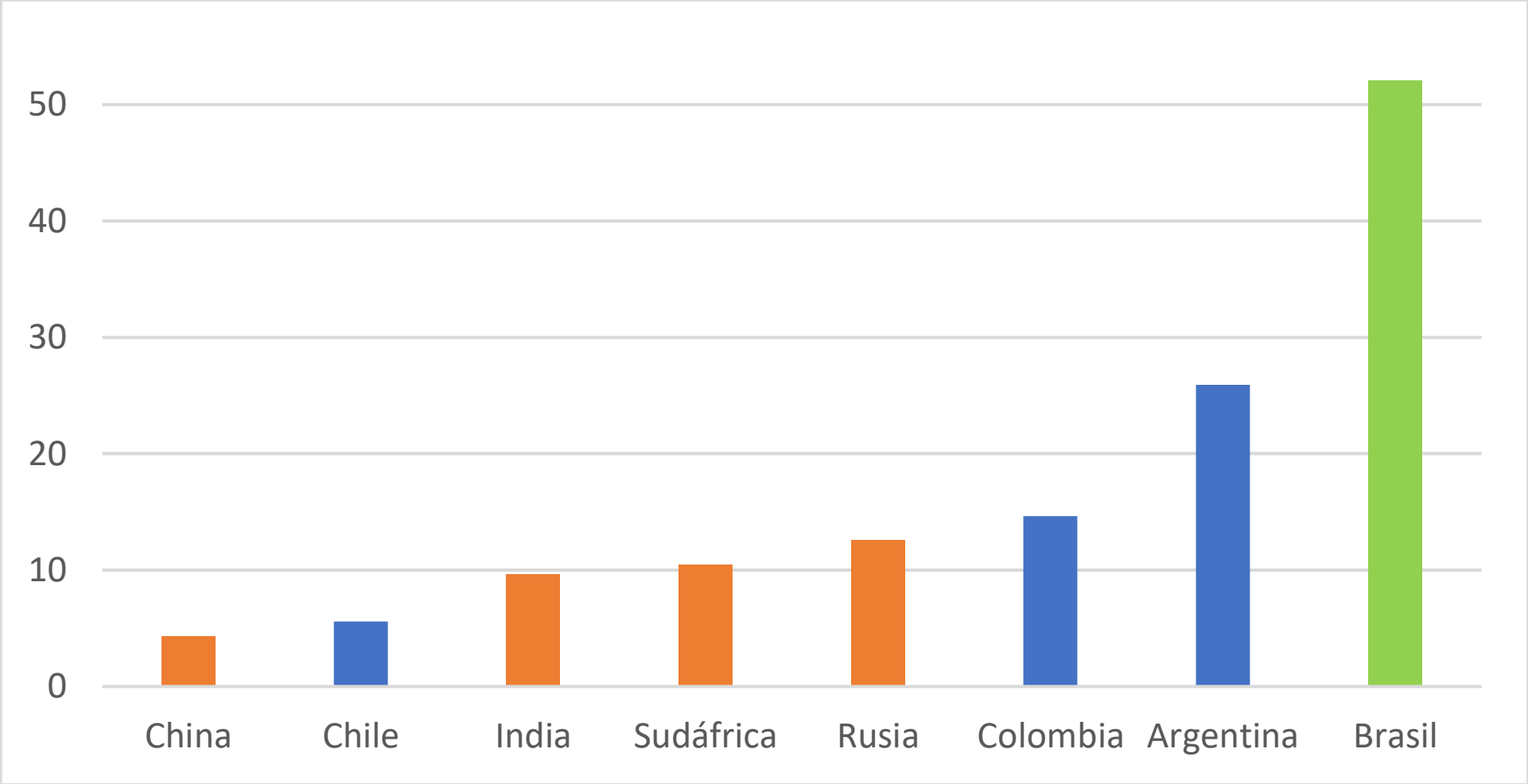

Fuente: elaboración propia a partir de (WB, 2018)

Estos altos tipos de interés dificultan la financiación de las empresas brasileñas, teniendo en cuenta que, según el Banco Mundial, alrededor del 45% de las compañías del país tienen que recurrir a créditos bancarios para poder realizar inversiones. Por otra parte, esto hace que las empresas inviertan más en productos financieros que en inversiones productivas, como se vio anteriormente.

Todo ello afecta al comportamiento de las compañías en los mercados de renta variable. La capitalización bursátil de Brasil ha estado por debajo de la media del grupo BRICS-Sudamérica en los últimos 10 años, no llegando esta al 50% del PIB, algo que sólo ocurre en otras 2 economías del grupo: Argentina y Rusia. Esta débil capitalización hace que las capacidades de financiación se vean reducidas, lo que profundiza en el estancamiento de la productividad.

**Gráfica 34.** Capitalización bursátil media (% PIB), 2008-2017.

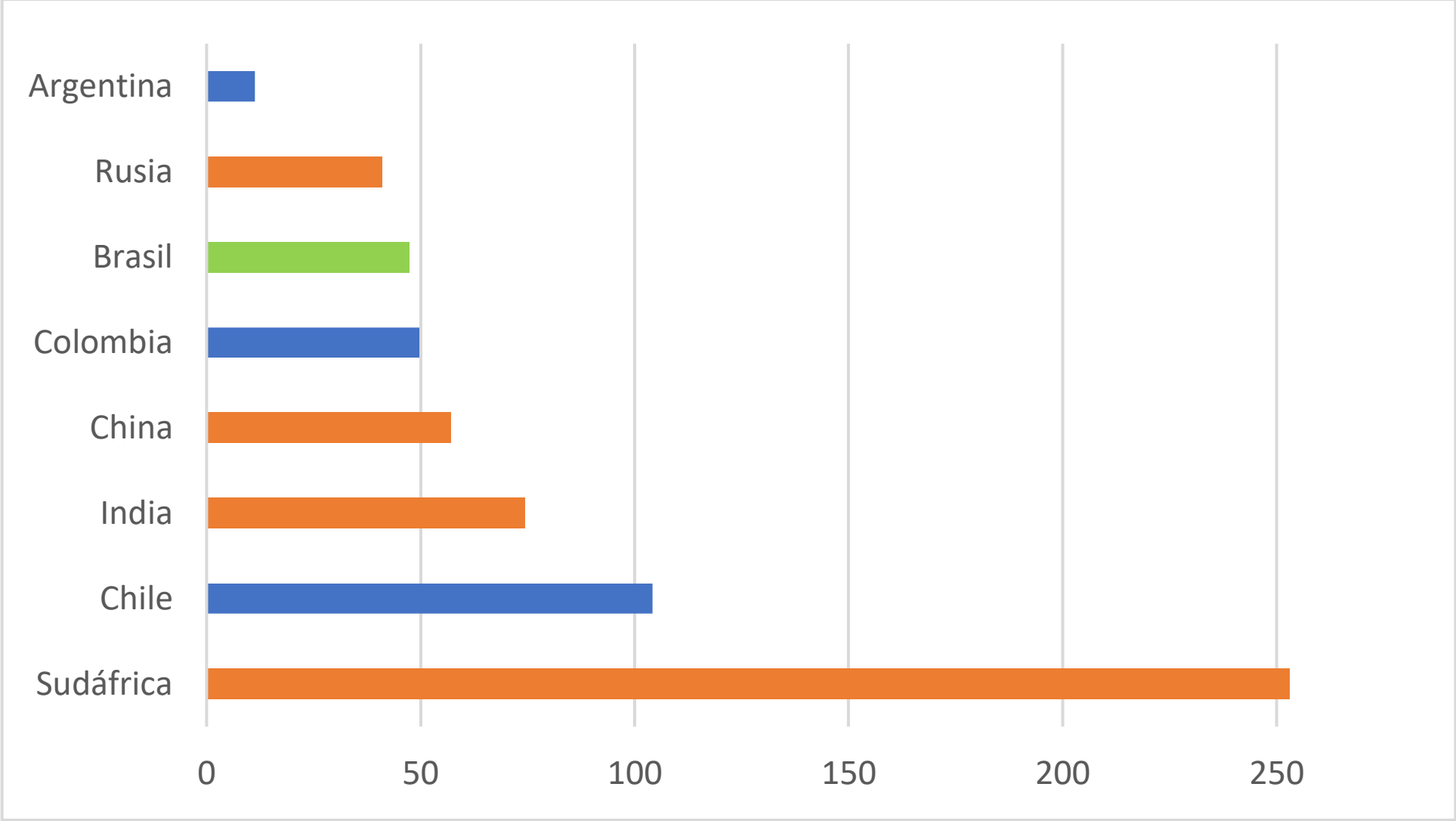

Fuente: elaboración propia a partir de (WB, 2018)

No obstante, la Bolsa de Sao Paulo ha "bendecido" la elección de Jair Bolsonaro como presidente. El índice BOVESPA alcanzó su máximo histórico 4 días después de la victoria de Bolsonaro en segunda vuelta, creciendo más de 5.000 puntos en este breve período (EFE, 2018b; Investing, 2018).

# 7. Análisis de la inserción externa de la economía.

## 7.1. Apertura externa.

Brasil, debido a su gran mercado interno y a la baja competitividad internacional de sus productos (excepto los productos mineros y agrícolas), nunca ha tenido una economía volcada hacia el exterior desde los tiempos coloniales y el tiempo de transición de economía de plantación azucarera-cafetera a la economía más diversificada que es hoy en día. Hoy en día, aunque su apertura externa haya mejorado con el paso de los años, la suma de importaciones y exportaciones sigue suponiendo menos de una cuarta parte de su PIB, algo inusual en unos países con tan alta inserción externa como los BRICS.

**Gráfica 35.** Grado de apertura externa (% PIB), 1970-2017.

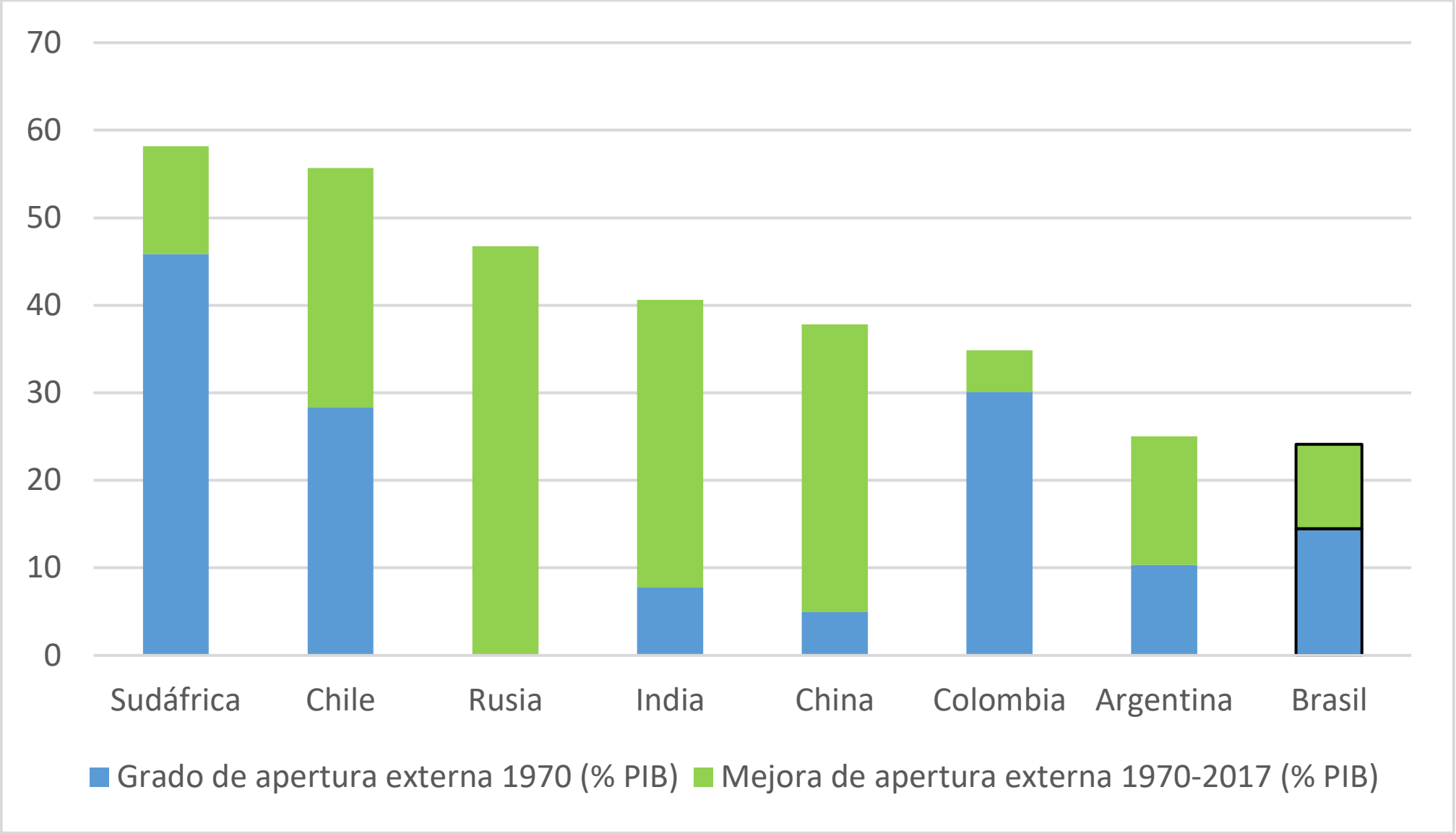

Fuente: elaboración propia a partir de (WB, 2018)

Actualmente presenta una baja apertura externa, y además ha sido el segundo país de los BRICS y potencias sudamericanas cuya apertura externa ha crecido menos en los últimos 50 años, sólo por detrás de Colombia. Esto nos muestra hasta qué punto la economía brasileña ha estado volcada hacia sí misma durante mucho tiempo, hasta que con la llegada del nuevo siglo las exportaciones y las importaciones comenzaron a aumentar a un ritmo hasta entonces nunca visto en Brasil.

Durante los años 80 y 90 las exportaciones se duplicaron, pero crecieron a un ritmo bastante lento en comparación con el resto de países. Sin embargo, a partir del año 2002, las exportaciones empezaron a aumentar a gran velocidad, y las importaciones les siguieron a partir de 2004, como podemos ver en la Gráfica 36. Las exportaciones llegaron a alcanzar su máximo en el año 2011, con un valor que suponía el 465% del del año 2000. A su vez, las importaciones lo alcanzaron en 2013, cuando su valor fue más de cuatro veces superior al de 2000. Este impresionante crecimiento se ha visto frenado por los años de crisis económica (a partir de 2014) y crisis política, que duran hasta la actualidad.

**Gráfica 36.** Evolución del valor del comercio exterior.

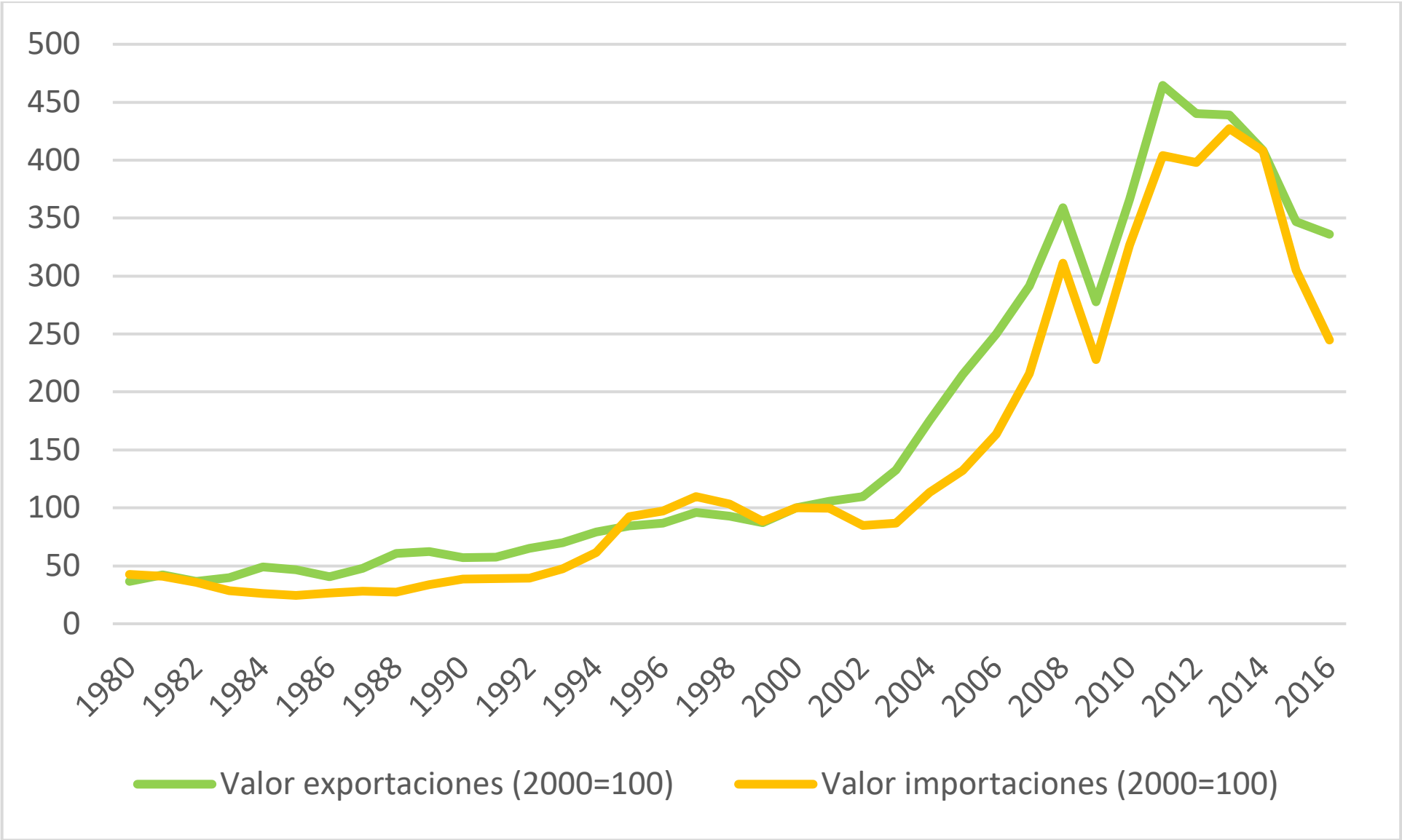

Fuente: elaboración propia a partir de (WB, 2018)

Surge ahora una pregunta: ¿por qué se dispararon las exportaciones a partir del año 2002 si, como hemos visto antes, los productos brasileros no son muy competitivos? Existen varios factores que influyeron en esta sustancial alza del comercio exterior. En primer lugar, la década de los 2000 es, hasta la crisis financiera de 2007, una época de crecimiento económico espectacular en prácticamente todo el mundo. En Brasil, el PIB per cápita (en PPA) subió un 21,4% entre 2000 y 2008, y el comercio internacional global pasó de suponer el 51% al 61% del Producto Mundial (WB, 2018). Un mayor comercio internacional impulsó a la economía brasileña a buscar oportunidades de negocio en el exterior, ampliando su capacidad exportadora.

En segundo lugar, siendo probablemente el factor más importante, es esta la época de lo que se conoce como el boom de las *commodities*. Como vemos en la Gráfica 37, los precios de los bienes no dejaron de subir desde 2002, prácticamente triplicándose entre 2002 y 2011. Este aumento de los precios se ha relacionado con diversas causas, pero la que más consenso tiene es el crecimiento de la demanda interna y las importaciones de las economías emergentes, especialmente de China y de India (Cypher, 2010; Gallagher & Porzecanski, 2009). Además, la consolidación de las economías del sudeste asiático tras la crisis financiera asiática iniciada en 1997 aumentó la demanda de materias primas para sostener las industrias manufactureras de estos países. Esta demanda no se veía satisfecha totalmente por la oferta existente, lo que hizo que los precios impulsaran las exportaciones de materias primas de muchos países, especialmente de los latinoamericanos. Esto planteó serias dudas sobre la fiabilidad de la hipótesis de Prebisch y Singer, que pronosticaba una bajada de los precios y un deterioro de la capacidad de importación de los países latinoamericanos.

Los precios se dispararon, especialmente los de los minerales y los productos agrícolas (destacando la soja y el maíz), productos hacia los que Brasil volcó su exportación. De esta forma, Asia ha pasado de ser el cuarto destino de las exportaciones brasileras (16% del total) a ser el principal destino, acumulando en 2016 el 41% del total.

**Gráfica 37.** Boom de las *commodities* e importaciones realizadas por China e India

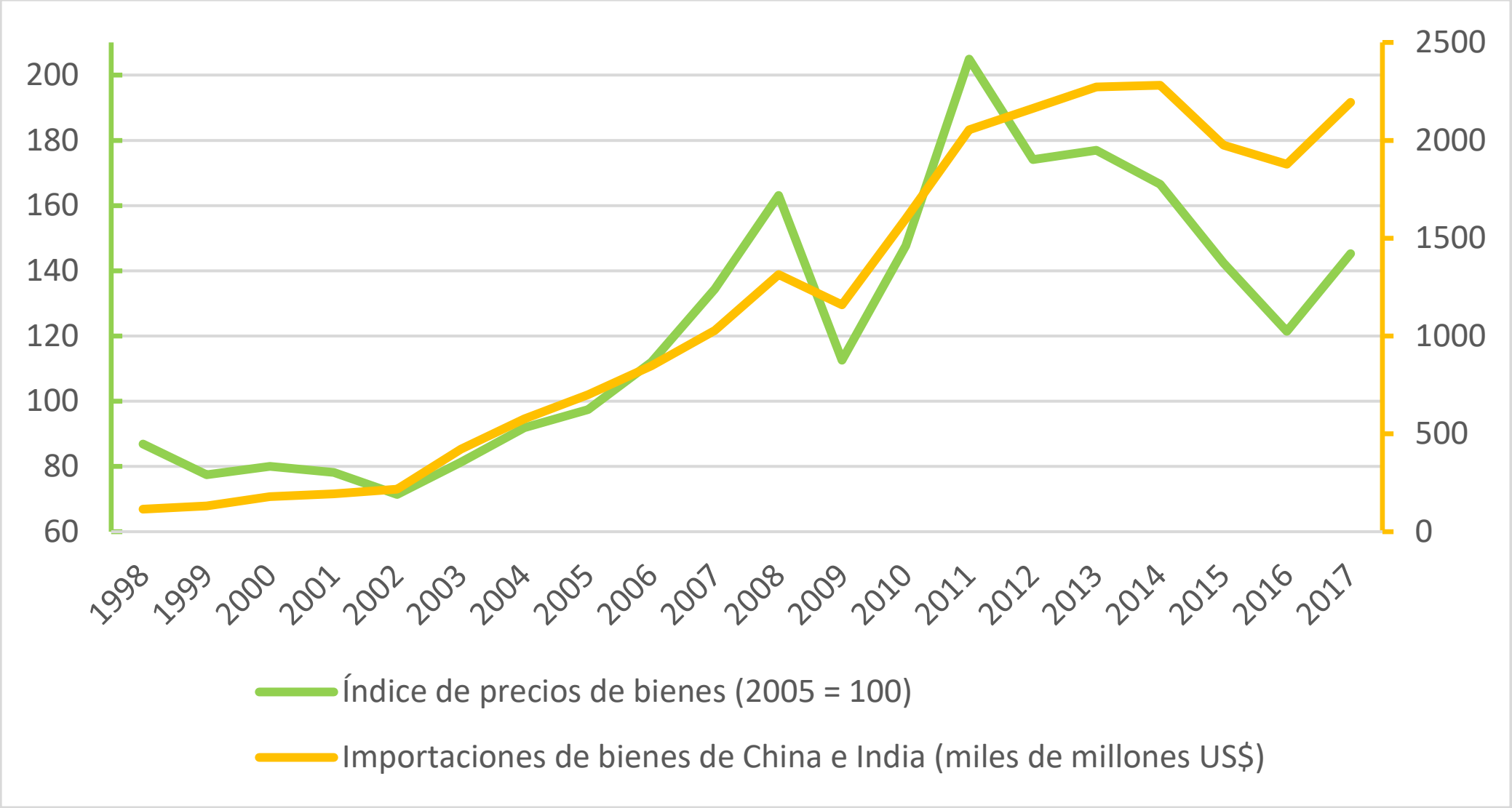

Fuente: elaboración propia a partir de (IMF, 2017b; WB, 2018)

Finalmente, como antes vimos, la devaluación del real brasileño durante 1999-2002 fue otro de los factores que permitió ese boom de las exportaciones brasileras.

Por otra parte, las importaciones crecieron debido a la activación económica del país durante esta época, además de por el incremento de la demanda interna derivada de la subida de los salarios mínimos y medios y de la salida de millones de personas de la pobreza.

## 7.2. Balanza de pagos.

El crecimiento de las exportaciones y la mejora de la balanza comercial no consiguió inclinar por mucho tiempo la balanza de pagos hacia los números positivos. Sólo entre 2003 y 2006 fue la balanza positiva, volviendo tras ello a un déficit de más de 200 mil millones de dólares con la llegada de la crisis de 2014.

Como vemos en la Gráfica 38, no podemos atribuir la totalidad de la responsabilidad de este déficit ni a la balanza por cuenta corriente ni a la cuenta financiera, ya que ambas se comportan prácticamente de manera idéntica en esta época.

**Gráfica 38.** Balanza de pagos (millones de US$)

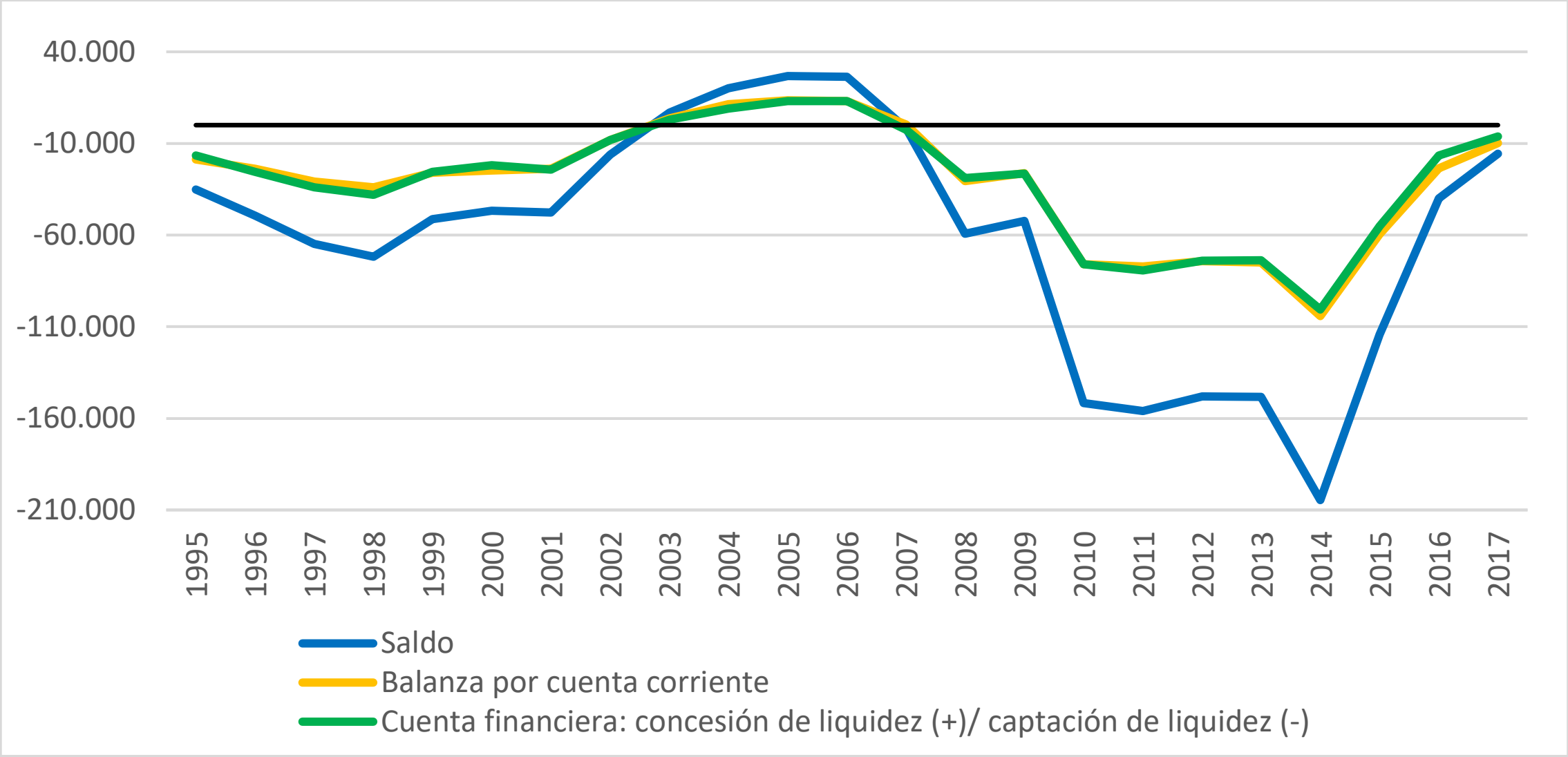

Fuente: elaboración propia a partir de (BCB, 2018b)

Por ello, veamos las subbalanzas por partes. En el caso de la balanza comercial, es positiva sólo entre 2003 y 2007, hundiéndose con la crisis de 2014. A su vez, la balanza de bienes ha sido la que ha impulsado la balanza por cuenta corriente en su conjunto, acumulando superávit desde 2001 (exceptuando en 2014). Por otro lado, la balanza de servicios no ha llegado nunca al equilibrio, especialmente a partir de 2004, donde el aumento de la demanda interna de servicios ha hecho que las importaciones de estos se disparen. De la misma manera, la balanza de rentas primarias lleva lastrando la balanza por cuenta corriente desde el comienzo de la serie, debiéndose principalmente a la repatriación de los dividendos y del rendimiento del capital extranjero invertido en Brasil, que ha llegado a extraer del país 70.500 millones de dólares en el año 2011.

**Gráfica 39.** Balanza por cuenta corriente (millones de US$)

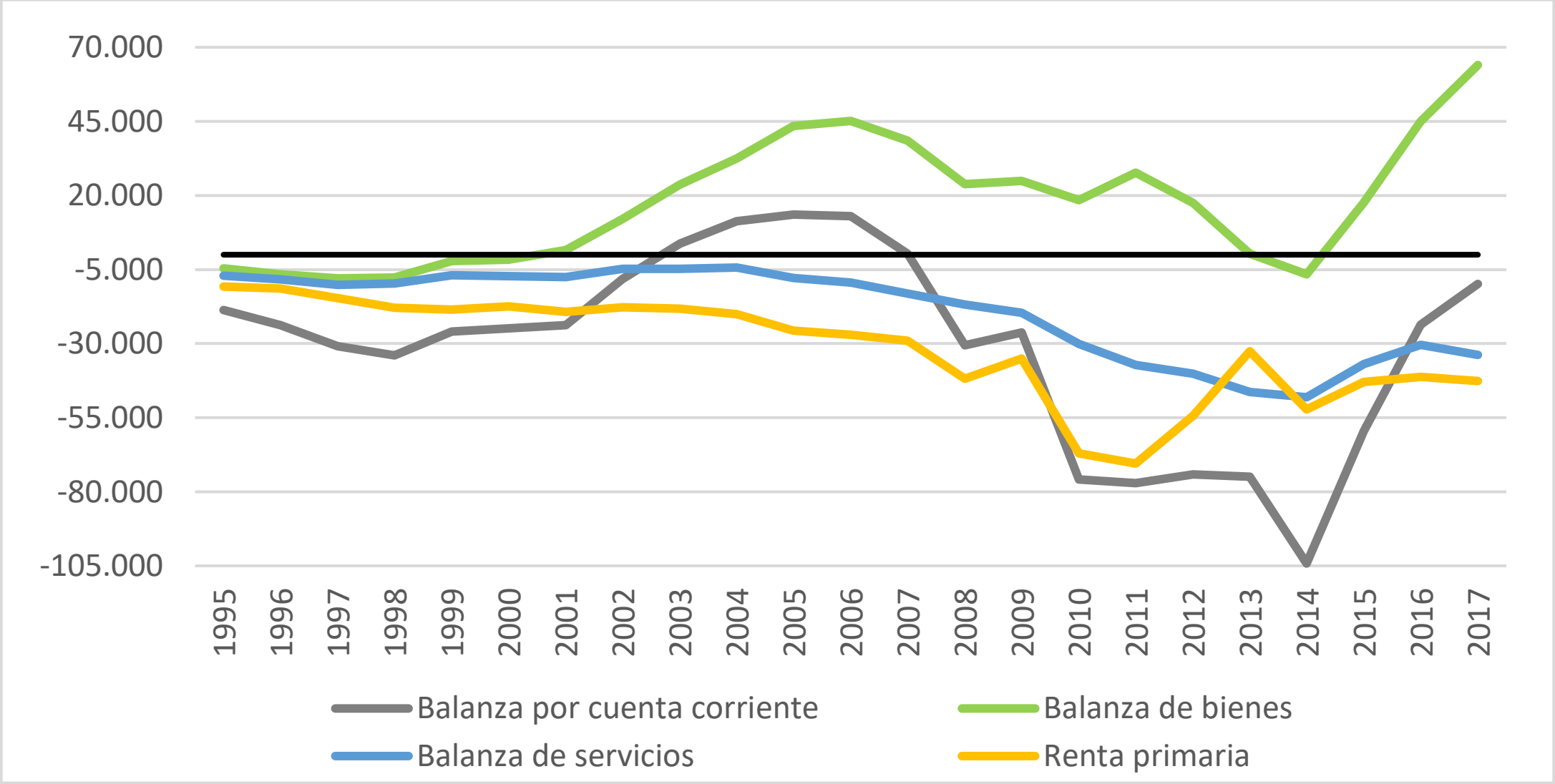

Fuente: elaboración propia a partir de (BCB, 2018b)

De igual forma, si analizamos la cuenta financiera de la balanza de pagos, encontramos que hay varios apartados que arrastran al déficit a la cuenta. Entre ellos, destacan los préstamos recibidos por Brasil [6], tanto por el Gobierno como por los agentes privados (recordemos que Santander es el quinto banco más importante del país), las inversiones en cartera (sobre todo títulos de renta fija) y, especialmente, la inversión extranjera.

**Gráfica 40.** Balanza financiera (millones de US$)

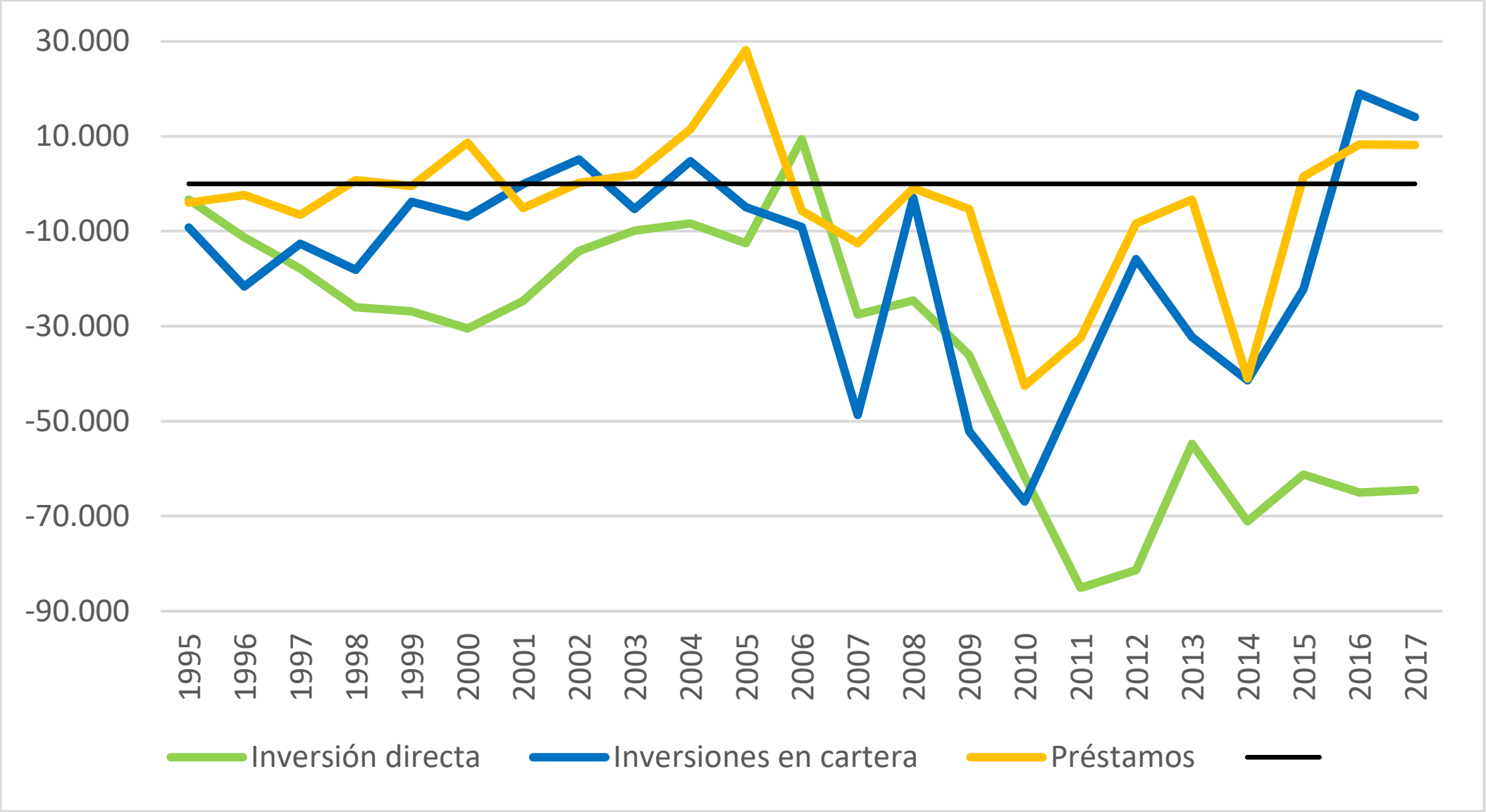

Fuente: elaboración propia a partir de (BCB, 2018b)

En conclusión, Brasil presenta un grave problema de necesidad constante de financiación exterior, llegando a un déficit financiero de 100 mil millones de dólares en 2014. Tras la crisis y la restructuración económica del país, acabar con esta necesidad de financiación ha sido uno de los objetivos de los gabinetes de Rousseff y Temer, llegando a un déficit financiero de sólo 6 mil millones en 2017.

## 7.3. Inserción comercial.

Como se ha visto anteriormente, las principales exportaciones de Brasil son las exportaciones de productos agrícolas, donde destaca la soja. En segundo lugar se encuentran las exportaciones de maquinaria y transportes, sobre todo de aviones, coches y barcos al resto de América. Cabe mencionar que en Brasil se encuentran las plantas de montaje de varias empresas aeronáuticas de primer nivel, además de contar con la mayor aeronáutica de América Latina (Jewell, 2017). En tercer lugar, los combustibles y minerales, con la predominancia de mineral de hierro y crudo mencionada anteriormente.

[6] El incremento que se aprecia en los préstamos en el año 2010 puede deberse a la necesidad de financiación para las inversiones necesarias para los Juegos Olímpicos de Río de Janeiro de 2016, cuya elección fue a finales de 2009.

**Gráfica 41.** Exportaciones de bienes por producto, 2016.

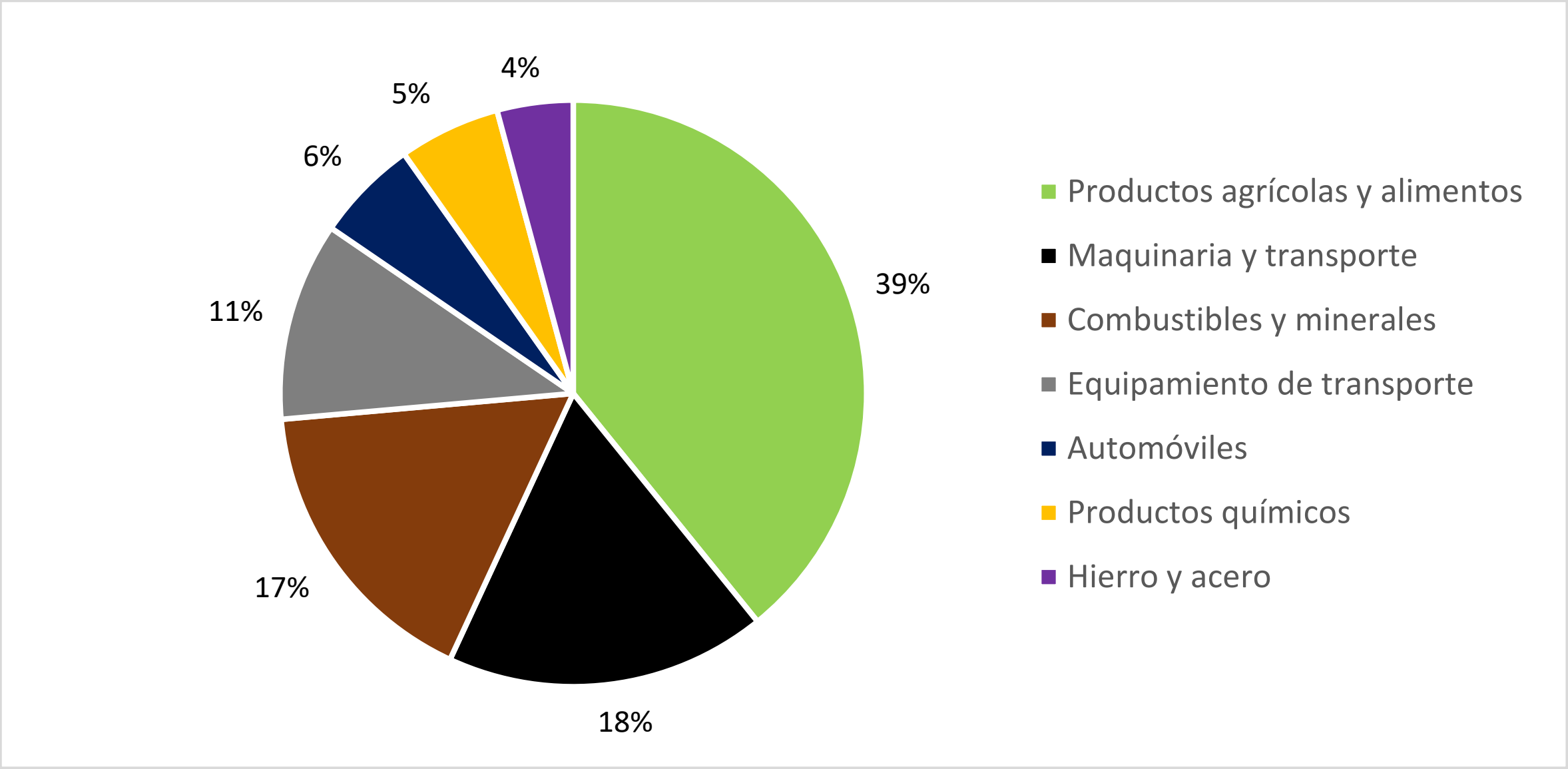

Fuente: elaboración propia a partir de (WTO, 2018)

Por la parte de las importaciones, destaca la demanda de maquinaria y transporte, en este caso de todos los productos intermedios de los transportes montados en Brasil. Tras ello le siguen los productos químicos, con una producción doméstica insuficiente para cubrir las necesidades de la industria y de una agricultura que usa de forma cada vez más intensiva los pesticidas y fertilizantes (OEC, 2018). Finalmente, la importación de petróleo ya refinado hace que los productos minerales y combustibles se sitúen en tercer lugar.

**Gráfica 42.** Importaciones de bienes por producto, 2016.

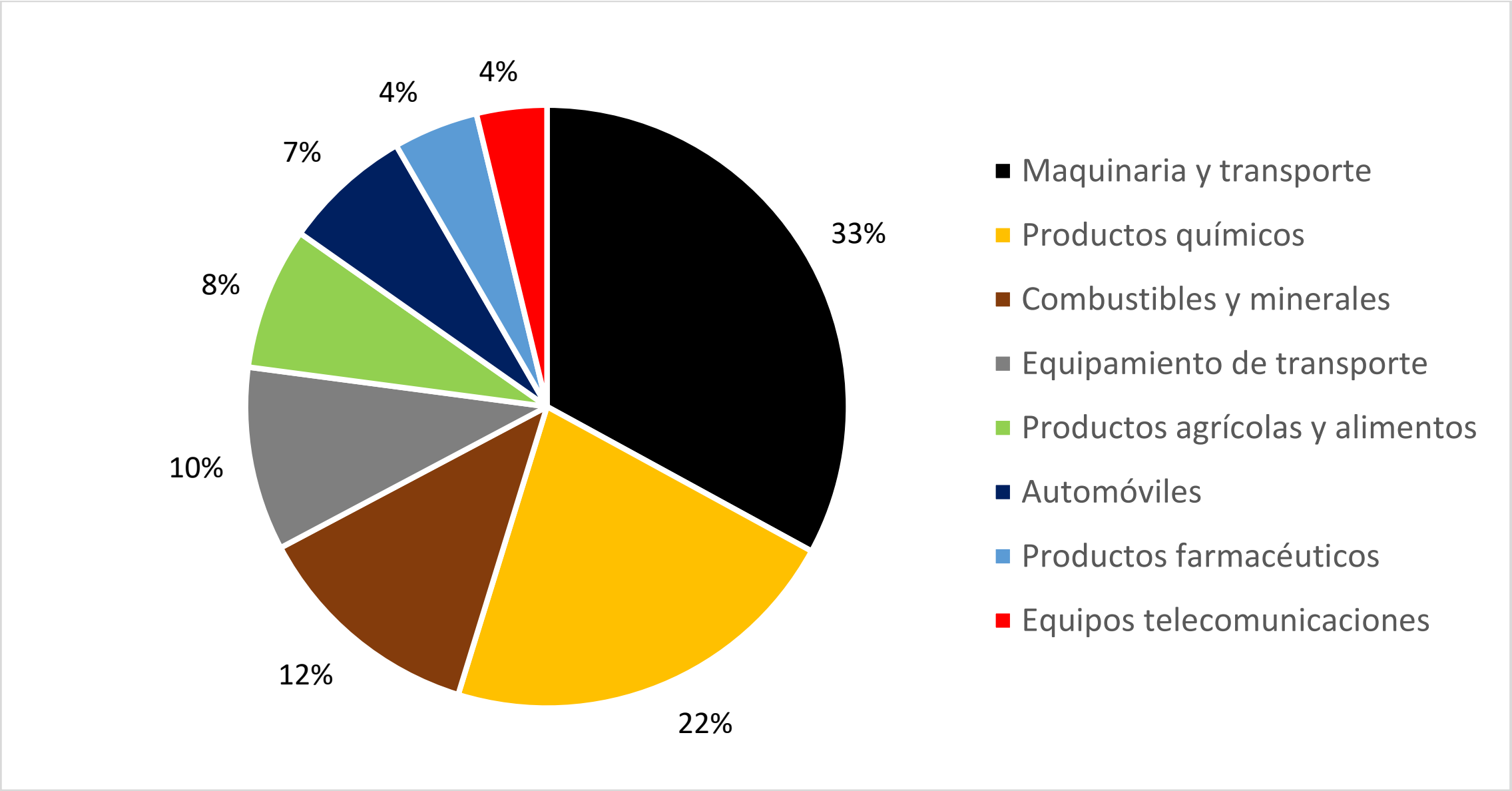

Fuente: elaboración propia a partir de (WTO, 2018)

## 7.4. Inserción productiva.

Con respecto a la Inversión Extranjera Directa (IED), debemos recordar los grandes flujos que recibe Brasil de inversión extranjera y que determinan el déficit de la cuenta financiera de la balanza de pagos. Desde que el país alcanzara el equilibrio de precios tras el Plan Real, la IED recibida se ha mantenido entre el 1 y el 5% del PIB, con una media del 2,7% entre 1994 y 2017.

En 2017, la IED se situó en un 3% del PIB, la segunda mejor cifra de los BRICS y de Sudamérica. Sin embargo, aunque no cabe duda de que la Inversión Extranjera permite el desarrollo del tejido productivo del país, lo que repercute positivamente en la economía brasileña, Brasil presenta una gran dependencia de la IED. También en 2017, fue el país donde la IED representó un mayor porcentaje sobre la inversión total del país, cercana a un 20%. Desde el año 2000, en los mejores momentos de la economía brasileña y con grandes incentivos a la inversión, esta tasa se ha mantenido en una media del 16%. Como resultado de esta dependencia, hay que tener en cuenta los beneficios de la IED en Brasil se repatrian, lo que provoca una fuga de capital constante hacia el extranjero. Así, la balanza de pagos brasilera se resiente tanto por la pérdida de los rendimientos de las inversiones (cuenta corriente) como por la continuidad de la necesidad de financiación externa, ya que si esos beneficios se mantuvieran en el país podrían ser reinvertidos domésticamente, reduciendo esa necesidad externa.

**Gráfica 43.** Dependencia de la Inversión Extranjera Directa, 2016-2017

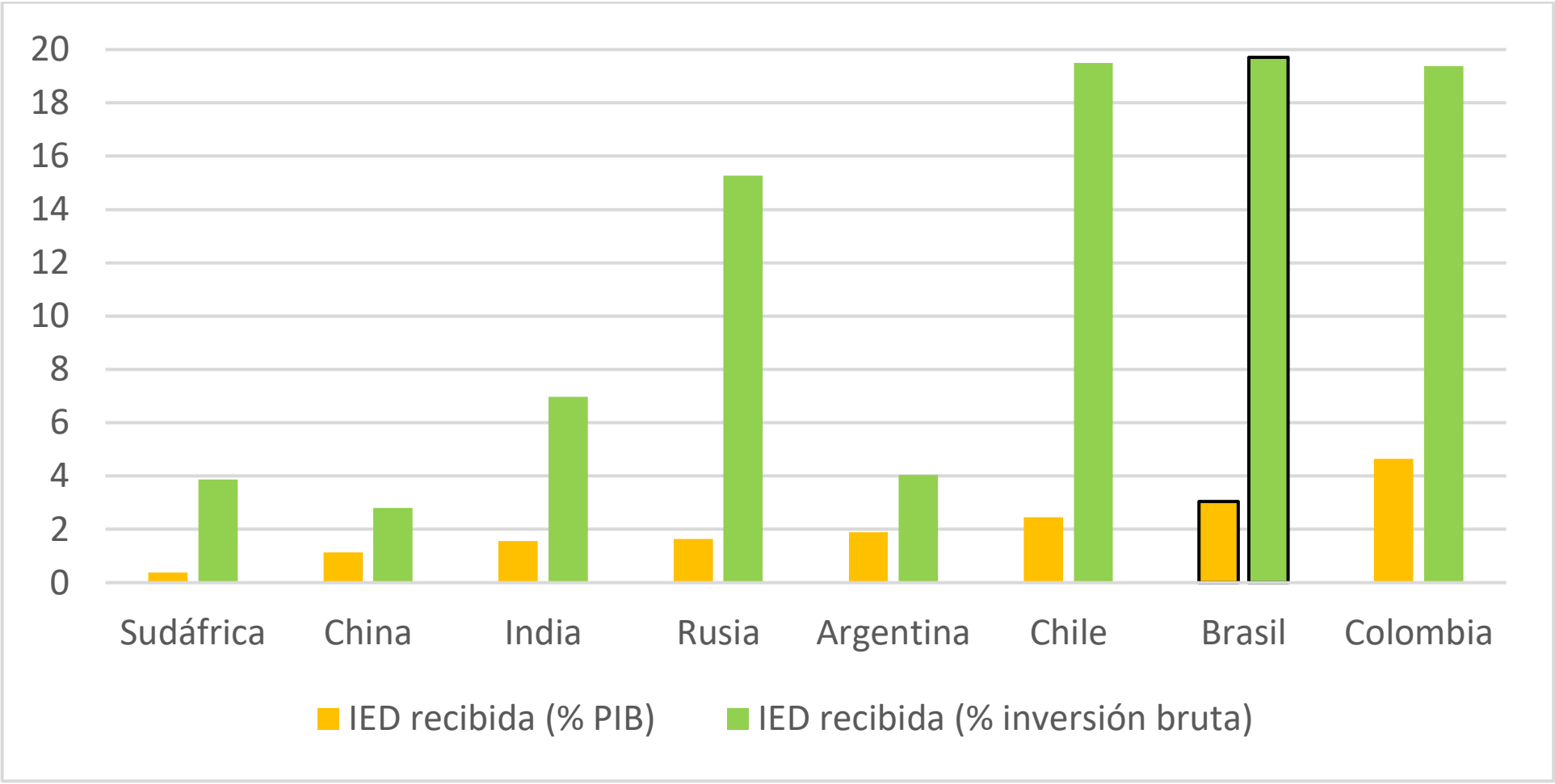

Fuente: elaboración propia a partir de (UNCTAD, 2018)

## 8. Análisis de la evolución de las condiciones de vida.

Para concluir, habiendo visto ya varias dimensiones de las condiciones de vida de las brasileñas y brasileños a lo largo de todo el informe, sólo queda ya recalcar la gran desigualdad que existe en el país, tanto en el ámbito urbano donde millones de persones viven en condiciones insalubres en las llamadas *favelas* como en el rural, donde unos pocos propietarios concentran la mayor parte de la tierra cultivable.

Si bien esas desigualdades entre propietarios y obreros son indiscutibles, también lo son los avances que ha experimentado el pueblo brasilero en los últimos años. Desde 2003, con un gobierno que tenía como uno de sus objetivos principales la reducción de la pobreza y del hambre, el mayor éxito ha sido el programa público Bolsa Família. Como vimos anteriormente, se trata de un ambicioso plan de transferencias públicas directas a las familias con menos recursos, siempre y cuando escolarizaran a sus hijos y les vacunaran.

Como podemos ver en la Gráfica 44, este programa ha supuesto un gran desembolso para el Estado brasileño, unos 145 mil millones de reales (39 mil millones de dólares) desde su inicio en 2004. El éxito en reducción de la pobreza es evidente: hasta 15 millones de familias lo han recibido en un año, y la población debajo del umbral de la pobreza ha bajado en 35 millones de personas (también por otros factores como la subida del salario mínimo y la mejora del empleo).

**Gráfica 44.** El impacto del programa Bolsa Família.

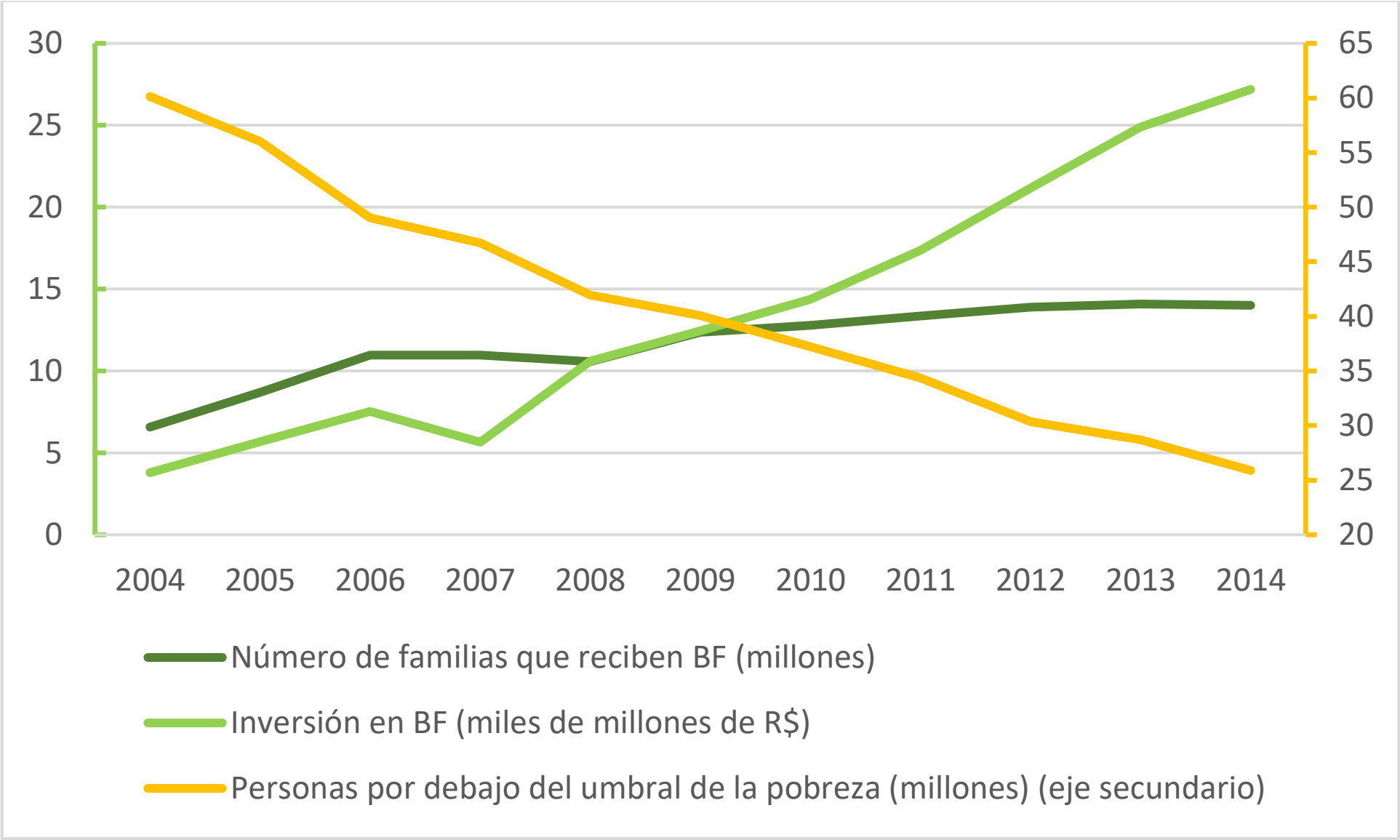

Fuente: elaboración propia a partir de (IPEA, 2018; MDS, 2018)

Pese a los éxitos, queda mucho por hacer. La desigualdad no sólo se refleja en la renta, afecta a todos los aspectos de la vida de los brasileños. En la actualidad, cerca del 20% de la población no tiene acceso a un saneamiento adecuado, el 12% de la población rural no tiene acceso ni siquiera a un baño, el 5% de los niños siguen sin escolarizar y todavía se producen muertes por falta de acceso a fuentes de agua. Además, es el segundo país del grupo con más tasa de analfabetismo, sólo por detrás de la India.

**Gráfica 45.** Tasa de analfabetismo (%), último año disponible 2014-2017.

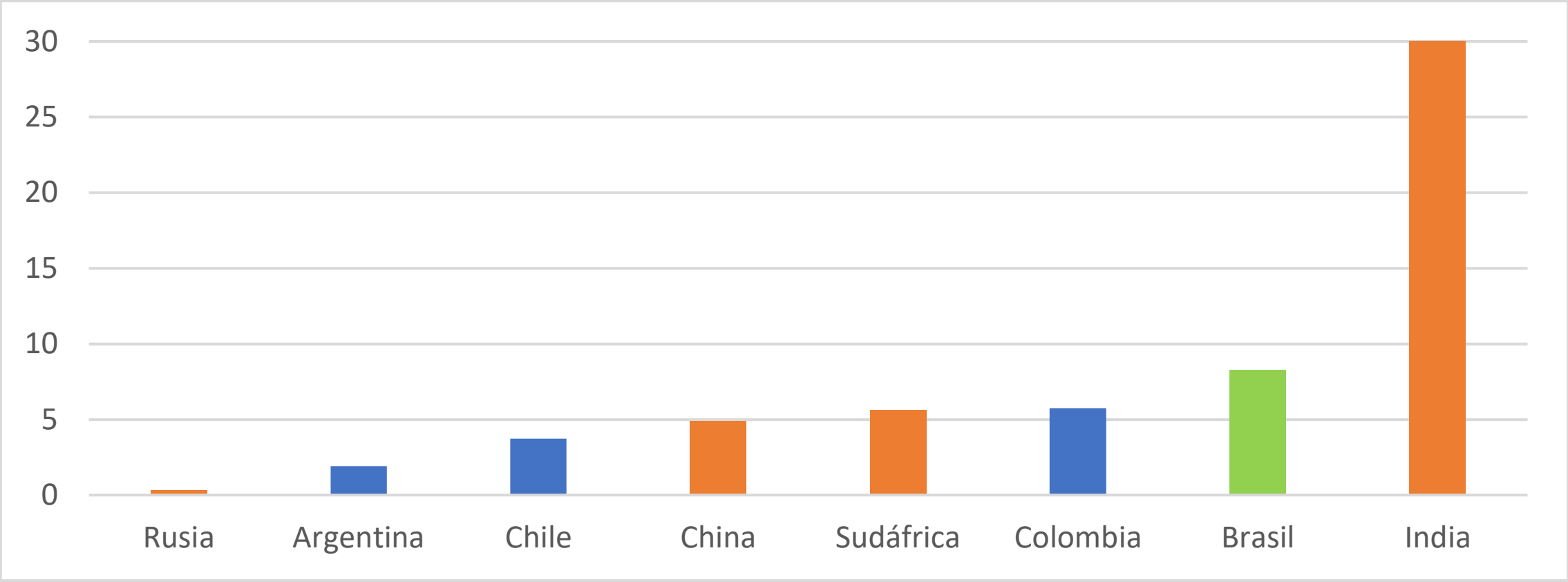

Fuente: elaboración propia a partir de (WB, 2018)

## 9. Conclusiones

A lo largo de este informe hemos visto la evolución de Brasil durante el último medio siglo. Se ha analizado su crecimiento demográfico y su evolución política reciente, de un gobierno progresista imputado por corrupción al futuro gobierno del ultraderechista Jair Bolsonaro. Hemos visto cómo han evolucionado sus dotaciones naturales, su progresiva independencia energética, la evolución de su medio agrario hacia una agricultura intensiva enfocada a la exportación y la necesidad de una reforma agraria que reduzca la desigualdad rural y que proteja el medio ambiente y el Amazonas de la amenaza del capital.

También se ha analizado la progresiva terciarización de la economía brasileña, y su preocupante estancamiento de la productividad de sus factores, tanto del capital como del trabajo, que hace que sus productos sean poco competitivos internacionalmente y sus empresas se hayan dedicado principalmente a abastecer el mercado interno más que a la exportación. Además, Brasil es uno de los países donde es más complicado montar un negocio, por el complejo entramado institucional y fiscal necesario para poder establecerlo.

A su vez, hemos visto la evolución de la distribución de la demanda del país, dado que ha aumentado tanto el gasto público como la demanda privada (por el aumento de los salarios y la salida de la pobreza de millones de personas), mientras que la inversión cada vez supone un menor porcentaje de la demanda, lo que profundiza el estancamiento de la productividad.

De estos sectores, destaca el sector público, que controla alguna de las grandes empresas clave del país: Petrobras, Eletrobras, Banco do Brasil, etc. Sin embargo, el gobierno acumula una deuda pública cada vez mayor, y mantiene un déficit público estructural debido a la baja recaudación y a un gasto público afectado por la complejidad institucional de la República Federal. El gobierno ha tenido éxito en algunas de sus políticas sociales, como la Bolsa Família, y la actuación del Banco Central desde los años 90 ha conseguido frenar una inflación que llegó al 3.000% anual. Con respecto al sector bancario, cabe decir que está muy concentrado, y un puñado de empresas controlan la inmensa mayoría del mercado.

Finalmente, se ha analizado la inserción externa de la economía brasileña. Las exportaciones del país han aumentado en los últimos años, pero sigue siendo un importador neto de servicios, y la fuga de capitales lastra estructuralmente la balanza por cuenta corriente. Así, Brasil se ha mantenido en una situación de necesidad de financiación externa, con un déficit en la cuenta financiera que se ha ido reduciendo en los últimos años y que permite pensar en la posibilidad de que el país alcance un saldo financiero positivo si el peso de la Inversión Extranjera se reduce.

Brasil se encuentra en una posición muy prometedora, con gran margen de mejora en las condiciones de vida de sus ciudadanos. Un país que es medio continente, el quinto estado más extenso del mundo, merece una mejor posición en los rankings económicos mundiales, y poder contestar a medio plazo el poder hegemónico de Estados Unidos en Sudamérica.

Para poder ocupar la posición geopolítica a la que Brasil aspira, es necesario que consolide su economía como motor indiscutible de Iberoamérica. Esto no lo podrá hacer si no se consigue controlar definitivamente la inflación sin que esto lastre otros aspectos de la economía, como hemos visto. Por ello, uno de los aspectos clave del futuro de Brasil será la actuación de su Banco Central.

Otro de los aspectos que es necesario mejorar para convertir al país carioca en el gigante que debería ser es la productividad. Tanto el trabajo como el capital han aumentado muy poco su productividad con respecto a las décadas anteriores, y el TFP ha bajado desde 1980. Para cambiar esta tendencia, Brasil ya está invirtiendo ingentes cantidades en I+D, pero se hace necesaria una modernización del sistema educativo, hasta ahora basado en conseguir sacar a los niños de las calles, para impartir una educación más técnica y profesional, para mejorar la productividad desde el momento en que los jóvenes entran al mercado de trabajo. Además, se debe facilitar el acceso de las empresas a financiación, ya que los tipos de interés de crédito hacen que sea imposible financiarse mediante el endeudamiento.

Finalmente, debemos destacar que en los últimos 15 años más de 30 millones de personas han salido de la pobreza absoluta en Brasil. No obstante, todavía más de 20 millones siguen estando en ella, y millones de personas se ven obligadas a vivir en *favelas*. Este es un claro problema de justicia que debe ser atendido correctamente por el nuevo gobierno brasilero: los programas sociales que han tenido éxito, como el *Bolsa Família,* no pueden ser retirados, sino que deben ser diversificados para atender a las distintas necesidades de las familias del país, e intensificado para poder llegar a esos 20 millones de personas que viven bajo el umbral de la pobreza a día de hoy. Para poder conseguir este objetivo sin aumentar el déficit fiscal, el gobierno federal debe conseguir reducir el déficit de los gobiernos estatales, además de aumentar la recaudación mediante una subida de los tipos impositivos máximos y un control eficaz del Ministerio de Fazenda contra el fraude fiscal y la corrupción. Sólo así se conseguirá tener un sistema tributario eficiente en la redistribución.

Todo ello, sin olvidarnos de la tragedia que ocurre diariamente en la selva amazónica. Diariamente, porque durante el período más intenso de deforestación se talaba cada día una superficie similar a la de Granada. Brasil tiene una gran responsabilidad: proteger el pulmón del mundo de los abusos del capital, haciendo que el crecimiento del país respete y proteja la naturaleza, y sea inclusivo para mejorar la vida de las personas. Este es el gran reto brasileño.

## Bibliografía

## Anexo Gráfico

**Figura 1.** Densidad de población en favelas de Brasil y en algunas de las ciudades más densamente pobladas del mundo (miles de habitantes por km$^2$), 2016-2017

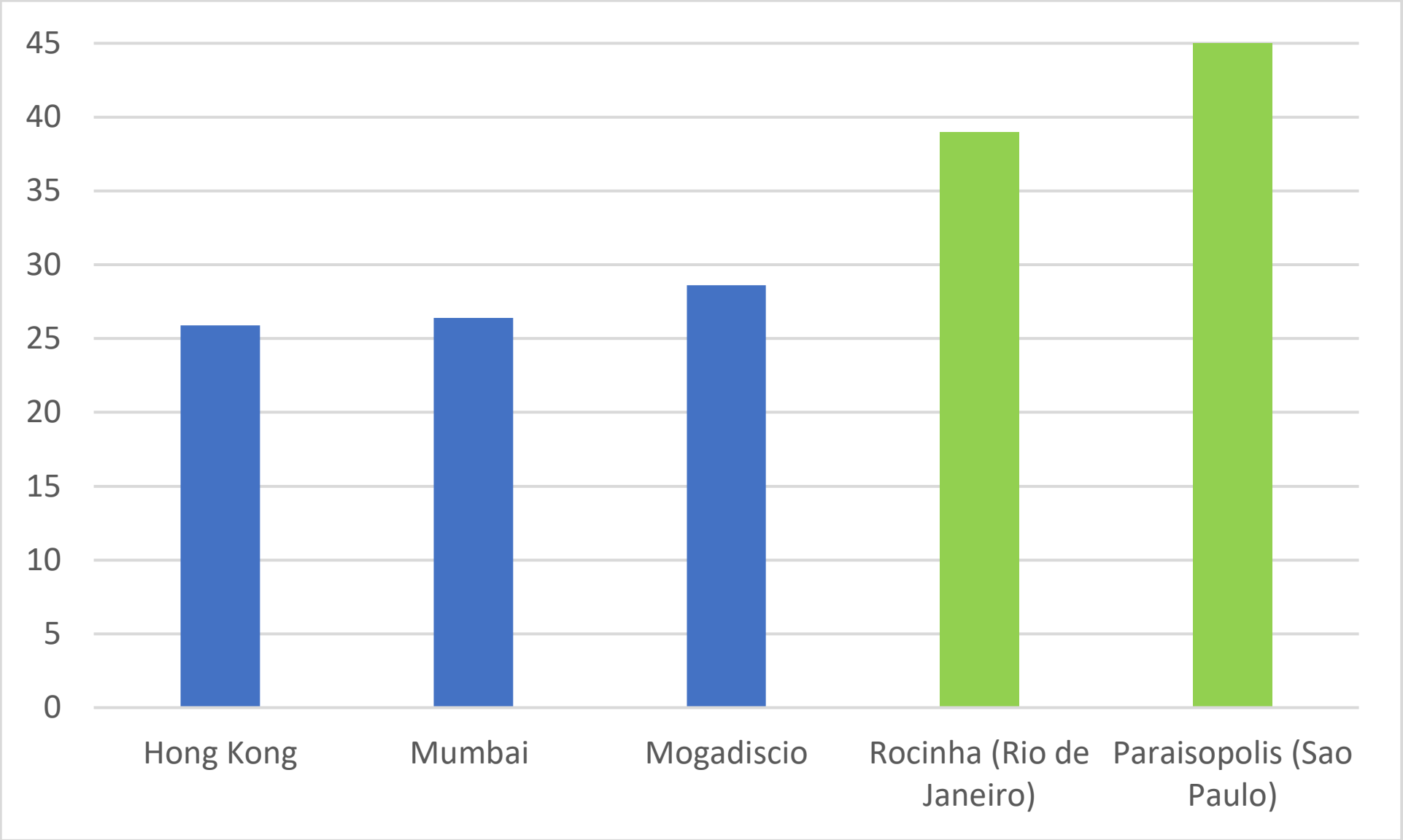

Fuente: elaboración propia a partir de (Demographia, 2018; IBGE, 2016)

**Figura 2.** Indicadores sociales durante los gobiernos de Brasil por partido

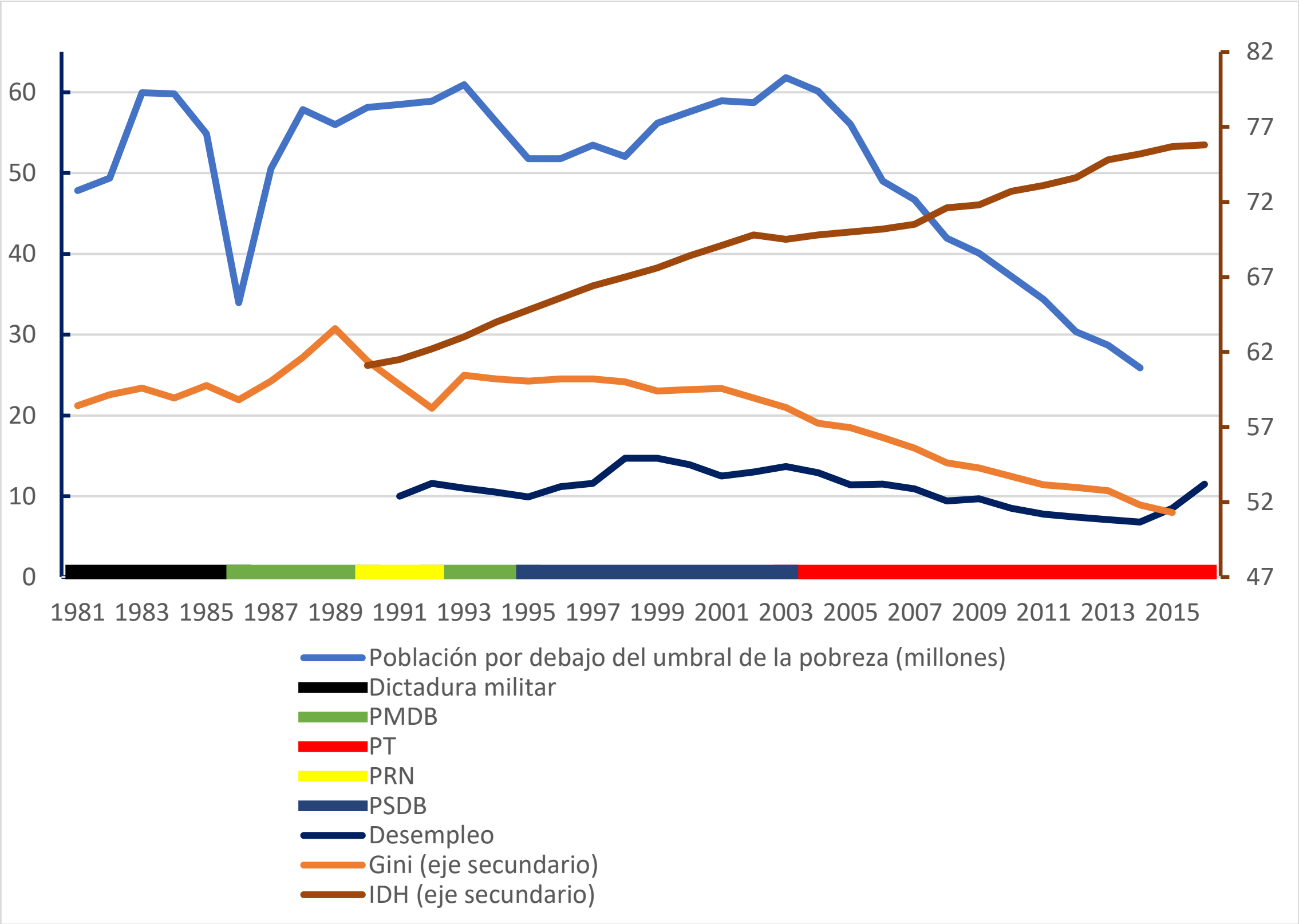

Fuente: elaboración propia a partir de (IPEA, 2018; UNDP, 2018; WB, 2018)

**Figura 3.** Movimiento de Trabajadores Rurales Sin Tierra (MST), 1986.

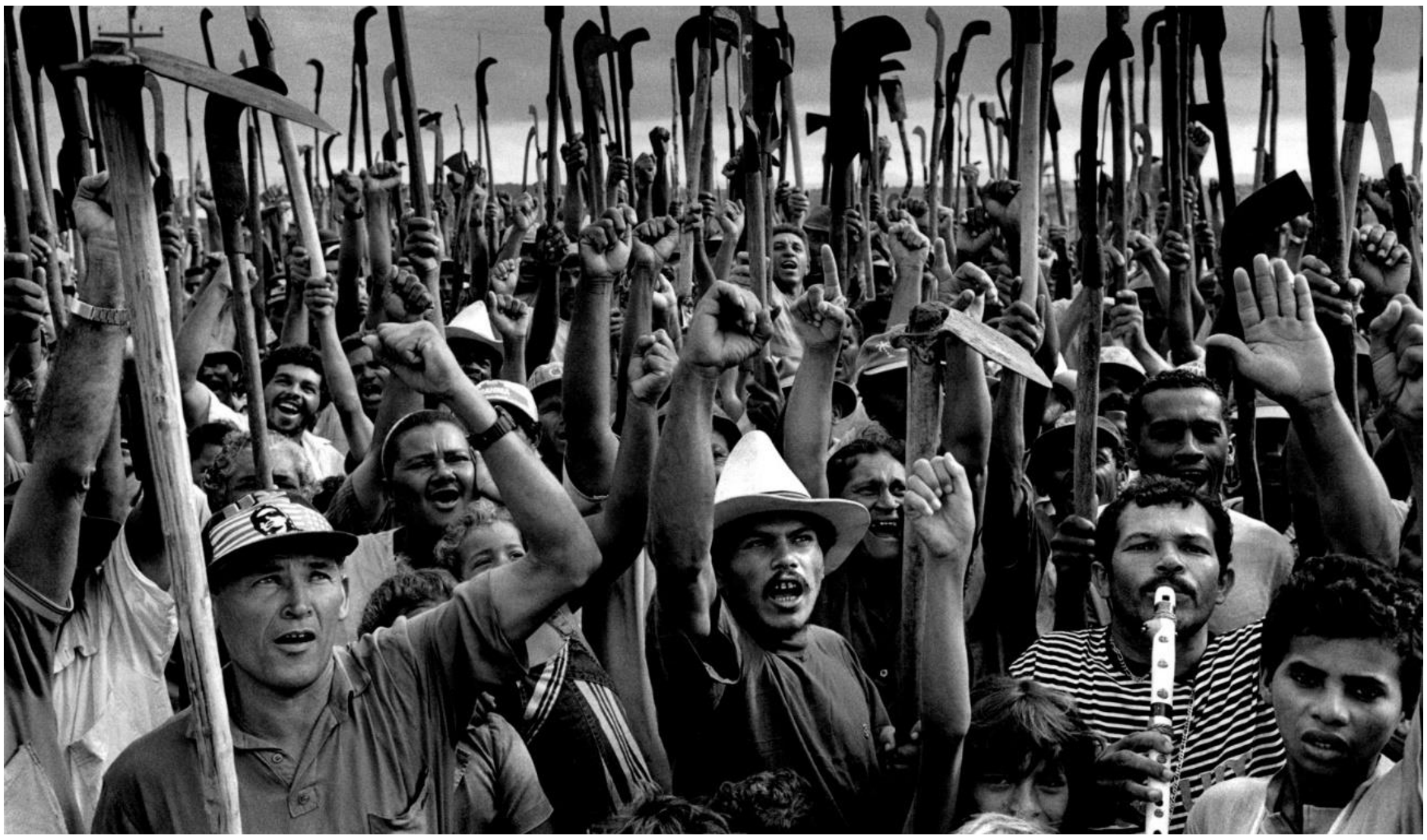

Fuente: fotografía de Sebastiao Salgado, en (Salgado & Wanick Salgado, 1986)

**Figura 4.** Consumo energético final, 2017.

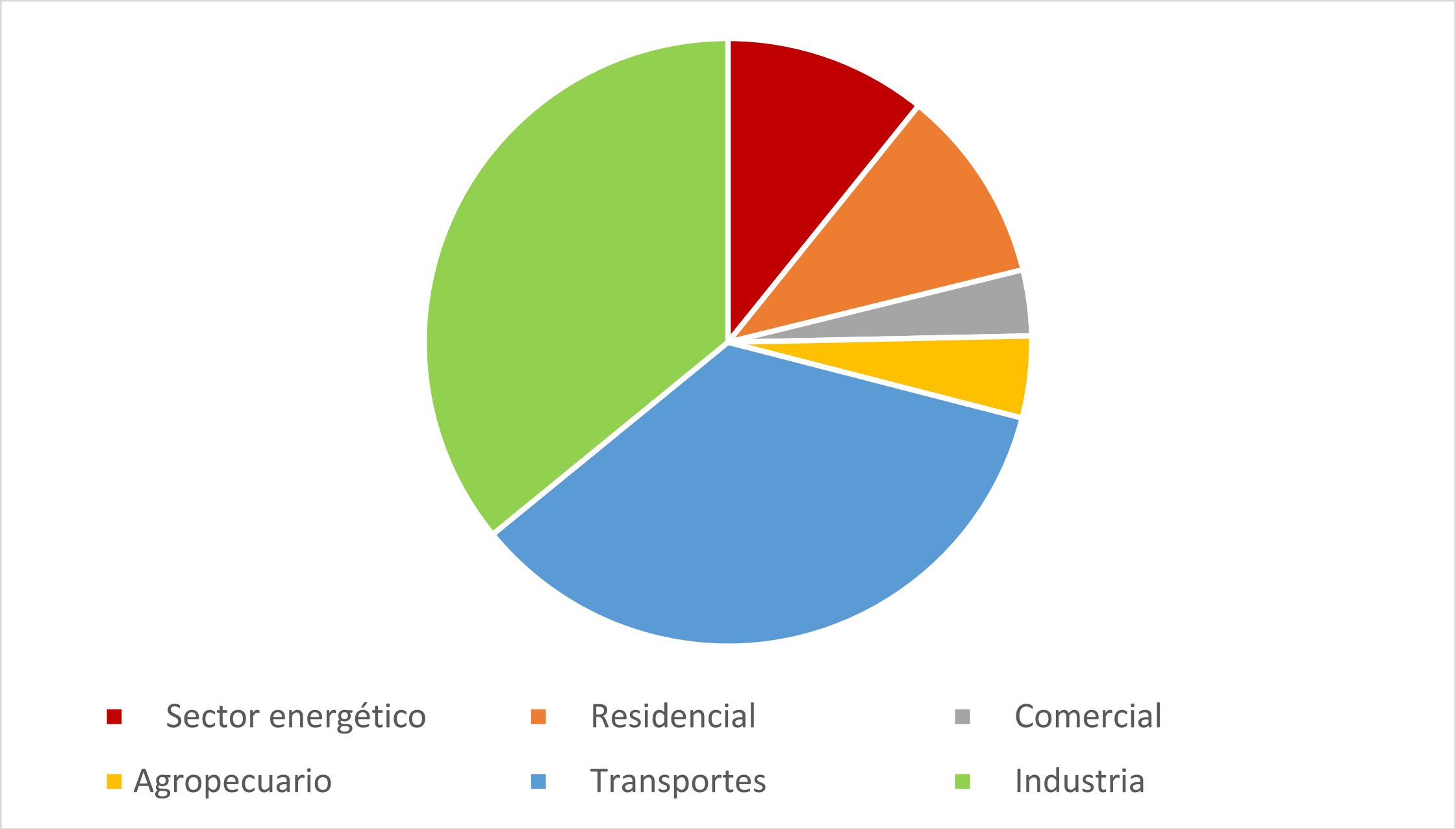

Fuente: elaboración propia a partir de (MME, 2018)

**Figura 5.** Deforestación de la Amazonia, 1988-2017. En verde, la selva actual; en rojo, el terreno talado y convertido, principalmente, en cultivos. La línea negra marca la antigua extensión (legal) de la selva amazónica.

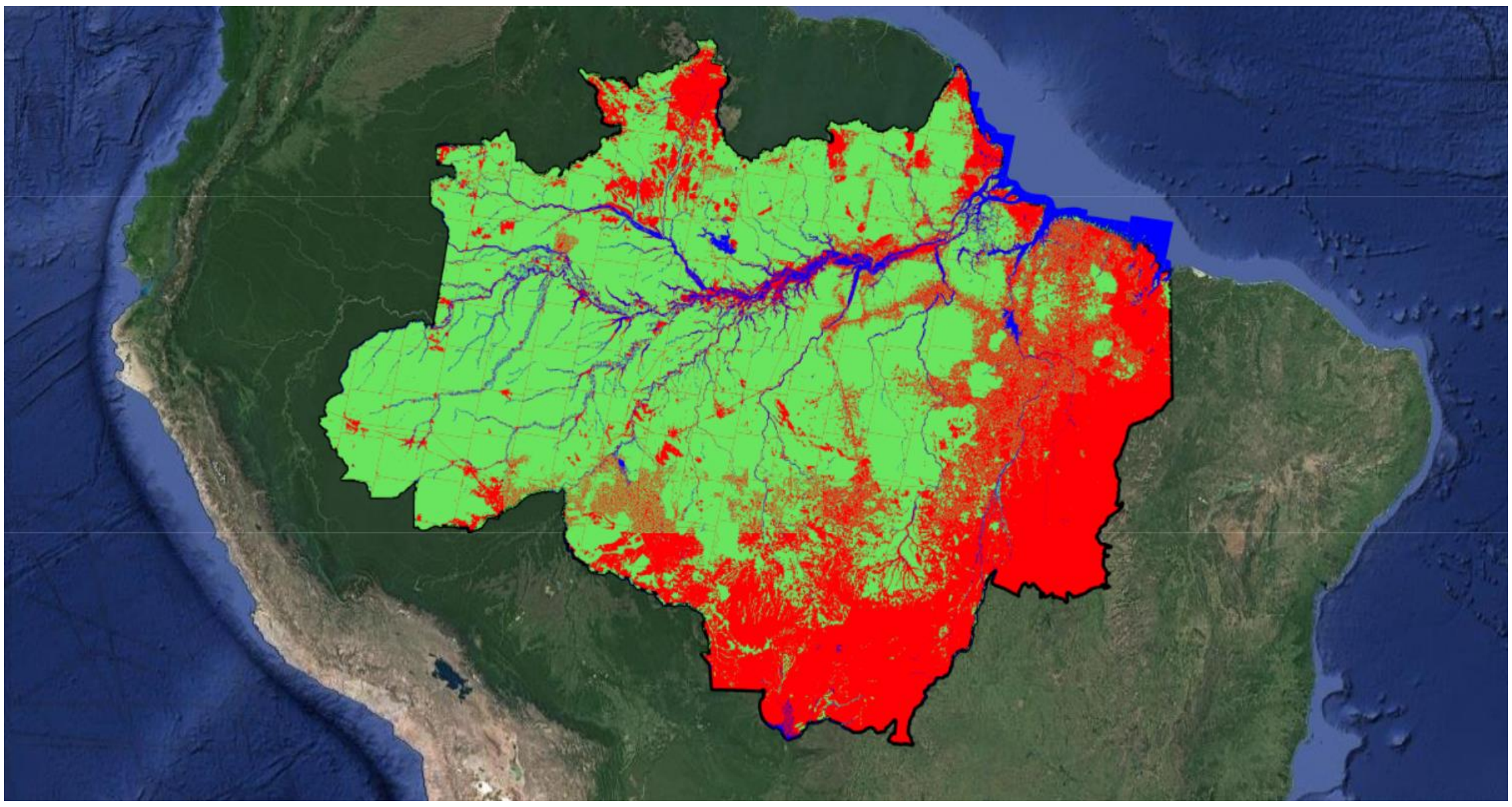

Fuente: (OBT, 2018)

**Figura 6.** Explotación aurífera de Serra Pelada, en 1986.

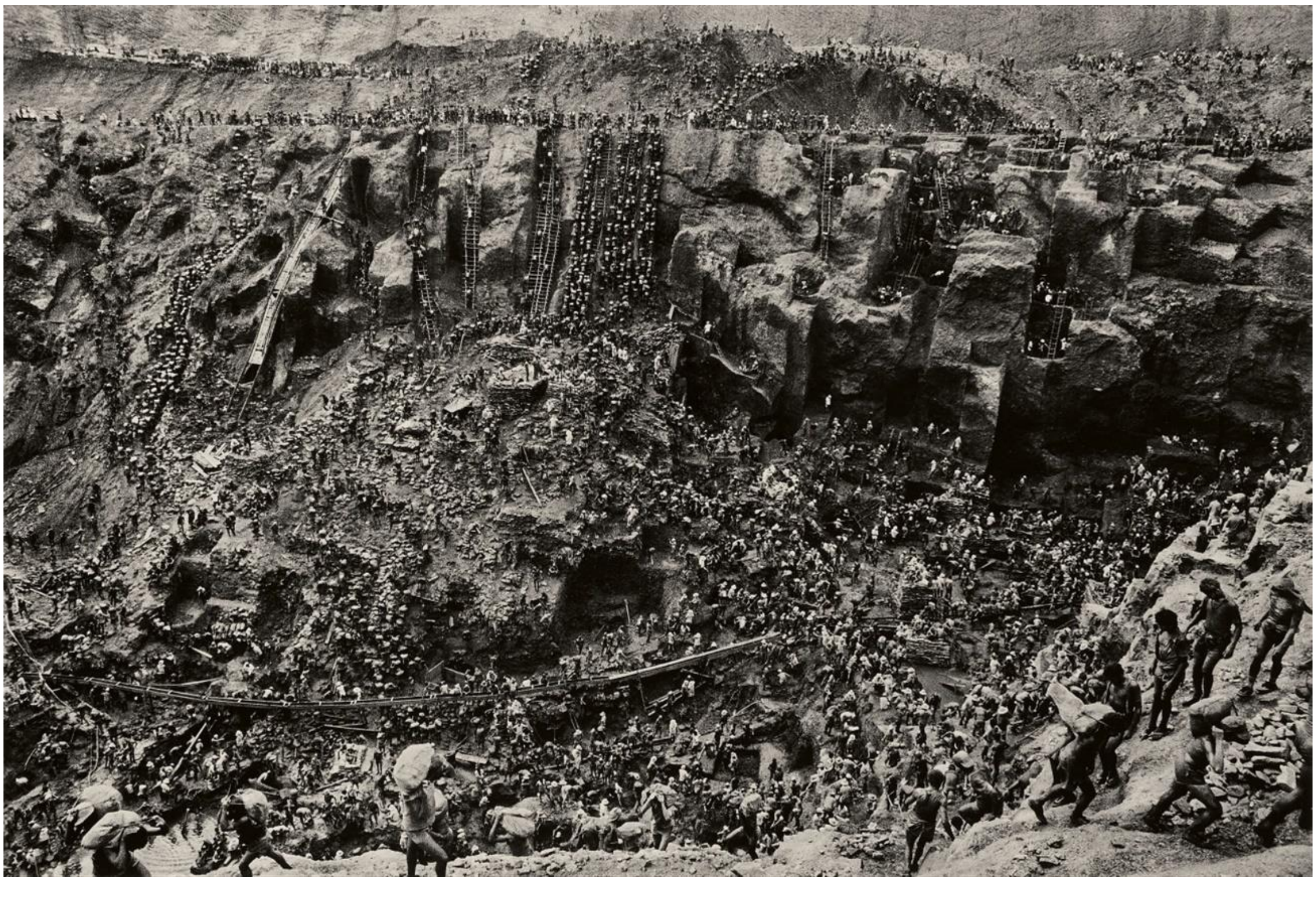

Fuente: fotografía de Sebastiao Salgado, en (Salgado & Wanick Salgado, 1986)

**Figura 7**. Gasto total en investigación y desarrollo (% PIB), último año disponible 2015-2017.

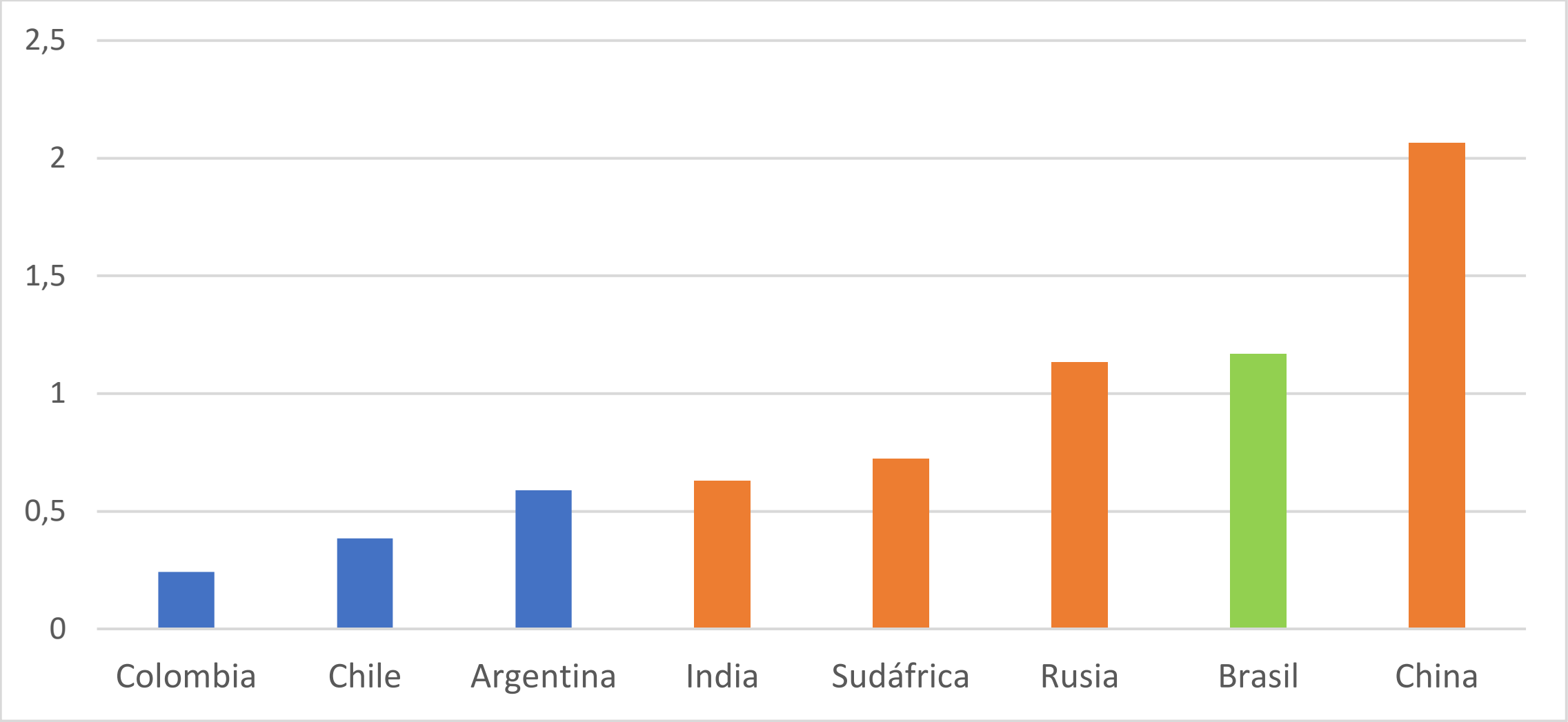

Fuente: elaboración propia a partir de (WB, 2018)

**Figura 8.** Evolución reciente de la productividad laboral y del salario medio

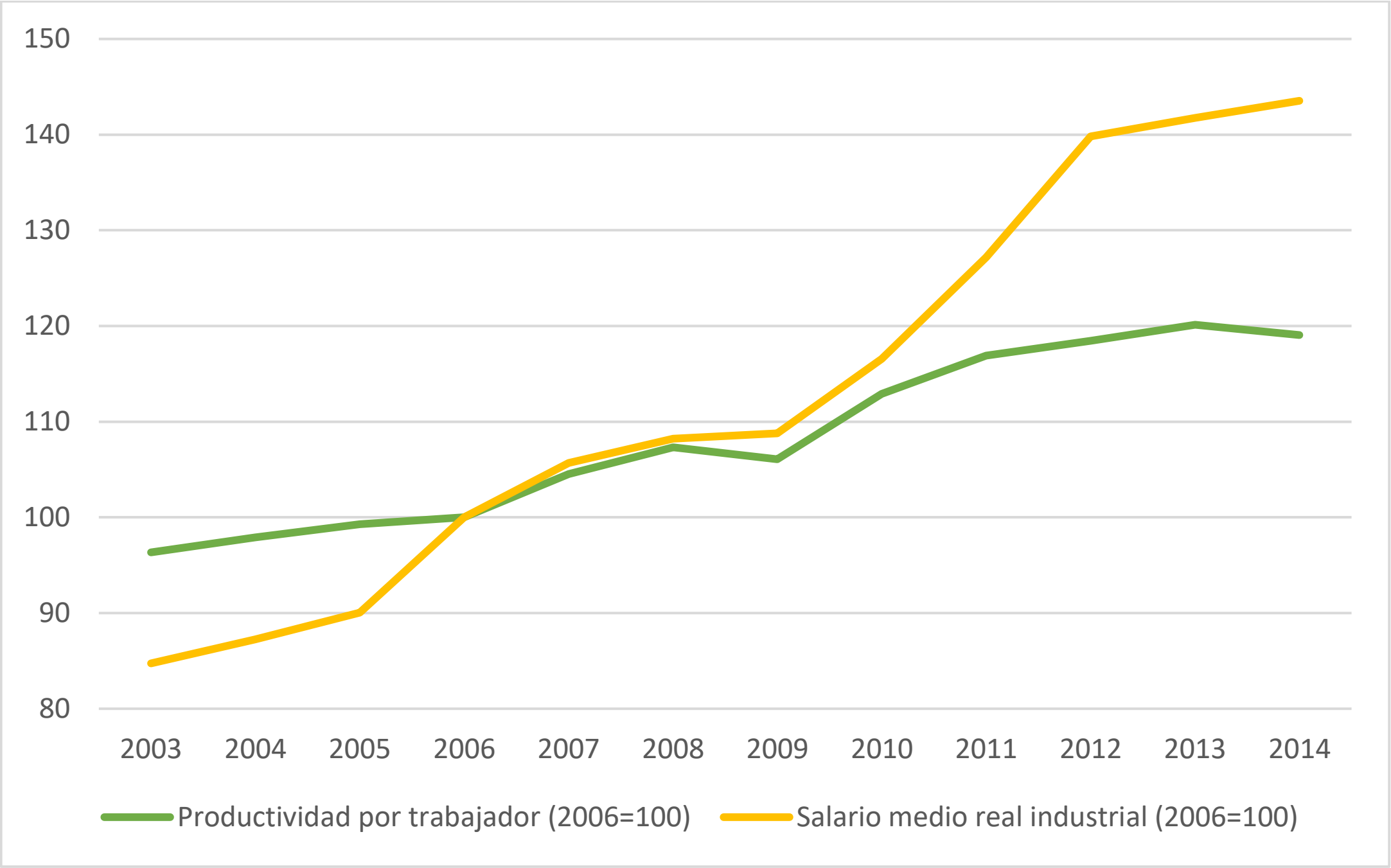

Fuente: elaboración propia a partir de (ILO, 2018b; IPEA, 2018)